\documentclass[preprint,showpacs,amsmath,amssymb,aps,prd,nofootinbib]{revtex4}
\usepackage{epsfig,graphicx,color,appendix}
\begin{document}
\begin{flushright}
KIAS-P14059
\end{flushright}
\def\CP{{\it CP}~}
\def\cp{{\it CP}}
\title{\mbox{}\\[10pt]
Flavored Peccei-Quinn symmetry}

\author{Y. H. Ahn}
\email{yhahn@kias.re.kr}
\affiliation{ School of Physics, KIAS, Seoul 130-722, Korea}




\begin{abstract}
\noindent In an attempt to uncover any underlying physics in the standard model (SM),
we suggest a $\mu$--$\tau$ power law in the lepton sector, such that relatively large 13 mixing angle with bi-large ones can be derived.
On the basis of this, we propose a neat and economical model for both the fermion mass hierarchy problem of the SM and a solution to the strong CP problem, in a way that no domain wall problem occurs, based on $A_{4}\times U(1)_{X}$ symmetry in a supersymmetric framework. Here we refer to the global $U(1)_X$ symmetry that can explain the above problems as ``flavored Peccei-Quinn symmetry". In the model, a direct coupling of the SM gauge singlet flavon fields responsible for spontaneous symmetry breaking to ordinary quarks and leptons, both of which are charged under $U(1)_X$, comes to pass through Yukawa interactions, and all vacuum expectation values breaking the symmetries are connected each other. So, the scale of Peccei-Quinn symmetry breaking is shown to be roughly located around $10^{12}$ GeV section through its connection to the fermion masses.

The model predictions are shown to lie on the testable regions in the very near future through on-going experiments for neutrino oscillation, neutrinoless double beta decay and axion.
We examine the model predictions, arisen from the $\mu$--$\tau$ power law, on leptonic $CP$ violation, neutrinoless double beta decay and atmospheric mixing angle, and show that the fermion mass and mixing hierarchies are in good agreement with the present data. Interestingly, we show the model predictions on the axion mass $m_a\simeq2.53\times10^{-5}$ eV and the axion coupling to photon $g_{a\gamma\gamma}\simeq1.33\times10^{-15}~{\rm GeV}^{-1}$. And subsequently the square of the ratio between them is shown to be 1 or 2 orders of magnitude lower than that of the conventional axion model.

\end{abstract}

\maketitle %
\section{Introduction}
The standard model (SM) of particle physics has been successful in describing phenomena until now, but it suffers from some problems which have not been solved yet, among which are the following:
the fine-tuning of the cosmological constant, the gauge hierarchy problem, the candidate for dark matter, the baryon asymmetry of the Universe, and the flavor puzzle associated with the fermion mass matrices and the strong charge parity (CP) problem. Surely the most pressing among them are the first and second problem. The gauge hierarch problem is solved if we introduce the supersymmetry (SUSY) which is the symmetry with respect to the replacement of bosons with fermions. All of the latter threes may be solved economically by implementing the seesaw mechanism~\cite{Minkowski:1977sc} for neutrino masses and Froggatt and Nielsen mechanism~\cite{Froggatt:1978nt} for quark mixing angles and masses. Various solutions to these problems have been proposed, inevitably leading to physics beyond the SM~\footnote{There is a recent summary on flavor puzzles in Ref.~\cite{Xing:2014sja}
.}.
The most elegant solution for the strong CP problem was proposed by Peccei and Quinn (PQ)~\cite{Peccei-Quinn}. When the PQ symmetry is broken spontaneously, a pseudo-Nambu-Goldstone boson appears, which is called an axion~\cite{Peccei-Quinn, axion}. The PQ mechanism has been invented to account for the small value of the QCD vacuum angle that is required to explain the observed bounds on the neutron electric dipole moment~\cite{Beringer:1900zz}. And its resulting axion is a strongly motivated particle candidate as dark matter.

In the absence of a fundamental theory, one has to adopt a model independent approach and search for symmetries that may explain the mixing pattern which in turn can shed light on the nature of fundamental theory for quarks and leptons. Flavor symmetry provides a promising framework for generating viable quark and lepton masses and mixings. Indeed implementing the see-saw mechanism with non-Abelian discrete symmetries~\cite{nonAbelian,Altarelli:2010gt} has been shown to lead quite naturally to ``near tribimaximal" neutrino mixing~\cite{Ahn:2012cg}, while Froggatt and Nielsen mechanism has been suggested for a hierarchical structure~\footnote{In Ref.~\cite{extraU(1)}
 authors described acceptable quark and lepton mass matrices based on anomalous $U(1)$ symmetry in a supersymmetric standard model.}. This fact has motivated an interest in non-Abelian finite groups with an Abelian flavor $U(1)$ symmetry as means to depict the flavor structure of leptons and quarks.
Since such discrete or continuous global symmetry is protected against violations by quantum gravity effects~\cite{Krauss:1988zc}, one can assume that this symmetry originates in a continuous gauge symmetry which is spontaneously broken.

In this work, we speculate on possible origin of the quark and lepton spectra that masses of successive particles increase by large factors by the introduction of global $U(1)_{X}$ symmetry with non-Abelian discrete $A_4$ symmetry~\footnote{E.Ma and G.Rajasekaran~\cite{Ma:2001dn} have introduced for the first time the $A_{4}$ symmetry to avoid the mass degeneracy of $\mu$ and $\tau$ under a $\mu$--$\tau$ symmetry~\cite{mutau}.}.
Moreover, we wish to discuss an automatic theory for strong CP invariance by the $U(1)_{X}$ symmetry which is anomalous in Lagrangian, like the PQ symmetry. So we will refer this $U(1)_X$ symmetry as ``flavored-PQ symmetry". We stress that the flavored-PQ symmetry $U(1)_{X}$ be better to be embedded in the non-Abelian $A_4$ finite group. First, the $U(1)_X$ symmetry is natural in that it is a part of a flavor symmetry, which explains the mass hierarchy of quarks and leptons. So the choice of $X$-quantum numbers could be in some sense unique. Second, the scale of PQ symmetry breaking can be coincident with that of $A_4$ symmetry breaking. Third, the $U(1)_X$ symmetry provides a neat and economical solution to the strong CP problem and its resulting axion. Fourth, the $U(1)_X$ symmetry introduced can remove the axionic domain wall problem if it is composed of two anomalous $U(1)$ symmetries~\cite{Barr:1982uj}. Thus we have a good motivation for considering the flavor-axion model in the framework of SUSY.

The goal of this work is to construct a minimalistic supersymmetric model based on $A_{4}\times U(1)_{X}$ symmetry with the following features:
\begin{description}
  \item[(i)] All the hat Yukawa couplings appearing in superpotential are complex numbers and of order unity. The right-handed Majorana neutrino and the top quark terms are only renormalizable, while non-renormalizable terms appear with successive powers of the flavon fields ${\cal F}_A=\Phi,\Theta,\Psi$ according to appropriate $A_{4}\times U(1)_{X}$ symmetry. Here the $U(1)_{X}$ symmetry (simultaneously, $A_{4}$ symmetry as well) is broken spontaneously by SM gauge singlet flavon field ${\cal F}_A$ which acquires vacuum expectation value (VEV) below a cutoff scale $\Lambda$ which corresponds to a mass of messenger field. By integrating out all heavy messenger fields, all effective Yukawa couplings become hierarchical, and the $U(1)_X$ charge assignments make them correspond to the measured fermion mass hierarchies.
  \item[(ii)] The $U(1)_X$ symmetry, which is responsible for both the fermion mass hierarchy of the SM and vacuum configuration, is composed of two anomalous $U(1)_X\equiv U(1)_{X_1}\times U(1)_{X_2}$ symmetries which are generated by the charges $X_1$ and $X_2$. When flavon fields ${\cal F}_A$ acquire VEVs, both lepton number $U(1)_L$ and $U(1)_{\rm PQ}$ appear to be broken. Actually, there are linear combinations of the two $U(1)_{X_i}$ symmetries, which are $U(1)_{\tilde{X}}\times U(1)_f$. Here the $U(1)_{\tilde{X}}$ symmetry has anomaly, while the $U(1)_f$, which corresponds to lepton number, is anomaly-free. Then the right-handed neutrinos acquire Majorana masses when $U(1)_{f}$ symmetry is broken with its breaking scale.
  \item[(iii)] Even though the flavon fields ${\cal F}_{A}$ are the SM gauge singlets, a direct coupling of ${\cal F}_{A}$ to the quarks and leptons is possible through Yukawa couplings. So, the $U(1)_{X}$ symmetry plays a role in the solution to the strong CP problem leading to the existence of a light axion. The mass scale of the $U(1)_{X}$ breaking is equivalent to the one of $A_{4}$ symmetry breaking. Thus, $\langle{\cal F}_A\rangle\neq0$ leads to $U(1)_{X}$ violation. All VEVs breaking the symmetries are connected each other. After the $X$-symmetry is broken spontaneously, axion $A$ appears as a pseudo-Nambu-Goldstone boson of the $X$-symmetry. Accordingly, the mass of the axion is given by $m_{A}\simeq m_{\pi}f_{\pi}/\langle{\cal F}_A\rangle$ with its decay constant $\langle{\cal F}_A\rangle\sim10^{12}$ GeV. Interestingly, the axion decay constant is constrained by its connection to the fermion masses, see Eqs.~(\ref{scaleLambda}-\ref{MassRangge}) and Eqs.~(\ref{AhnMass}-\ref{fA1}).
  \item[(iv)] The flavored-PQ symmetry $U(1)_{X}$ is spontaneously broken at a scale much higher than the electroweak scale. And the explicit breaking of the $U(1)_{\tilde{X}}$ by the chiral anomaly effect further breaks it down to $Z_{N}$ discrete symmetry, where $N$ is the color anomaly number. At the QCD phase transition, the $Z_{N}$ symmetry is spontaneous broken, and which gives rise to a domain wall problem~\cite{Sikivie:1982qv}.
Such domain wall problem can be overcome by the two anomalous axial $U(1)$ symmetries, $U(1)_{X_1}\times U(1)_{X_2}$, when $N_1$ and $N_2$ are relative prime~\cite{Barr:1982uj}.
\end{description}
The rest of this paper is organized as follows.
In section II we address a special pattern of lepton sector in a model independent way which follows a $\mu$--$\tau$ power law under which certain elements associated with the muon and tau flavors in mass matrices are distinguished. And furthermore we consider a renormalizable ultraviolet (UV) complete theory above a new physics scale where among the fermion operators only the heavy neutrino and top quark operators are renormalizable. We argue that this is a plausible way to depict leptonic mixing pattern.
In section III, according to the $\mu$--$\tau$ power law and the UV completion textures, we construct a minimalistic SUSY model for quarks and leptons based on $A_4\times U(1)_X$ symmetry. Here we show that the observed hierarchy in the masses and mixings of quarks and leptons, which is one of the most puzzling features of nature, can be obtained in a natural way. Especially, we show explicitly symmetry breaking scales, explore what values of the low energy CP phases can predict a value for the mass hierarchy of neutrino and investigate the observables that can be tested in the current and the next generation of experiments.
Since an observation of neutrinoless double beta ($0\nu\beta\beta$)-decay and a sufficiently accurate measurement of its half-life can provide information on lepton number violation, the Majorana vs.\ Dirac nature of neutrinos, and the neutrino mass scale and hierarchy, we show that our model is experimentally testable in the near future. In section IV, we study the higher order corrections in our framework and show that a direct extension to the lepton and quark sectors can lead, apart from negligible terms, to would-be nontrivial next leading contributions for Majorana neutrino and down-type quark mass matrices, both of which could be well controlled, so that both a light neutrino mass matrix can remain leading order term and the Cabibbo-Kobayashi-Maskawa (CKM) matrix is reproduced.
Section V is dedicated to the study of the strong CP invariance and its resulting axion. We demonstrate how the domain wall problem can be overcome and show model predictions on the axion mass and axion-photon coupling. We give our conclusions in section V.

\section{hint for a fundamental theory}
Let us address a special pattern of lepton sector as a hint for a fundamental theory.
In the weak eigenstate basis, the Yukawa interactions in both neutrino and charged lepton sectors and the charged gauge interaction can be written as
 \begin{eqnarray}
 -{\cal L} &=& \frac{1}{2}\overline{\nu_{L}}{\cal M}_{\nu}(\nu_{L})^c +\overline{\ell_{L}}{\cal M}_{\ell}\ell_{R}
  + \frac{g}{\sqrt{2}}W^{-}_{\mu} ~\overline{\ell_{L}}\gamma^{\mu}\nu_{L}  + {\rm h.c.} ~.
 \label{lagrangianA}
 \end{eqnarray}
In the charged lepton mass basis, {\it i.e.} ${\cal M}_{\ell}={\rm diag}(m_{e},m_{\mu},m_{\tau})$, the neutrino mass matrix has the form
  \begin{eqnarray}
 \mathcal{M}_{\nu}&\equiv&{\left(\begin{array}{ccc}
 m_{ee} & m_{e\mu} &  m_{e\tau} \\
 m_{e\mu} &  m_{\mu\mu} &  m_{\mu\tau} \\
 m_{e\tau} & m_{\mu\tau} &  m_{\tau\tau}
 \end{array}\right)}=U_{\nu}\mathcal{M}^{d}_{\nu}U^{T}_{\nu}~,
 \label{mnu1}
 \end{eqnarray}
where $\mathcal{M}^{d}_{\nu}={\rm diag}(m_{\nu_1},m_{\nu_2},m_{\nu_3})$.
Then in this mass eigenstates basis the Pontecorvo-Maki-Nakagawa (PMNS) leptonic mixing matrix~\cite{PDG} at low energies is visualized in the charged weak interaction terms : $U_{\rm PMNS}=U_{\nu}$. And in the standard parametrization of the leptonic mixing matrix  $U_{\rm PMNS}$, it is expressed in terms of three mixing angles, $\theta_{12}, \theta_{13}, \theta_{23}$, and three \cp-odd phases (one $\delta_{CP}$ for the Dirac neutrino and two $\varphi_{1,2}$ for the Majorana neutrino) as
 \begin{eqnarray}
  U_{\rm PMNS}=
  {\left(\begin{array}{ccc}
   c_{13}c_{12} & c_{13}s_{12} & s_{13}e^{-i\delta_{CP}} \\
   -c_{23}s_{12}-s_{23}c_{12}s_{13}e^{i\delta_{CP}} & c_{23}c_{12}-s_{23}s_{12}s_{13}e^{i\delta_{CP}} & s_{23}c_{13}  \\
   s_{23}s_{12}-c_{23}c_{12}s_{13}e^{i\delta_{CP}} & -s_{23}c_{12}-c_{23}s_{12}s_{13}e^{i\delta_{CP}} & c_{23}c_{13}
   \end{array}\right)}P_{\nu}~,
 \label{PMNS}
 \end{eqnarray}
where $s_{ij}\equiv \sin\theta_{ij}$, $c_{ij}\equiv \cos\theta_{ij}$ and $P_{\nu}$ is the phase matrix in which particles are Majorana ones.
\begin{table}[h]
\caption{\label{exp} The global fit of three-flavor oscillation parameters at the best-fit (BF) and $3\sigma$ level~\cite{Forero:2014bxa}. NO = normal neutrino mass ordering; IO = inverted mass ordering. And this ($^{\ast}$) is a local minimum in the first octant of $\theta_{23}$.}
\begin{ruledtabular}
\begin{tabular}{cccccccccccc} &$\theta_{13}[^{\circ}]$&$\delta_{CP}[^{\circ}]$&$\theta_{12}[^{\circ}]$&$\theta_{23}[^{\circ}]$&$\Delta m^{2}_{\rm Sol}[10^{-5}{\rm eV}^{2}]$&$\Delta m^{2}_{\rm Atm}[10^{-3}{\rm eV}^{2}]$\\
\hline
BF $\begin{array}{ll}
\hbox{NO}\\
\hbox{IO}
\end{array}$&$\begin{array}{ll}
8.80 \\
8.91
\end{array}$&$\begin{array}{ll}
241.2 \\
266.4
\end{array}$&$34.63$&$\begin{array}{ll}
48.85~(43.11)^\ast \\
49.20
\end{array}$&$7.60$
 &$\begin{array}{ll}
2.48 \\
2.38
\end{array}$ \\
\hline
$3\,\sigma$$\begin{array}{ll}
\hbox{NO}\\
\hbox{IO}
\end{array}$&$\begin{array}{ll}
7.65\rightarrow9.87 \\
7.77\rightarrow9.92
\end{array}$&$0\rightarrow360$&~$31.82\rightarrow37.76$&$\begin{array}{ll}
38.76\rightarrow53.31 \\
39.41\rightarrow53.13
\end{array}$
 &$7.11\rightarrow8.18$&$ \begin{array}{ll}
                           2.30\rightarrow2.65 \\
                           2.20\rightarrow2.54
                          \end{array}$\\
\end{tabular}
\end{ruledtabular}
\end{table}
The large values of the solar ($\theta_{12}$) and atmospheric ($\theta_{23}$) mixings as well as the non-zero but relatively large reactor mixing angle ($\theta_{13}$) are consequences of a nontrivial structure of the neutrino mass matrix $\mathcal{M}_{\nu}$ in the charged lepton basis, as indicated in Table~\ref{exp}. The very different structure of leptonic mixings compared to the quark ones for all possible neutrino mass orderings indicates an unexpected texture of the mass matrix and may provide important clues to our understanding of the physics of fundamental constituents of matter. Even nothing is known on the physics related to the leptonic CP violation, the measurements of non-vanishing 13 mixing, $\theta_{13}$, opens up the possibilities for searching for CP violation in neutrino oscillation experiments. It needs a new paradigm to explain the peculiar structure of lepton sector compared to the quark one.

After the relatively large reactor angle $\theta_{13}$ measured in Daya Bay~\cite{An:2012eh} and RENO~\cite{Ahn:2012nd} including Double Chooz, T2K and MINOS experiments~\cite{Other}, the recent analysis based on global fits~\cite{Capozzi:2013csa,GonzalezGarcia:2012sz,Forero:2014bxa} of the neutrino oscillations enters into a new phase of precise determination of mixing angles and mass squared differences, indicating that the tri-bimaximal mixing (TBM)~\cite{Harrison:2002er} for three flavors should be corrected in the lepton sector: especially, in the most recent analysis~\cite{Forero:2014bxa} their allowed ranges at $1\sigma$ best-fit $(3\sigma)$ from global fits are given by Table~\ref{exp},
where $\Delta m^{2}_{\rm Sol}\equiv m^{2}_{\nu_2}-m^{2}_{\nu_1}$, $\Delta m^{2}_{\rm Atm}\equiv m^{2}_{\nu_3}-m^{2}_{\nu_1}$ for the normal mass ordering (NO), and  $\Delta m^{2}_{\rm Atm}\equiv m^{2}_{\nu_1}-m^{2}_{\nu_3}$ for the inverted one (IO).

In the limit of reactor mixing angle $\theta_{13}\rightarrow0$ and atmospheric mixing angle $\theta_{23}\rightarrow45^{\circ}$, the neutrino mass matrix reflects the $\mu$--$\tau$ symmetric form: $m_{e\mu}=m_{e\tau}$ and $m_{\mu\mu}=m_{\tau\tau}$ in Eq.~(\ref{mnu1}).
In a basis where charged leptons are mass eigenstates, a simple way to address the $\mu$--$\tau$ symmetry~\cite{mutau} (interchange symmetry of the second and third generation of the leptonic fields; $m_{e\mu}=m_{e\tau}$ and $m_{\mu\mu}=m_{\tau\tau}$ in neutrino sector and $m_\mu=m_\tau$ in charged lepton sector) is to postulate that both the charged leptons and the neutrinos follow a $\mu$--$\tau$ symmetry:
 \begin{eqnarray}
 \mathcal{M}_{\ell}&=& {\left(\begin{array}{ccc}
 A_{\ell} & 0 & 0 \\
 0 & C_{\ell} & 0 \\
 0 & 0 & C_{\ell}
 \end{array}\right)}~,\qquad\mathcal{M}_{\nu}= {\left(\begin{array}{ccc}
 A_{\nu} & B_{\nu} & B_{\nu} \\
 B_{\nu} & C_{\nu} & D_{\nu} \\
 B_{\nu} & D_{\nu} & C_{\nu}
 \end{array}\right)}~.
 \label{MuTau}
 \end{eqnarray}
Surely the muon and tau lepton masses are so different~\cite{PDG}, as well as the 13 mixing angle has non-zero value~\cite{An:2012eh, Ahn:2012nd, Other}, that such a symmetry could therefore not be realized in nature.

In this work, we consider two ansatzs in order to describe the present and future lepton and quark sector. First, we consider that the elements of the neutrino and charged lepton mass matrices, in a basis where the charged lepton mass matrix is diagonal, follow a power law. According to this law, certain elements associated with the flavors $\mu$ and $\tau$ in both ${\cal M}_{\nu}$ and ${\cal M}_{\ell}$ are distinguished. We will call this the ``$\mu$--$\tau$ power'' on lepton masses.
Assigning the distinctions to each $\mu$ and $\tau$ flavor, the charged lepton and neutrino mass matrices are written as
 \begin{eqnarray}
 \mathcal{M}_{\ell}&=& {\left(\begin{array}{ccc}
 A_{\ell} & 0 & 0 \\
 0 & C_{\ell}x^{2}_{2} & 0 \\
 0 & 0 & C_{\ell}x^{2}_{3}
 \end{array}\right)}~,\qquad\mathcal{M}_{\nu}= {\left(\begin{array}{ccc}
 A_{\nu} & B_{\nu}y_{2} & B_{\nu}y_{3} \\
 B_{\nu}y_{2} & C_{\nu}y^{2}_{2} & D_{\nu}y_{3}y_{2} \\
 B_{\nu}y_{3} & D_{\nu}y_{3}y_{2} & C_{\nu}y^{2}_{3}
 \end{array}\right)}~,
 \label{MuTauPower}
 \end{eqnarray}
which presents that the $\mu$--$\tau$ symmetry is explicitly broken. It is clear from the above discussion in the limit of $y_{2,3}\rightarrow1$ and $x_{2,3}\rightarrow1$ exact $\mu$--$\tau$ symmetry is recovered.
The mass ratio between $m_{\mu}$ and $m_{\tau}$ can be expressed in terms of the Cabbibo parameter $\lambda \equiv \sin\theta_{\rm C}$
 \begin{eqnarray}
 \frac{m_{\mu}}{m_{\tau}}=\frac{({\cal M}_{\ell})_{22}}{({\cal M}_{\ell})_{33}}=\left(\frac{x_{2}}{x_{3}}\right)^2\equiv\lambda^{2}~.
 \label{power1}
 \end{eqnarray}
And in terms of the neutrino mass matrix elements, ratios associated with $\mu$ and $\tau$ flavors are written as
 \begin{eqnarray}
 \frac{m_{e\mu}}{m_{e\tau}}=\frac{y_2}{y_3}~,\qquad \frac{m_{\mu\mu}}{m_{\mu\tau}}=\frac{C_{\nu}y_2}{D_{\nu}y_3}~,\qquad \frac{m_{\mu\tau}}{m_{\tau\tau}}=\frac{D_{\nu}y_2}{C_{\nu}y_3}~,\qquad \frac{m_{\mu\mu}}{m_{\tau\tau}}=\left(\frac{y_2}{y_3}\right)^2~.
 \label{power2}
 \end{eqnarray}
Both Eqs.~(\ref{power1}) and (\ref{power2}) indicate that the ``$\mu$--$\tau$ power'' has a relationship between two quantities associated with $\mu$ and $\tau$ flavors and the matrix elements  vary as a power of some attribute of those flavors, where the distinctions $y_{2}$ and $y_{3}$ are taken as real and positive parameters (which will be shown below Eq.~(\ref{Flavor0})).

As second ansatz, we consider the renormalizable ultraviolet (UV) complete theory above a new physics scale. For neutrinos, it leads to  a number of independent ${\cal O}(1)$ parameters, which is of the form~\cite{Harari:1978yi,Democra}
 \begin{eqnarray}
 \mathcal{M}^{0}_{\nu}= A_{\nu}{\left(\begin{array}{ccc}
 {\cal O}(1) & {\cal O}(1) & {\cal O}(1) \\
 {\cal O}(1) & {\cal O}(1) & {\cal O}(1) \\
 {\cal O}(1) & {\cal O}(1) & {\cal O}(1)
 \end{array}\right)}.
 \label{Demo}
 \end{eqnarray}
The above matrix seems that the masses and mixing angles of neutrinos are expected to be of order ${\cal O}(1)$.
On the other hand, above the new physics scale among charged fermion operators only the top quark operator seems to be dominated by the (3,3) matrix element, which is of the form~\cite{Harari:1978yi}
 \begin{eqnarray}
 \mathcal{M}^{0}_{Q}= A_{t}{\left(\begin{array}{ccc}
 0 & 0 & 0 \\
 0 & 0 & 0 \\
 0 & 0 & 1
 \end{array}\right)}.
 \label{DemoQuark}
 \end{eqnarray}
This may provide a hint of why the mass of top quark is uniquely big compared with those of other fermions.

Now, as a good example, considering flavored structure $\Delta\mathcal{M}_{\nu}$ to the democratic matrix $\mathcal{M}^{0}_{\nu}$, leading to TBM pattern, the mass matrix is given by
 \begin{eqnarray}
 \mathcal{M}^{1}_{\nu}= A_{\nu}{\left(\begin{array}{ccc}
 1+2a_{\nu} & 1-a_{\nu} & 1-a_{\nu} \\
 1-a_{\nu} & 1+\frac{a_{\nu}}{2}+\frac{3}{2}b_{\nu} & 1+\frac{a_{\nu}}{2}-\frac{3}{2}b_{\nu} \\
 1-a_{\nu} & 1+\frac{a_{\nu}}{2}-\frac{3}{2}b_{\nu} & 1+\frac{a_{\nu}}{2}+\frac{3}{2}b_{\nu}
 \end{array}\right)}= U_{0} {\left(\begin{array}{ccc}
 3A_{\nu}a_{\nu} & 0 & 0 \\
 0 & 3A_{\nu} & 0 \\
 0 & 0 & 3A_{\nu}b_{\nu}
 \end{array}\right)}U^{T}_{0}.
 \label{DemoFlavor}
 \end{eqnarray}
Here the diagonalizing matrix, so-called TBM mixing matrix~\cite{Harrison:2002er}, $U_{0}$ is given by
 \begin{eqnarray}
 U_{0}= {\left(\begin{array}{ccc}
 \sqrt{\frac{2}{3}} & \frac{1}{\sqrt{3}} & 0 \\
 -\frac{1}{\sqrt{6}} & \frac{1}{\sqrt{3}} & -\frac{1}{\sqrt{2}} \\
 -\frac{1}{\sqrt{6}} & \frac{1}{\sqrt{3}} & \frac{1}{\sqrt{2}}
 \end{array}\right)}.
 \label{TB}
 \end{eqnarray}
While the matrix in Eq.~(\ref{DemoQuark}) may give a hint for hierarchical pattern of charged fermion masses, the above neutrino mass matrix in Eq.~(\ref{DemoFlavor}) would provide a clue of the mildness of neutrino masses due to the matrix having a democratic form given by Eq.~(\ref{Demo}).
According to the $\mu$--$\tau$ power law, the above matrix in Eq.~(\ref{DemoFlavor}) is modified in a way that muon and tau flavors are distinguished, leading to naturally non-zero $\theta_{13}$, to
 \begin{eqnarray}
 \mathcal{M}_{\nu}&=& A_{\nu}{\left(\begin{array}{ccc}
 1+2a_{\nu} & (1-a_{\nu})\,y_{2} & (1-a_{\nu})\,y_{3} \\
 (1-a_{\nu})\,y_{2} & (1+\frac{a_{\nu}}{2}+\frac{3}{2}b_{\nu})\,y^{2}_{2} & (1+\frac{a_{\nu}}{2}-\frac{3}{2}b_{\nu})\,y_{2}y_{3} \\
 (1-a_{\nu})\,y_{3} & (1+\frac{a_{\nu}}{2}-\frac{3}{2}b_{\nu})\,y_{2}y_{3} & (1+\frac{a_{\nu}}{2}+\frac{3}{2}b_{\nu})\,y^{2}_{3}
 \end{array}\right)}.\label{Flavor0}
 \end{eqnarray}
As expected, in the limit $y_{2,3}\rightarrow1$ the neutrino mass matrix recovers the TBM mixing pattern. And small deviations of $y_{2,3}$ from unity guarantee the small but relatively large value of $\theta_{13}$.

Now one can count the physical parameters in the $\mu$--$\tau$ power mass matrix in Eq.~(\ref{MuTauPower}) or Eq.~(\ref{Flavor0}).
A general $3\times3$ mixing matrix contains 3 moduli and 6 phases and can be written as
$U=e^{i\Omega}\,P\tilde{U}Q$ where $Q\equiv{\rm diag}(1,e^{i\xi_{2}},e^{i\xi_{3}})$ and $P\equiv{\rm diag}(1,e^{-i\zeta_{2}},e^{-i\zeta_{3}})$ are diagonal phase matrices, and $\tilde{U}$ is a unitary ``CKM-like" matrix containing 1 phase and 3 mixing angles, with an overall phase $\Omega$. Then the leptonic PMNS mixing matrix can be expressed as $U_{\rm PMNS}=V^{\ell\dag}_{L}U_{\nu}=\tilde{V}^{\ell\dag}_{L}P^{\ast}_{\ell}P_{\nu}\tilde{U}_{\nu}Q_{\nu}$
which contains 6 mixing angles and 8 phases, while it should have physical 3 mixing angles and 1 Dirac and 2 Majorana phases as indicated in Eq.~(\ref{PMNS}). This can be achieved by choosing $P_{\ell}=P_{\nu}$ in a basis where the charged lepton mass matrix is diagonal. Letting $\arg(y_{2})=\zeta_{2}$ and $\arg(y_{3})=\zeta_{3}$,
the parameters $y_{2}, y_{3}$ appearing in the $\mu$--$\tau$ power mass matrix can always be chosen to be real and positive. Therefore, the $\mu$--$\tau$ power mass matrix contains 9 physical parameters $A_{\nu}, |B_{\nu}|, |C_{\nu}|,|D_{\nu}|$, $\arg(B_{\nu})$, $\arg(C_{\nu})$, $\arg(D_{\nu})$, $y_{2}$ and $y_{3}$ in Eq.~(\ref{MuTauPower}) for 9 observables $\theta_{23},\theta_{13},\theta_{12}$, $\delta_{CP}, \varphi_{1},\varphi_{2}$ (mixing parameters), and  $m_{\nu_1}, m_{\nu_2}, m_{\nu_3}$ (mass eigenvalues). By considering the $\mu$--$\tau$ power flavored symmetry like as Eq.~(\ref{Flavor0}), one can reduce physical degree of freedoms more: there are 7 physical parameters $A_{\nu}, |a_{\nu}|, |b_{\nu}|$, $\arg(a_{\nu}), \arg(b_{\nu})$, $y_{2}$ and $y_{3}$, which in turn can lead to
any light neutrino mass pattern, {\it i.e.} normal mass hierarchy, inverted one or quasi-degenerate one (remember that there are 5 neutrino oscillation observables $\theta_{12},\theta_{13},\theta_{23},\Delta m^{2}_{\rm Atm}, \Delta m^{2}_{\rm Sol}$).
Note that the $\mu$--$\tau$ power mass matrix leads naturally to a non-zero $\theta_{13}$. Moreover, as will be seen later, by embedding a specific flavor model to Lagrangian the $\mu$--$\tau$ power mass matrix can contain only 5 physical parameters (see Eqs.~(\ref{mass matrix},\ref{Numass1})) and lead to TBM-like one, which would not provide all possible neutrino mass pattern unlike Eq.~(\ref{Flavor0}), because it has a neutrino mass sum-rule $1/m_{\nu_1}-1/m_{\nu_3}=2/m_{\nu_2}$ in the limit $y_{2,3}\rightarrow1$ (which is guaranteed by the small value of $\theta_{13}$).

We believe that this approach is very important to take a step forward in understanding the mixing patterns for large leptonic and small quark mixings as well as the origin of the fermion mass hierarchies (mildness of neutrino masses and the strongly hierarchical charged fermion masses).

\section{flavor $A_{4}\times U(1)_{X}$ symmetry}
Unless flavor symmetries are assumed, particle masses and mixings are generally undetermined in the SM gauge theory. To understand the present fermion mass hierarchy with the large leptonic mixing and small quark mixing data, we introduce the non-Abelian discrete $A_{4}$ flavor symmetry which is mainly responsible for the peculiar mixing patterns with an additional continuous global symmetry $U(1)_{X}$ which is mainly for vacuum configuration as well as for  describing mass hierarchies of leptons and quarks. Moreover, the spontaneous breaking of $U(1)_{X}$ realizes the existence of the Nambu-Goldstone (NG) mode (called axion) and provides an elegant solution of the strong CP problem. Therefore, we refer this global $U(1)$ symmetry to ``flavored-PQ symmetry".  Then the symmetry group for matter fields (leptons and quarks), flavon fields and driving fields is $A_{4}\times U(1)_{X}$, whose quantum numbers are assigned in Table~\ref{DrivingRef} and \ref{reps}. In addition, there is a continuous $U(1)_{R}$ symmetry, containing the usual $R$-parity as a subgroup, that is classified as three sectors: driving fields $+2$, flavon fields and Higgs fields $0$, and matter fields $+1$. And the other superpotential term $\kappa_{\alpha}L_{\alpha}H_{u}$ and the terms violating the lepton and baryon number symmetries are not allowed by this $U(1)_{R}$ symmetry~\footnote{In addition, higher-dimensional supersymmetric operators like $Q_{i}Q_{j}Q_{k}L_{l}$ ($i,j,k$ must not all be the same) are not allowed either, and stabilizing proton.}.

To impose the $A_{4}$ flavor symmetry on our model properly, apart from the usual two Higgs doublets $H_{u,d}$ responsible for electroweak symmetry breaking, which are invariant under $A_{4}$ ({\it i.e.} flavor singlets $\mathbf{1}$ with no $T$-flavor), the scalar sector is extended by introducing two types of new scalar multiplets, flavon fields~\footnote{These flavon fields are responsible for the spontaneous breaking of the flavor symmetry.} $\Phi_{T},\Phi_{S},\Theta,\tilde{\Theta}, \Psi, \tilde{\Psi}$ that are $SU(2)$-singlets and driving fields $\Phi^{T}_{0},\Phi^S_{0},\Theta_{0},\Psi_{0}$ that are associated to a nontrivial scalar potential in the symmetry breaking sector: we take the flavon fields $\Phi_{T},\Phi_{S}$ to be $A_{4}$ triplets, and $\Theta,\tilde{\Theta},\Psi,\tilde{\Psi}$ to be $A_{4}$ singlets with no $T$-flavor ($\mathbf{1}$ representation), respectively, that are $SU(2)$-singlets, and driving fields $\Phi_{0}^{T},\Phi_{0}^{S}$ to be $A_{4}$ triplets and $\Theta_{0}, \Psi_{0}$ to be an $A_{4}$ singlet.
Moreover, due to the assignment of quantum numbers under $A_{4}\times U(1)_{X}\times U(1)_{R}$ the usual superpotential term $\mu H_{u}H_{d}$ is not allowed, while the next leading order operator is allowed
 \begin{eqnarray}
  \frac{g_{T}}{\Lambda}(\Phi^{T}_{0}\Phi_{T})_{{\bf 1}}H_{u}H_{d}
 \label{muterm}
 \end{eqnarray}
which promotes the $\mu$-term $\mu_{\rm eff}\equiv g_{T}\langle\Phi^{T}_{0}\rangle\, v_{T}/\Lambda$ of the order of $m_{S}\,v_{T}/\Lambda$ ($\langle\Phi^{T}_{0}\rangle$: the VEV of the scalar components of the driving field, $m_{S}$: soft SUSY breaking mass). Here the supersymmetry of the model is assumed broken by all possible holomorpic soft terms which are invariant under $A_{4}\times U(1)_{X}\times U(1)_{R}$ symmetry, where the soft breaking terms are already present at the scale relevant to flavor dynamics.

In the lepton sector the $A_{4}$ model giving non-zero $\theta_{13}$ as well as bi-large mixings, $\theta_{23}, \theta_{12}$, works as follows. According to both the $\mu$--$\tau$ power law in Eqs.~(\ref{power1}) and (\ref{power2}) and the UV completion textures in Eqs.~(\ref{Demo}) and (\ref{DemoQuark}), one can assign charged-leptons to the three inequivalent singlet representations of $A_{4}$: we assign the left-handed charged leptons denoted as $L_e,\,L_\mu,\,L_\tau$, the electron flavor to the ${\bf 1}$ ($T$-flavor 0), the muon flavor to the ${\bf 1}'$  ($T$-flavor $+$1), and the tau flavor to the ${\bf 1}''$ ($T$-flavor $-$1), while the right-handed charged leptons denoted as $e^c,\,\mu^c,\,\tau^c$, the electron flavor to the ${\bf 1}$ ($T$-flavor 0), the muon flavor to the ${\bf 1}''$  ($T$-flavor $-$1), and the tau flavor to the ${\bf 1}'$ ($T$-flavor $+$1). On the other hand, for the quark flavors we assign the left-handed quark $SU(2)_{L}$ doublets denoted as $Q_{1}$, $Q_{2}$ and $Q_{3}$ to the ${\bf 1}$, ${\bf 1}''$ and ${\bf 1}'$, respectively, while the right-handed up-type quarks are assigned as $u^{c}$, $c^{c}$ and $t^{c}$ to the ${\bf 1}$, ${\bf 1}'$ and ${\bf 1}''$ under $A_{4}$, respectively, and the right-handed down-type quark SM gauge singlet $D^{c}=\{d^{c}, s^{c}, b^{c}\}$ to the ${\bf 3}$ under $A_{4}$.

Finally, the additional symmetry $U(1)_{X}$ is imposed, which is a continuous global symmetry and under which matter fields, flavon fields, and driving fields carry their own $X$-charges. The $U(1)_{X}$ invariance forbids renormalizable Yukawa couplings for the light families, but would allow them through effective nonrenormalizable couplings suppressed by $({\cal F}/\Lambda)^n$ with $n$ being positive integers. Then the gauge singlet flavon field ${\cal F}$ is activated to dimension-4(3) operators with different orders~\cite{Ahn:2014zja}
 \begin{eqnarray}
  c_{0}\,{\cal OP}_{4}\,({\cal F})^{0}+c'_{1}\,{\cal OP}_{3}\,({\cal F})^{1}+c_{1}\,{\cal OP}_{4}\,\left(\frac{{\cal F}}{\Lambda}\right)^1+c_{2}\,{\cal OP}_{4}\,\left(\frac{{\cal F}}{\Lambda}\right)^{2}+c_{3}\,{\cal OP}_{4}\,\left(\frac{{\cal F}}{\Lambda}\right)^{3}+...
 \label{flavon}
 \end{eqnarray}
where ${\cal OP}_{4(3)}$ is a dimension-$4(3)$ operator, and all the coefficients $c_{i}$ and $c'_{i}$ are complex numbers with absolute value of order unity. Even with all couplings being of order unity, hierarchical masses for different flavors can be naturally realized. The flavon field ${\cal F}$ is a scalar field which acquires a VEV and breaks spontaneously the flavored-PQ symmetry $U(1)_{X}$. Here $\Lambda$, above which there exists unknown physics, is the scale of flavor dynamics, and is associated with heavy states which are integrated out. The effective theory below $\Lambda$ is rather simple, while the full theory will have many heavy states.
So, in our framework, the hierarchy $\langle H_{u,d}\rangle=v_{u,d}\ll\Lambda$ is maintained, and below the scale $\Lambda$ the higher dimensional operators express the effects from the unknown physics.
Since the Yukawa couplings are eventually responsible for the fermion masses they must be related in a very simple way at a large scale in order for intermediate scale physics to produce all the interesting structure in the fermion mass matrices.

Here we recall that $A_{4}$ is the symmetry group of the tetrahedron and the finite groups of the even permutation of four objects having four irreducible representations: its irreducible representations are ${\bf 3}, {\bf 1}, {\bf 1}' , {\bf 1}''$ with ${\bf 3}\otimes{\bf 3}={\bf 3}_{s}\oplus{\bf 3}_{a}\oplus{\bf 1}\oplus{\bf 1}'\oplus{\bf 1}''$, and ${\bf 1}'\otimes{\bf 1}'={\bf 1}''$. The details of the $A_{4}$ group are shown in Appendix~\ref{A4group}.
Let $(a_{1}, a_{2}, a_{3})$ and $(b_{1}, b_{2}, b_{3})$ denote the basis vectors for two ${\bf 3}$'s. Then we have
 \begin{eqnarray}
  (a\otimes b)_{{\bf 3}_{\rm s}} &=& \frac{1}{\sqrt{3}}(2a_{1}b_{1}-a_{2}b_{3}-a_{3}b_{2}, 2a_{3}b_{3}-a_{2}b_{1}-a_{1}b_{2}, 2a_{2}b_{2}-a_{3}b_{1}-a_{1}b_{3})~,\nonumber\\
  (a\otimes b_c)_{{\bf 3}_{\rm a}} &=& i(a_{3}b_{2}-a_{2}b_{3}, a_{2}b_{1}-a_{1}b_{2}, a_{1}b_{3}-a_{3}b_{1})~,\nonumber\\
  (a\otimes b)_{{\bf 1}} &=& a_{1}b_{1}+a_{2}b_{3}+a_{3}b_{2}~,\nonumber\\
  (a\otimes b)_{{\bf 1}'} &=& a_{1}b_{2}+a_{2}b_{1}+a_{3}b_{3}~,\nonumber\\
  (a\otimes b)_{{\bf 1}''} &=& a_{1}b_{3}+a_{2}b_{2}+a_{3}b_{1}~.
  \label{A4reps}
 \end{eqnarray}
Under $A_{4}\times U(1)_{X}\times U(1)_{R}$, the driving, flavon, and Higgs fields are assigned as in Table~\ref{DrivingRef}.
\begin{table}[h]
\caption{\label{DrivingRef} Representations of the driving, flavon, and Higgs fields under $A_4 \times U(1)_{X}$ with $U(1)_{R}$.}
\begin{ruledtabular}
\begin{tabular}{cccccccccccccccc}
Field &$\Phi^{T}_{0}$&$\Phi^{S}_{0}$&$\Theta_{0}$&$\Psi_{0}$&\vline\vline&$\Phi_{S}$&$\Phi_{T}$&$\Theta$&$\tilde{\Theta}$&$\Psi$&$\tilde{\Psi}$&\vline\vline&$H_{d}$&$H_{u}$\\
\hline
$A_4$&$\mathbf{3}$&$\mathbf{3}$&$\mathbf{1}$&$\mathbf{1}$&\vline\vline&$\mathbf{3}$&$\mathbf{3}$&$\mathbf{1}$&$\mathbf{1}$&$\mathbf{1}$&$\mathbf{1}$&\vline\vline&$\mathbf{1}$&$\mathbf{1}$\\
$U(1)_{X}$&$0$&$4p$&$4p$&$0$&\vline\vline&$-2p$&$0$&$-2p$&$-2p$&$-q$&$q$&\vline\vline&$0$&$0$\\
$U(1)_R$&$2$&$2$&$2$&$2$&\vline\vline&$0$&$0$&$0$&$0$&$0$&$0$&\vline\vline&$0$&$0$\\
\end{tabular}
\end{ruledtabular}
\end{table}

\subsection{Vacuum configuration}
\noindent Now, let us first investigate the vacuum configuration. Indeed, the VEV pattern of the flavons is determined dynamically, in which the vacuum alignment problem can be solved by the supersymmetric driving field method~\cite{Altarelli:2005yx}~\footnote{There is another generic way for the vacuum alignment problem by extending the model with a spacial extra dimension~\cite{Altarelli:2005yp}.}. In order to make a non-trivial scalar potential in the SUSY breaking sector, we introduce driving fields $\Phi^{T}_{0},\Phi^{S}_{0},\Theta_{0},\Psi_{0}$ whose have the representation of $A_{4}\times U(1)_{X}$ as in Table~\ref{DrivingRef}.
The leading order superpotential dependent on the driving fields, which is invariant under the flavor symmetry $A_{4}\times U(1)_{X}$, is given by
 \begin{eqnarray}
 W_{v} &=& \Phi^{T}_{0}\left(\tilde{\mu}\,\Phi_{T}+\tilde{g}\,\Phi_{T}\Phi_{T}\right)+\Phi^{S}_{0}\left(g_{1}\,\Phi_{S}\Phi_{S}+g_{2}\,\tilde{\Theta}\Phi_{S}\right)\nonumber\\
 &+& \Theta_{0}\left(g_{3}\,\Phi_{S}\Phi_{S}+g_{4}\,\Theta\Theta+g_{5}\,\Theta\tilde{\Theta}+g_{6}\,\tilde{\Theta}\tilde{\Theta}\right)+\Psi_{0}\left(g_{7}\,\Psi\tilde{\Psi}+\mu^{2}_{\Psi}\right)\,,
 \label{potential}
 \end{eqnarray}
where the fields $\Psi$ and $\tilde{\Psi}$ charged by $-q,q$, respectively, are ensured by the $U(1)_{X}$ symmetry extended to a complex $U(1)$ due to the holomorphy of the supepotential.
Note here that the model implicitly has two $U(1)_{X}\equiv U(1)_{X_1}\times U(1)_{X_2}$ symmetries which are generated by the charges $X_{1}=-2p$ and $X_{2}=-q$, which will be discussed more in section~\ref{U1-Axion}~\footnote{In the model there are three $U(1)$ symmetries, $U(1)_L$ (lepton number) (or $U(1)_{B-L}$), $U(1)_{\rm PQ}$ and $U(1)_{Y}$ except for $U(1)_{R}$ and $U(1)_B$ (baryon number). All of these threes are finally broken. When flavon fields acquire VEVs, both $U(1)_L$ and $U(1)_{\rm PQ}$ appear to be broken. Actually, there are linear combinations of the two $U(1)_{X_i}$ symmetries, which are $U(1)_{\tilde{X}}\times U(1)_f$. Here the $U(1)_{\tilde{X}}$ symmetry has anomaly, while the $U(1)_f$, which corresponds to lepton number, is anomaly-free. See the superpotential~(\ref{lagrangian}), (\ref{lagrangianU}) and (\ref{lagrangianD}).}.
Since there is no fundamental distinction between the singlets $\Theta$ and $\tilde{\Theta}$ as indicated in Table~\ref{DrivingRef}, we are free to define $\tilde{\Theta}$ as the combination that couples to $\Phi^{S}_{0}\Phi_{S}$ in the superpotential $W_{v}$~\cite{Altarelli:2005yx}. At the leading order there are no terms involving the Higgs fields $H_{u,d}$, while the next leading order the effective $\mu$-term arises $\Phi^{T}_{0}\Phi_{T}H_{u}H_{d}/\Lambda$ in Eq.~(\ref{muterm}). And it is evident that at the leading order the scalar supersymmetric $W(\Phi_{T}\Phi_{S})$ terms are absent due to different $U(1)_{X}$ quantum number, which is crucial for relevant vacuum alignments in the model to produce the present lepton and quark mixings. It is interesting that at the leading order the electroweak scale does not mix with the potentially large scales $v_{S},v_{T},v_{\Theta}$ and $v_{\Psi}$.

In the SUSY limit, the vacuum configuration is obtained by the $F$-terms of all fields being required to vanish. The vacuum alignment of the flavon $\Phi_{T}$ is determined by
 \begin{eqnarray}
 \frac{\partial W_{v}}{\partial\Phi^{T}_{01}}&=&\tilde{\mu}\,\Phi_{T1}+\frac{2\tilde{g}}{\sqrt{3}}\left(\Phi^{2}_{T1}-\Phi_{T2}\Phi_{T3}\right)=0~,\nonumber\\
 \frac{\partial W_{v}}{\partial\Phi^{T}_{02}}&=&\tilde{\mu}\,\Phi_{T3}+\frac{2\tilde{g}}{\sqrt{3}}\left(\Phi^{2}_{T2}-\Phi_{T1}\Phi_{T3}\right)=0~,\nonumber\\
 \frac{\partial W_{v}}{\partial\Phi^{T}_{03}}&=&\tilde{\mu}\,\Phi_{T2}+\frac{2\tilde{g}}{\sqrt{3}}\left(\Phi^{2}_{T3}-\Phi_{T1}\Phi_{T2}\right)=0~.
 \label{potential1}
 \end{eqnarray}
From this set of three equations, we can obtain the supersymmetric vacuum for $\Phi_{T}$
 \begin{eqnarray}
\langle\Phi_{T}\rangle=\left(\frac{v_{T}}{\sqrt{2}},\,0,\,0\right)\,,\qquad \text{with}\,\, v_{T}=-\sqrt{\frac{3}{2}}\,\frac{\tilde{\mu}}{\tilde{g}}\,,
 \label{vevdirection1}
 \end{eqnarray}
where $\tilde{g}$ is a dimensionless coupling.
The minimization equations for the vacuum configuration of $\Phi_{S}$ and $(\Theta,\tilde{\Theta})$ are given by
 \begin{eqnarray}
 \frac{\partial W_{v}}{\partial\Phi^{S}_{01}}&=&\frac{2g_{1}}{\sqrt{3}}\left(\Phi_{S1}\Phi_{S1}-\Phi_{S2}\Phi_{S3}\right)+g_{2}\Phi_{S1}\tilde{\Theta}=0~,\nonumber\\
 \frac{\partial W_{v}}{\partial\Phi^{S}_{02}}&=&\frac{2g_{1}}{\sqrt{3}}\left(\Phi_{S2}\Phi_{S2}-\Phi_{S1}\Phi_{S3}\right)+g_{2}\Phi_{S3}\tilde{\Theta}=0~,\nonumber\\
 \frac{\partial W_{v}}{\partial\Phi^{S}_{03}}&=&\frac{2g_{1}}{\sqrt{3}}\left(\Phi_{S3}\Phi_{S3}-\Phi_{1}\Phi_{S2}\right)+g_{2}\Phi_{S2}\tilde{\Theta}=0~,\nonumber\\
 \frac{\partial W_{v}}{\partial\Theta_{0}}&=&g_{3}\left(\Phi_{S1}\Phi_{S1}+2\Phi_{S2}\Phi_{S3}\right)+g_{4}\Theta^{2}+g_{5}\Theta\tilde{\Theta}+g_{6}\tilde{\Theta}^{2}=0~.
 \label{potential2}
 \end{eqnarray}
And from  Eq.~(\ref{potential2}), we can get the supersymmetric vacuua for the fields $\Phi_{S},\Theta,\tilde{\Theta}$
 \begin{eqnarray}
 \langle\Phi_{S}\rangle=\frac{1}{\sqrt{2}}\left(v_{S},v_{S},v_{S}\right)\,,\quad\langle\Theta\rangle=\frac{v_{\Theta}}{\sqrt{2}}\,,\quad\langle\tilde{\Theta}\rangle=0\,,\qquad\text{with}\,\,v_{\Theta}=v_{S}\sqrt{-3\frac{g_{3}}{g_{4}}}\,,
 \label{vevdirection2}
 \end{eqnarray}
where $v_{\Theta}$ is undetermined. As can be seen in Eq.~(\ref{vevdirection2}), the VEVs $v_{\Theta}$ and $v_{S}$ are naturally of the same order of magnitude (here the dimensionless parameters $g_{3}$ and $g_{4}$ are the same order of magnitude).

Finally, the minimization equation for the vacuum configuration of $\Psi$ is given by
 \begin{eqnarray}
 \frac{\partial W_{v}}{\partial\Psi_{0}}&=&g_{7}\Psi\tilde{\Psi}+\mu^{2}_{\Psi}=0~,
 \label{potential3}
 \end{eqnarray}
where $\mu_{\Psi}$ is the $U(1)_{X}$ breaking scale and $g_{7}$ is a dimensionless coupling.
And from  Eq.~(\ref{potential3}), we can get the supersymmetric vacuua for the fields $\Psi,\tilde{\Psi}$
 \begin{eqnarray}
 \langle\Psi\rangle=\langle\tilde{\Psi}\rangle=\frac{v_{\Psi}}{\sqrt{2}}\,,\qquad\text{with}\,\,v_{\Psi}=\mu_{\Psi}\sqrt{\frac{-2}{g_{7}}}\,.
 \label{vevdirection3}
 \end{eqnarray}
We see that the global minima of the potential are located at Eqs.~(\ref{vevdirection1}), (\ref{vevdirection2}) and (\ref{vevdirection3}). The vacuum configuration of the driving fields in the SUSY limit is given in Appendix~\ref{driving}.
As can be seen in Eqs.~(\ref{vevdirection2}) and (\ref{vevdirection3}), in the SUSY limit there exist flat directions along which the scalar fields $\Phi_{S}, \Theta$ and $\Psi,\tilde{\Psi}$ do not feel the potential.
The SUSY-breaking effect lifts up the flat directions and corrects the VEV of the driving field $\Psi_0$, leading to soft SUSY-breaking mass terms (here we do not specify a SUSY breaking mechanism in this work).
The full scalar potential is given by
 \begin{eqnarray}
  V_{total}=\sum_{i}\left|\frac{\partial W}{\partial\varphi_{i}}\right|^{2}+V_{soft}+V_{D}~,
  \label{Vtotal}
 \end{eqnarray}
where $\varphi_{i}=\{\Phi^{T}_{0},\Phi^{S}_{0},\Theta_{0},\Psi_{0},\Phi_{T},\Phi_{S},\Theta,\tilde{\Theta},\Psi,\tilde{\Psi}\}$ stand for all the scalar fields, $V_{soft}$ and $V_{D}$ represent soft- and D-terms for the fields charged under the gauge group. Since all the soft SUSY breaking parameters in $V_{soft}$ are expected to be of order $m_{S}$ which is much smaller than the mass scales involved in $W_{v}$, it makes sense to minimize $V_{total}$ in the SUSY limit and to explain soft breaking effects subsequently.

By including generic soft SUSY breaking terms, which originate from another sector of the theory, neutral under the action of gauge group and under $A_{4}\times U(1)_{X}$, one can introduce a set of generic soft SUSY breaking terms by promoting the coupling constant of the theory to constant superfields with non-vanishing auxiliary components~\cite{Luty:2005sn}. Since all soft SUSY breaking parameters are of order $m_{S}$, all the VEVs appearing in Eq.~(\ref{Drdirection1}) can be of order $m_{S}$.
And, by adding a soft SUSY breaking mass term to the scalar potential one can execute $\langle\tilde{\Theta}\rangle=0$ for the scalar field $\tilde{\Theta}$ with $m^{2}_{\tilde{\Theta}}>0$. Since there are flat directions in the SUSY limit, by taking $m^{2}_{\Phi_{S}}, \,m^{2}_{\Theta}, \,m^{2}_{\Psi}, \,m^{2}_{\tilde{\Psi}}<0$, $v_{\Theta}$ and $v_\Psi$ roll down toward its true minimum from a large scale, which we assume to be stabilized far away from the origin by one-loop radiative corrections in the SUSY broken phase.  Then the vacuum alignment is taken as the absolute minimum.

Under $A_{4}\times U(1)_{X}\times U(1)_{R}$, the matter fields are assigned as in Table~\ref{reps}.
\begin{table}[h]
\caption{\label{reps} Representations of the matter fields under $A_4 \times U(1)_{X}$ with $U(1)_{R}$.}
\begin{ruledtabular}
\begin{tabular}{ccccccccccc}
Field &$Q_{1},~Q_{2},~Q_{3}$&$D^c$&$u^c,~c^c,~t^c$&$L_{e},~L_{\mu},~L_{\tau}$&$e^c,~\mu^c,~\tau^c$&$N^{c}$\\
\hline
$A_4$&$\mathbf{1}$, $\mathbf{1}''$, $\mathbf{1^{\prime}}$&$\mathbf{3}$&$\mathbf{1}$, $\mathbf{1}'$, $\mathbf{1^{\prime\prime}}$&$\mathbf{1}$, $\mathbf{1^{\prime}}$, $\mathbf{1^{\prime\prime}}$&$\mathbf{1}$, $\mathbf{1^{\prime\prime}}$, $\mathbf{1^\prime}$&$\mathbf{3}$\\
$U(1)_{X}$&$(3q-r,2q-r,-r)$ &$r+2p$&$(r+5q,r+2q,r)$& $ -p $ & $(8q+p, 4q+p, 2q+p)$& $p$\\
$U(1)_R$&$1$ &~$1$~&~$1$& $ 1 $ & $1$~~~& $1$\\
\end{tabular}
\end{ruledtabular}
\end{table}
In the following superpotential, the matter fields interact with $X$ fields and have some $X$ charges.
\subsection{Lepton sector}
\noindent The superpotential for Yukawa interactions in the lepton sector, which is invariant under  $SU(2)_L\times U(1)_{Y}\times A_{4}\times U(1)_{X}\times U(1)_{R}$, is given at leading order by
 \begin{eqnarray}
 W_{\ell\nu} &=& \hat{y}^{\nu}_{1}\,L_{e}(N^{c}\Phi_{T})_{{\bf 1}}\frac{H_{u}}{\Lambda}+\hat{y}^{\nu}_{2}\,L_{\mu}(N^{c}\Phi_{T})_{{\bf 1}''}\frac{H_{u}}{\Lambda}+\hat{y}^{\nu}_{3}\,L_{\tau}(N^{c}\Phi_{T})_{{\bf 1}'}\frac{H_{u}}{\Lambda}\nonumber\\
&+& \frac{1}{2}(\hat{y}_{\Theta}\Theta+\hat{y}_{\tilde{\Theta}}\tilde{\Theta}) (N^{c}N^{c})_{{\bf 1}}+\frac{\hat{y}_{R}}{2} (N^{c}N^{c})_{{\bf 3}_{s}}\Phi_{S}\nonumber\\
&+&y_{e}\,L_{e}\,e^{c}\,H_{d}+y_{\mu}\,L_{\mu}\,\mu^{c}\,H_{d}+y_{\tau}\,L_{\tau}\,\tau^{c}\,H_{d}\,.
 \label{lagrangian}
 \end{eqnarray}
Because of the chiral structure of weak interactions, bare fermion masses are not allowed in the SM. Fermion masses arise through Yukawa interactions~\footnote{Since the right-handed neutrinos having a mass scale much above the weak interaction scale are complete singlets of the SM gauge symmetry, it can possess bare SM invariant mass terms. However, the flavored-PQ symmetry $U(1)_{X}$ guarantees the absence of a bare mass term $M\,N^{c}N^{c}$.}.
Since the $U(1)_{X}$ quantum numbers are assigned appropriately to the matter fields content as in Table \ref{reps}, the Yukawa couplings of charged leptons appearing in the superpotential $W_{\ell\nu}$ are a function of flavon field $\Psi$, {\it i.e.} $y_{e,\mu,\tau}=y_{e,\mu,\tau}(\Psi)$:
\begin{eqnarray}
y_{e}&=&\hat{y}_{e}\left(\frac{\Psi}{\Lambda}\right)^{8}\,,\quad\qquad
y_{\mu}=\hat{y}_{\mu}\left(\frac{\Psi}{\Lambda}\right)^{4}\,,\quad\qquad
y_{\tau}=\hat{y}_{\tau}\left(\frac{\Psi}{\Lambda}\right)^{2}\,.
 \label{YukawaWl}
\end{eqnarray}
Here the couplings $\hat{y}_{e,\mu,\tau}$ are complex numbers and of order unity, {\it i.e.} $1/\sqrt{10}\lesssim|\hat{y}_{e,\mu,\tau}|\lesssim\sqrt{10}$, while the neutrino Yukawa couplings are given as
\begin{eqnarray}
\hat{y}^{\nu}_{1}&\approx& \hat{y}^{\nu}_{2}\approx \hat{y}^{\nu}_{3}\approx{\cal O}(1)\,,\qquad \hat{y}_{\Theta}\approx \hat{y}_{\tilde{\Theta}}\approx \hat{y}_{R}\approx{\cal O}(1)\,.
 \label{YukawaWnu}
\end{eqnarray}
Since the fields associated with the superpotential (\ref{lagrangian}) are charged under $U(1)_{X}$, it is expected that all the hat neutrino Yukawa couplings appearing in the superpotential (\ref{lagrangian}) are of order unity and complex numbers.

In the above leptonic Yukawa superpotential, the right-handed Majorana neutrino terms are associated with an $A_{4}$ singlet $\Theta$ and an $A_{4}$ triplet $\Phi_{S}$ flavon fields both of which are the SM gauge singlets. So, below the cutoff scale $\Lambda$, the Majorana neutrino mass terms comprise an exact TBM pattern, which will be shown later. We note that the flavon field $\Phi_{T}$ derives dimension-5 operators in the Dirac neutrino sector, while the flavon field $\Psi$ derives higher dimensional operators with the $U(1)_{X}$ flavor symmetry responsible for the hierarchical charged lepton masses as shown in Eq.~(\ref{YukawaWl}). Imposing the continuous global $U(1)_{X}$ symmetry in Table~\ref{reps} explains the absence of the Yukawa terms $LN^{c}\Phi_{S}$ and $N^{c}N^{c}\Phi_{T}$ as well as does not allow the interchange between $\Phi_{T}$ and $\Phi_{S}$, both of which transform differently under $U(1)_{X}$, so that bi-large $\theta_{12},\theta_{23}$ mixings with a non-zero $\theta_{13}$ mixing for the leptonic mixing matrix could be obtained after seesawing (which will be shown later).

Especially, since the field $\Phi_{T}$ is not charged under the $U(1)_{X}$, nontrivial next-to-leading order operators could be generated via $\Phi_{T}$. So, we will show that, after flavor symmetry breaking, the next leading operators can contribute to the Majorana neutrino sector (see more details in Sec.~\ref{corrects}), while there are no new structures contributing to the Dirac neutrino and charged-lepton sectors after symmetry breaking.
It is very crucial to notify the followings, which guarantees the superpotential for the Dirac neutrino and charged lepton sectors in Eq.~(\ref{lagrangian}):
(i) in the charged lepton sector higher dimensional operators including $(\Phi_{T}\Phi_{T})_{{\bf 1}', {\bf 1}''}$, that is, $\hat{y}_{\alpha\beta}\left(\frac{\Psi}{\Lambda}\right)^nL_{\alpha}\beta^cH_{d}(\Phi_{T}\Phi_{T})_{{\bf 1}', {\bf 1}''}/\Lambda^2$ where $\alpha\neq\beta=e,\mu,\tau$ and $n\geq1$ (integer), are all vanishing due to the VEV alignment $\langle\Phi_{T}\rangle\sim v_{T}(1,0,0)$, (ii) since higher dimensional operators involving $(\Phi_{T}\Phi_{T})_{{\bf 3}_s}$ or $(\Phi_{T}\Phi_{T})_{{\bf 1}}$ have the same direction as $\Phi_{T}$, the corrections to the charged lepton sector appear as an order of $1/\Lambda^4$ and absorbed into a redefinition of the leading order terms, (iii) in the Dirac neutrino sector higher dimensional operators driven by the $\Phi_T$ field, that is, $\hat{y}_{\alpha}L_{\alpha}[N^c(\Phi_{T})^n]_{{\bf 1}, {\bf 1}', {\bf 1}''}H_u/\Lambda^n$ with $n\geq2$ (integer), are absorbed into a redefinition of the leading order terms due to the same reasons addressed in the previous cases, and (iv) higher dimensional operators via the insertions of $\Psi\tilde{\Psi}/\Lambda^2$ and $H_uH_d/\Lambda^2$ are all absorbed into the leading order terms and redefined, on the other hand, (v) higher dimensional operators including $(\Phi_{S}\Phi_{S})_{{\bf 1}, {\bf 1}', {\bf 1}'', {\bf 3}}$ or $\Theta$ are forbidden by the $U(1)_{X}$ symmetry. Note that the other higher dimensional operators invariant under $A_{4}\times U(1)_X$ are vanishing due to $U(1)_{R}$ symmetry. Therefore, the unwanted off-diagonal entries in the charged lepton and Dirac neutrino mass matrices, as will be shown in Eqs.~(\ref{ChL1},\ref{Ynu1}), are all vanishing or absorbed into a redefinition of the leading order terms, while there will be new structure corrections to the Majorana neutrino sector due to next-to-leading order operators whose contributions could be below the percent level as will be seen in Eqs.~(\ref{MRCorr},\ref{DMR2}).

As mentioned before, the model has two $U(1)$ symmetries which are generated by the charges $X_{1}\equiv-2p$ and $X_{2}\equiv-q$.
The $A_{4}$ flavor symmetry along with the flavored PQ symmetry $U(1)_{X_1}$ is spontaneously broken by two $A_{4}$-triplets $\Phi_{T},\Phi_{S}$ and by a singlet $\Theta$ in Table~\ref{reps}.
And the $U(1)_{X_2}$ symmetry is spontaneously broken by $\Psi,\tilde{\Psi}$, whose scale is denoted as $\mu_{\Psi}$, and the VEV of $\Psi$ (scaled by the cutoff $\Lambda$) is assumed as
 \begin{eqnarray}
 \frac{\langle\Psi\rangle}{\Lambda}\equiv\lambda~.
 \label{Cabbibo}
 \end{eqnarray}
Here the parameter $\lambda$ stands for the Cabbibo parameter.
We take the $A_{4}$ symmetry breaking scale and the $U(1)_{X_2}$ breaking scale to be much above the electroweak scale in our scenario, i.e., $\langle\Psi\rangle,\langle\tilde{\Psi}\rangle,\langle\Theta\rangle,\langle\Phi_{T}\rangle,\langle\Phi_{S}\rangle\gg\langle H_{u,d}\rangle$. We assume that the electroweak symmetry is broken by some mechanism, such as radiative effects when SUSY is broken. As discussed in the previous section, the fields $\Phi_{T},\Phi_{S},\Theta, \tilde{\Theta}$ and $\Psi,\tilde{\Psi}$ develop VEVs along the directions
 \begin{eqnarray}
  \langle\Phi_T\rangle&=&\frac{1}{\sqrt{2}}\left(v_{T},0,0\right),\qquad\qquad\qquad\qquad\langle\Phi_{S}\rangle=\frac{1}{\sqrt{2}}\left(v_{S},v_{S},v_{S}\right),\nonumber\\
  \langle\Theta\rangle&=&\frac{v_{\Theta}}{\sqrt{2}},\qquad\qquad\qquad\langle\tilde{\Theta}\rangle=0,\qquad\qquad\qquad\langle\Psi\rangle=\langle\tilde{\Psi}\rangle=\frac{v_{\Psi}}{\sqrt{2}}\,.
 \label{vev}
 \end{eqnarray}
Even these VEVs could be slightly perturbed by higher dimensional operators contributing to the driving superpotential, their corrections to the lepton and quark mass matrices are absorbed into the leading order terms and redefined due to the same VEV directions as in Eq.~(\ref{NewVEV}), or can be kept small enough and negligible, which will be shown in Sec.~\ref{VEVcorrects}.

Once the scalar fields $\Phi_{S}, \Theta, \tilde{\Theta},\Psi$ and $\tilde{\Psi}$ get VEVs, the flavor symmetry $U(1)_{X}\times A_{4}$ is spontaneously broken~\footnote{If the symmetry $U(1)_{X}$ is broken spontaneously, the Goldstone modes would be axions. See more details in section~\ref{U1-Axion}.}.
After electroweak and flavor symmetry breaking, the mass terms and the charged gauge interactions in a weak eigenstate basis are simply expressed as
 \begin{eqnarray}
 -{\cal L}_{mW} &=& \frac{1}{2}\overline{N^{c}_{R}}M_{R}N_{R}
 +\overline{\nu_{L}}m_{D}N_{R}+\overline{\ell_{L}}{\cal M}_{\ell}\ell_{R}+\frac{g}{\sqrt{2}}W^{-}_{\mu}\overline{\ell_{L}}\gamma^{\mu}\nu_{L}+\text{h.c.}\nonumber
 \\
 &=& \frac{1}{2} \begin{pmatrix} \overline{\nu_L} & \overline{N^{c}_R} \end{pmatrix} \begin{pmatrix} 0 & m_D \\ m_D^T & M_R \end{pmatrix} \begin{pmatrix} \nu^{c}_L \\ N_R \end{pmatrix} + \overline{\ell_{L}}{\cal M}_{\ell}\ell_{R}+\frac{g}{\sqrt{2}}W^{-}_{\mu}\overline{\ell_{L}}\gamma^{\mu}\nu_{L}+\text{h.c.},
 \label{lagrangianA}
 \end{eqnarray}
where $g$ is the SU(2) coupling constant.

We first consider the charged lepton sector. After the breaking of the flavor symmetries and electroweak symmetry, with the VEV alignment in Eq.~(\ref{vev}), the mass matrix of charged leptons is given by
 \begin{eqnarray}
 {\cal M}_{\ell}&=& {\left(\begin{array}{ccc}
 y_{e} & 0 &  0 \\
 0 & y_{\mu} & 0 \\
 0 & 0 & y_{\tau}
 \end{array}\right)}v_{d}={\left(\begin{array}{ccc}
 \hat{y}_{e}\lambda^8 & 0 &  0 \\
 0 & \hat{y}_{\mu}\lambda^4 & 0 \\
 0 & 0 & \hat{y}_{\tau}\lambda^2
 \end{array}\right)}v_{d}~.
 \label{ChL1}
 \end{eqnarray}
Recalling that the hat Yukawa couplings are all of order unity and complex numbers.
And the corresponding charged lepton masses are given by
 \begin{eqnarray}
 m_{\tau}&\equiv&|y_{\tau}|\,v_{d}=\lambda^2\,|\hat{y}_{\tau}|\,v_{d}\,,\quad
 m_{\mu}\equiv|y_{\mu}|\,v_{d}=\lambda^4\,|\hat{y}_{\mu}|\,v_{d}\,,\quad
 m_{e}\equiv|y_{e}|\,v_{d}=\lambda^8\,|\hat{y}_{e}|\,v_{d}\,,
 \label{ChL2}
 \end{eqnarray}
where $\langle H_{d}\rangle\equiv v_{d}=v\cos\beta/\sqrt{2}$ with $v\simeq246$ GeV.
These results are in a good agreement with the empirical charged lepton mass ratios calculated from the measured values~\cite{PDG}:
 \begin{eqnarray}
 \frac{m_{e}}{m_{\tau}}\simeq2.9\times10^{-4}\,,\qquad \frac{m_{\mu}}{m_{\tau}}\simeq5.9\times10^{-2}\,.
 \label{MlRatio}
 \end{eqnarray}

On the other hand, the Dirac and Majorana neutrino mass terms read
 \begin{eqnarray}
 m_{D}&=&{\left(\begin{array}{ccc}
 \hat{y}^{\nu}_{1} &  0 &  0 \\
 0 &  0 &  \hat{y}^{\nu}_{2}   \\
 0 &  \hat{y}^{\nu}_{3}  &  0
 \end{array}\right)}\frac{v_{T}}{\sqrt{2}\Lambda}v_{u}
 =\hat{y}^{\nu}_{1}{\left(\begin{array}{ccc}
 1 &  0 &  0 \\
 0 &  0 &  y_{2}   \\
 0 &  y_{3}  &  0
 \end{array}\right)}\frac{v_{T}}{\sqrt{2}\Lambda}v_{u}, \label{Ynu1}\\
 M_{R}&=&{\left(\begin{array}{ccc}
 1+\frac{2}{3}\tilde{\kappa}\,e^{i\phi} &  -\frac{1}{3}\tilde{\kappa}\,e^{i\phi} &  -\frac{1}{3}\tilde{\kappa}\,e^{i\phi} \\
 -\frac{1}{3}\tilde{\kappa}\,e^{i\phi} &  \frac{2}{3}\tilde{\kappa}\,e^{i\phi} &  1-\frac{1}{3}\tilde{\kappa}\,e^{i\phi}\\
 -\frac{1}{3}\tilde{\kappa}\,e^{i\phi} &  1-\frac{1}{3}\tilde{\kappa}\,e^{i\phi} &  \frac{2}{3}\tilde{\kappa}\,e^{i\phi}
 \end{array}\right)}M~,
 \label{MR1}
 \end{eqnarray}
where $\langle H_{u}\rangle\equiv v_{u}=v\sin\beta/\sqrt{2}$, and
 \begin{eqnarray}
 y_{2}\equiv\frac{\hat{y}^{\nu}_{2}}{\hat{y}^{\nu}_{1}}~,\quad y_{3}\equiv\frac{\hat{y}^{\nu}_{3}}{\hat{y}^{\nu}_{1}}~,\quad\tilde{\kappa}\equiv\sqrt{\frac{3}{2}}\left|\hat{y}_{R}\frac{v_{S}}{M}\right|~,\quad\phi\equiv\arg\left(\frac{\hat{y}_{R}}{\hat{y}_{\Theta}}\right)~~\text{with}~M\equiv \left|\hat{y}_{\Theta}\,\frac{v_{\Theta}}{\sqrt{2}}\right|~.
 \label{MR2}
 \end{eqnarray}
Note here that due to the magnitude of $\hat{y}^{\nu}_{i}$ being of order unity, in other words ${\cal O}(y_{2})\simeq{\cal O}(y_{3})\simeq{\cal O}(1)$, the $\mu-\tau$ symmetry is broken, which leads to non-zero $\theta_{13}$ after seesawing.

A crucial point is that, by redefining the light neutrino field $\nu_{L}$ as $P_{\nu}\nu_{L}$ and transforming $\ell_{L}\rightarrow P_{\nu}\ell_{L}$, $\ell_{R}\rightarrow P_{\nu}\ell_{R}$, one can always make the Yukawa couplings $\hat{y}^{\nu}_{1}, y_{2}, y_{3}$ real and positive.
Then from Eqs.~(\ref{Ynu1}) and (\ref{MR2}) the light neutrino mass matrix formed by seesaw formula, ${\cal M}_{\nu}=-m_D \, M_R^{-1} \, m^T_D$, leads to the following $\mu$--$\tau$ power mass matrix:
 \begin{eqnarray}
  {\cal M}_{\nu}
   = m_{0}\,e^{i\pi}
   {\left(\begin{array}{ccc}
   1+2F & (1-F)\,y_{2} & (1-F)\,y_{3} \\
   (1-F)\,y_{2} & (1+\frac{F+3\,G}{2})\,y^{2}_{2} & (1+\frac{F-3\,G}{2})\,y_{2}\,y_{3}  \\
   (1-F)\,y_{3} & (1+\frac{F-3\,G}{2})\,y_{2}\,y_{3} & (1+\frac{F+3\,G}{2})\,y^2_{3}
   \end{array}\right)},
\label{mass matrix}
 \end{eqnarray}
where
 \begin{eqnarray}
 m_{0}\equiv \left|\frac{\hat{y}^{\nu2}_{1}\upsilon^{2}_{u}}{6M}\right|\left(\frac{v_{T}}{\Lambda}\right)^2,\quad F=\left(\tilde{\kappa}\,e^{i\phi}+1\right)^{-1},\quad G=\left(\tilde{\kappa}\,e^{i\phi}-1\right)^{-1}.
 \label{Numass1}
 \end{eqnarray}
It is diagonalized by the transformation
 \begin{eqnarray}
  U^{\dag}_{\rm PMNS}\,{\cal M}_{\nu}\,U^{\ast}_{\rm PMNS}= {\rm Diag.}(m_{\nu_1},m_{\nu_2},m_{\nu_3})~.
\label{mass matrix2}
 \end{eqnarray}
As is well-known, because of the observed hierarchy $|\Delta m^{2}_{\rm Atm}|\equiv |m^{2}_{\nu_3}-m^{2}_{\nu_1}|\gg\Delta m^{2}_{\rm Sol}\equiv m^{2}_{\nu_2}-m^{2}_{\nu_1}>0$, and the requirement of a Mikheyev-Smirnov-Wolfenstein resonance for solar neutrinos, there are two possible neutrino mass spectra: (i) the normal mass ordering (NO) $m_{\nu_1}<m_{\nu_2}<m_{\nu_3}$, and (ii) the inverted mass ordering (IO) $m_{\nu_3}<m_{\nu_1}<m_{\nu_2}$.
In the limit $y^\nu_2=y^\nu_3$ ($y_{2}\rightarrow y_{3}$), the mass matrix in Eq.~(\ref{mass matrix}) acquires a $\mu$--$\tau$ symmetry that leads to $\theta_{13}=0$ and $\theta_{23}=-\pi/4$. Moreover, in the limit  $y^\nu_1=y^\nu_2=y^\nu_3$ ($y_{2}, y_{3}\rightarrow1$), the  mass matrix~(\ref{mass matrix}) gives the TBM angles and their corresponding mass eigenvalues
 \begin{eqnarray}
 &&\sin^{2}\theta_{12}=\frac{1}{3}\,,~\quad\qquad\sin^{2}\theta_{23}=\frac{1}{2}\,,~~\qquad\sin\theta_{13}=0\,,\nonumber\\
 &&m_{\nu_1}= 3\,m_{0}\,|F|~,\qquad m_{\nu_2}=3\,m_{0}~,\qquad m_{\nu_3}= 3\,m_{0}\,|G|~.
 \label{TBM1}
 \end{eqnarray}
These mass eigenvalues are disconnected from the mixing angles~\cite{Ahn:2011i}. Note here that the light neutrino mass matrix in Eq.~(\ref{mass matrix}) contains 5 physical parameters ($m_{0}, y_{2},y_{3},\tilde{\kappa},\phi$), leading to a neutrino mass sum-rule~\footnote{The flavor symmetry models giving an exact TMB mixing pattern have neutrino mass sum-rules~\cite{sumrule}, which are different from our model due to in general $y_{2,3}\neq1$.} $1/m_{\nu_1}-1/m_{\nu_3}=2/m_{\nu_2}$ in the limit $y_{2,3}\rightarrow1$, while the neutrino mass matrix in Eq.(\ref{Flavor0}) has 7 physical parameters. However, it is in general expected that deviations of $y_2, y_3$ from unity, leading to recent neutrino data, {\it i.e.} $\theta_{13}\neq0$, and in turn opening a possibility to search for $CP$ violation in neutrino oscillation experiments.
These deviations generate relations between mixing angles and mass eigenvalues.
Therefore Eq.~(\ref{mass matrix}) directly indicates that there could be deviations from the exact TBM if the Dirac neutrino Yukawa couplings do not have the same magnitude, and the light neutrino masses are all of same order
 \begin{eqnarray}
 m_{\nu_1}\simeq m_{\nu_2}\simeq m_{\nu_3}\simeq {\cal O}(m_{0})~.
 \label{}
 \end{eqnarray}

Before discussing on quarks and axions, let us consider the constraints on the $X$-symmetry (or PQ symmetry) breaking scale implied by the fermion mass scales in the model. From the overall scale of the light neutrino mass in Eq.~(\ref{Numass1}) the scale of the heavy neutrino, which is connected to the PQ symmetry breaking scale via the axion decay constant in Eq.~(\ref{fA1}), is expected to be
\begin{eqnarray}
  M\simeq5\times10^{12}\left(\frac{{\rm eV}}{m_{0}}\right)\left|\hat{y}^{\nu}_{1}\frac{v_{T}}{\Lambda}\right|^{2}\sin^2\beta~{\rm GeV}~.
 \label{scaleLambda}
\end{eqnarray}
As shown in Eq.~(\ref{vevdirection2}), the scale of $M$ is expected as ${\cal O}(v_{\Theta})\sim{\cal O}(v_{S})\sim{\cal O}(M)$. And Eq.~(\ref{scaleLambda}) shows that the value of $\hat{y}^{\nu}_{1}v_{T}/\Lambda$ depends on the magnitude $M$ once $m_{0}$ is determined: the smaller the ratio $v_{T}/\Lambda$, the smaller becomes the leptogenesis (seesaw) scale~\footnote{Moreover, the overall scale of the light neutrino mass $m_{0}$ is closely related with a successful leptogenesis~\cite{Khlopov, review}, constraints of lepton flavor violation and $0\nu\beta\beta$-decay rate through the seesaw formula as well as the CKM mixing matrix, therefore it is very important to fit the parameters $v_{T}/\Lambda$ and $M$.}. The value of $v_{T}/\Lambda$ is also related to the $\mu$-term in Eq.~(\ref{muterm}): when soft SUSY breaking terms are included into the flavon potential, the driving fields attain VEVs, and in turn the magnitude of $\mu$-term is expected to be $200~{\rm GeV}\lesssim\mu_{\rm eff}\lesssim1$ TeV for $m_S\sim{\cal O}(10)$ TeV and $v_{T}/\Lambda\sim0.05$. For example, when the Yukawa coupling $\hat{y}^{\nu}_{1}$ being of order of unity, {\it i.e.}, $1/\sqrt{10}\lesssim|\hat{y}^{\nu}_{1}|\lesssim\sqrt{10}$, and $\sin\beta\simeq1$ due to Eq.~(\ref{tanbeta}) are considered, the scale $M$ should be close to
\begin{eqnarray}
  7.5\times10^{11}\lesssim M\,[{\rm GeV}]\lesssim7.5\times10^{13}\qquad\text{for}~\frac{v_{T}}{\Lambda}\simeq0.05\,.
 \label{scaleM}
\end{eqnarray}
Since the values of $v_{T}/\Lambda$ and $v_{S}/\Lambda$ are closely associated with the CKM mixing matrix and the down-type quark masses, respectively, see Eq.~(\ref{MDMD}), their values should lie in the ranges
\begin{eqnarray}
  \frac{v_{T}}{\Lambda}\sim{\cal O}(0.1)\,,\qquad \frac{v_{S}}{\Lambda}\lesssim\frac{v_{\Theta}}{\Lambda}\sim\lambda^2<\frac{v_{\Psi}}{\Lambda}=\lambda<1\,.
 \label{MassRangge}
\end{eqnarray}
Here the first term is derived from the requirement that the term should fit its size down to generate the correct CKM matrix in Eq.~(\ref{MDMD}) as well as the $\mu$-term in Eq.~(\ref{muterm}), and the second one comes from Eqs.~(\ref{vevdirection2}) and (\ref{Cabbibo}), and  $v_{\Theta}=v_{\Psi}N_1/N_2\sqrt{1+\kappa^2}$ with $N_1=3$, $N_2=17$ and $\kappa\equiv v_{S}/v_{\Theta}$ (see also its related parameter $\tilde{\kappa}$ in Eq.~(\ref{MR2})), which will be shown in Eq.~(\ref{AhnMass}). With the assumptions $\hat{y}_{\Theta}\simeq\hat{y}^{\nu}_1$, $\kappa\simeq0.5$ and $v_{T}/\Lambda\simeq0.05$, the neutrino overall scale $m_0\simeq(1-5)\times10^{-2}$ eV gives $10^{11}\lesssim v_{\Theta}[{\rm GeV}]\lesssim4\times10^{12}$. Thus it is very likely that the PQ symmetry breaking scale roughly lies in $7\times10^{11}\lesssim v_{\Psi}[{\rm GeV}]\lesssim2.8\times10^{13}$.

In conclusion, all VEVs (scaled by the cutoff $\Lambda$) breaking the symmetries are connected each other: (i) the VEV $v_{T}$ is correlated with both the $\mu$-term in Eq.~(\ref{muterm}) and the overall scale of light neutrino mass through the seesaw formula, Eq.~(\ref{Numass1}), and its size scaled by the cutoff $\Lambda$ is crucial for generating the correct CKM matrix. (ii) The scale between $v_{S}=\kappa\,v_{\Theta}$ and $v_{T}$ is determined by the overall scale of light neutrino mass through the seesaw formula. (iii) The VEV $v_{\Psi}$ (scaled by the cutoff $\Lambda$), which is defined as the Cabbibo parameter in Eq.~(\ref{Cabbibo}), is connected to the scale $v_{\Theta}$ or $v_{S}$ via the axion constraints, Eqs.~(\ref{AhnMass}-\ref{fA1}), in turn thereby the cutoff scale $\Lambda$ is determined.

\subsubsection{Light neutrino Phenomenology}
\noindent After the observation of a non-zero mixing angle $\theta_{13}$ in the Daya Bay~\cite{An:2012eh} and RENO~\cite{Ahn:2012nd} experiments, the Dirac CP-violating phase $\delta_{CP}$ and a precise measurement of the atmospheric mixing angle $\theta_{23}$ are the next observables on the agenda of neutrino oscillation experiments. We explore what values of the low energy CP phases can predict a value for the mass hierarchy of neutrino (normal or inverted mass ordering)
and investigate the observables that can be tested in the current and the next generation of experiments: the rate of $0\nu\beta\beta$-decay via the effective mass $|({\cal M}_{\nu})_{ee}|$ (the modulus of the $ee$-entry of the effective neutrino mass matrix) at $90\%$ C.L shows upper bounds :
 \begin{eqnarray}
  |({\cal M}_{\nu})_{ee}| &<&0.12-0.25\,{\rm eV},\qquad (^{136}\text{Xe-based experiments~\cite{Gando:2012zm, Auger:2012ar}})\nonumber\\
  |({\cal M}_{\nu})_{ee}| &<&0.20-0.40\,{\rm eV},\qquad (^{76}\text{Ge-based experiments~\cite{Agostini:2013mzu, KlapdorKleingrothaus:2000sn, Aalseth:2004wf}}).
 \label{expLNV}
 \end{eqnarray}
Current $0\nu\beta\beta$-decay experimental upper limits and the reach of near-future experiments are collected for example in Ref.~\cite{Schwingenheuer:2012zs}. Recently, there are two interesting measurements on the sum of the light neutrino masses $\sum_{i=1}^{3} m_{\nu_i}$, (i) the first one given by Planck Collaboration~\cite{Ade:2013zuv} is subject to the cosmological bounds $\sum_{i}m_{\nu_i}<0.23$ eV at $95\%$ CL (Planck-I, derived from the combination Planck + WMAP low-multipole polarization + high resolution CMB + baryon acoustic oscillations (BAO), assuming a standard $\Lambda$CDM cosmological model) and $\sum_{i}m_{\nu_i}<0.66$ eV at $95\%$ CL (Planck-II, derived from the data without BAO~\cite{Ade:2013zuv}), and (ii) the other one from the South Pole Telescope (SPT) Collaboration~\cite{SPTC} states a $3\,\sigma$ preference for positive neutrino masses, and the median value is 
 \begin{eqnarray}
 \sum_{i}m_{\nu_i}=0.32\pm0.11~\text{eV}.
  \label{SPT}
 \end{eqnarray}

We perform a numerical analysis using the linear algebra tools Ref.~\cite{Antusch:2005gp}. The Daya Bay~\cite{An:2012eh} and RENO~\cite{Ahn:2012nd} experiments have accomplished the measurement of all three neutrino mixing angles $\theta_{12}$, $\theta_{23}$, and $\theta_{13}$, associated with three kinds of neutrino oscillation experiments. Global fit values and $3\sigma$ intervals for the neutrino mixing angles and the neutrino mass-squared differences~\cite{GonzalezGarcia:2012sz}
are listed in Table~\ref{exp}~\footnote{The model parameter spaces constrained by the global analysis in Table I are slightly different from those of Ref.~\cite{Ahn:2014zja} where the global analysis by Ref.~\cite{GonzalezGarcia:2012sz} were used.}.
The mass matrices $m_{D}$ in Eq.~(\ref{Ynu1}) and $M_{R}$ in Eq.~(\ref{MR1}) contain seven parameters : $y_{1}(\equiv v_{T}\hat{y}^{\nu}_{1}/\sqrt{2}\Lambda),v_{u}, \,M,\,y_{2},\,y_{3},\,\tilde{\kappa},\,\phi$. The first three ($y_{1}$, $M,$ and $v_{u}$) lead to the overall neutrino scale parameter $m_{0}$. The next four ($y_2,\,y_3,\,\tilde{\kappa},\,\phi$) give rise to the deviations from TBM as well as the CP phases and corrections to the mass eigenvalues (see Eq.~(\ref{TBM1})).
In our numerical analysis, we take $M=10^{12}$ GeV and~\footnote{As noticed in Eq.~(\ref{tanbeta}), in our model small values of $\tan\beta=v_{u}/v_{d}$ are preferred.} $\tan\beta=5$ (see Eq.~(\ref{scaleM}) and Eq.~(\ref{tanbeta})), for simplicity, as inputs.
Then the effective neutrino mass matrix in Eq.~(\ref{mass matrix}) contains only the five parameters $m_{0},y_{2},y_{3},\tilde{\kappa},\phi$, which can be determined from the experimental results of three mixing angles, $\theta_{12},\theta_{13},\theta_{23}$, and the
two mass squared differences, $\Delta m^{2}_{\rm Sol}=m^{2}_{\nu_2}-m^{2}_{\nu_1}, \Delta m^{2}_{\rm Atm}=|m^{2}_{\nu_3}-m^{2}_{\nu_1}|$. In addition, the effective neutrino mass $|{\cal M}_{ee}|$ and the CP phases $\delta_{CP},\varphi_{1,2}$ can be predicted after determining the model parameters. Scanning all the parameter spaces by putting the experimental constraints in Table~\ref{exp} with the above input parameters, we obtain for the normal mass ordering (NO)
 \begin{align}
  &\tilde{\kappa} \in [0.17,0.73] ,
  && y_{2} \in [1.0,1.25],
  \hspace{6em} & y_{3} \in [1.0,1.25],
  \nonumber\\
  & m_{0}/(10^{-2}{\rm eV}) \in [1.5,5.3],
  && \phi \in [96^{\circ},114^{\circ}]\cup[246^{\circ},266^{\circ}];
  \label{input1}
 \end{align}
 for the inverted mass ordering (IO)
 \begin{align}
  &\tilde{\kappa} \in [0.17,0.63] ,
  && y_{2} \in [0.80,1.16],
  \hspace{6em} y_{3} \in [0.82,1.17],
  \nonumber\\
  & m_{0}/(10^{-2}{\rm eV}) \in [2.3,5.9],
  && \phi \in [93^{\circ},104^{\circ}]\cup[255^{\circ},267^{\circ}] .
  \label{input2}
 \end{align}

First, the magnitude of the CP-violating effects is determined by the invariant $J_{CP}$ associated with the Dirac CP-violating phase
 \begin{eqnarray}
J_{CP}\equiv-{\rm Im}[U^{\ast}_{e1}U_{e3}U_{\tau1}U^{\ast}_{\tau3}]=\frac{1}{8}\sin2\theta_{12}\sin2\theta_{13}\sin2\theta_{23}\cos\theta_{13}\sin\delta_{CP}.
 \end{eqnarray}
Here $U_{\alpha j}$ is an element of the PMNS matrix in Eq.~(\ref{PMNS}), with $\alpha=e,\mu,\tau$
corresponding to the lepton flavors and $j=1,2,3$ corresponding to the light neutrino mass eigenstates.
Due to the precise measurement of $\theta_{13}$, which is relatively large, it may now be possible to put constraints on the Dirac phase $\delta_{CP}$ which will be obtained in the long baseline  neutrino oscillation experiments T2K, NO$\nu$A, etc. (see, Ref.~\cite{PDG}). However, the current large uncertainty on $\theta_{23}$ is at present limiting the information that can be extracted from the $\nu_{e}$ appearance measurements. Precise measurements of all the mixing angles are needed to maximize the sensitivity to the leptonic CP violation.
\begin{figure}[b]
\begin{minipage}[h]{7.5cm}
\epsfig{figure=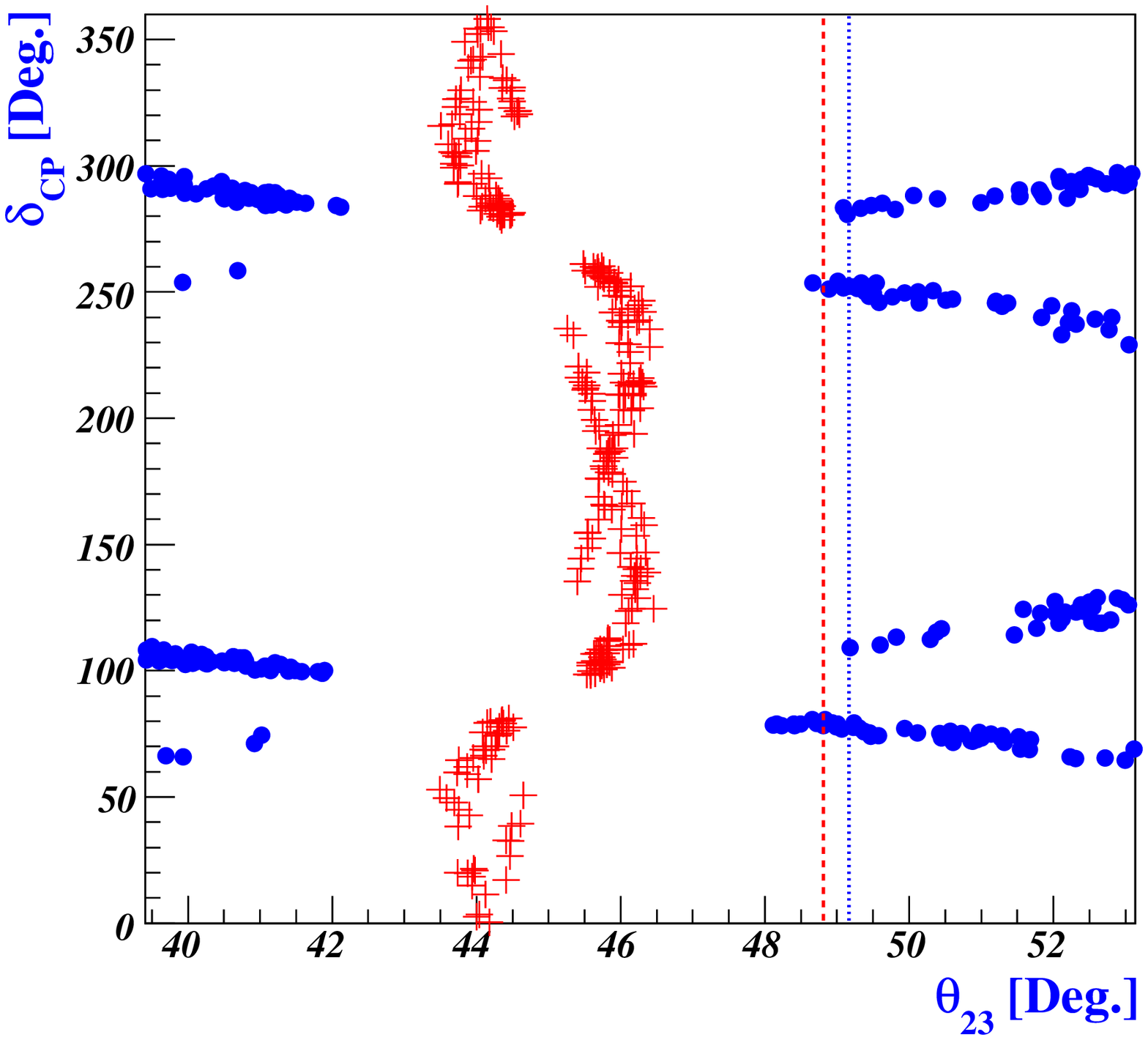,width=8.0cm,angle=0}
\end{minipage}
\hspace*{1.0cm}
\begin{minipage}[h]{7.5cm}
\epsfig{figure=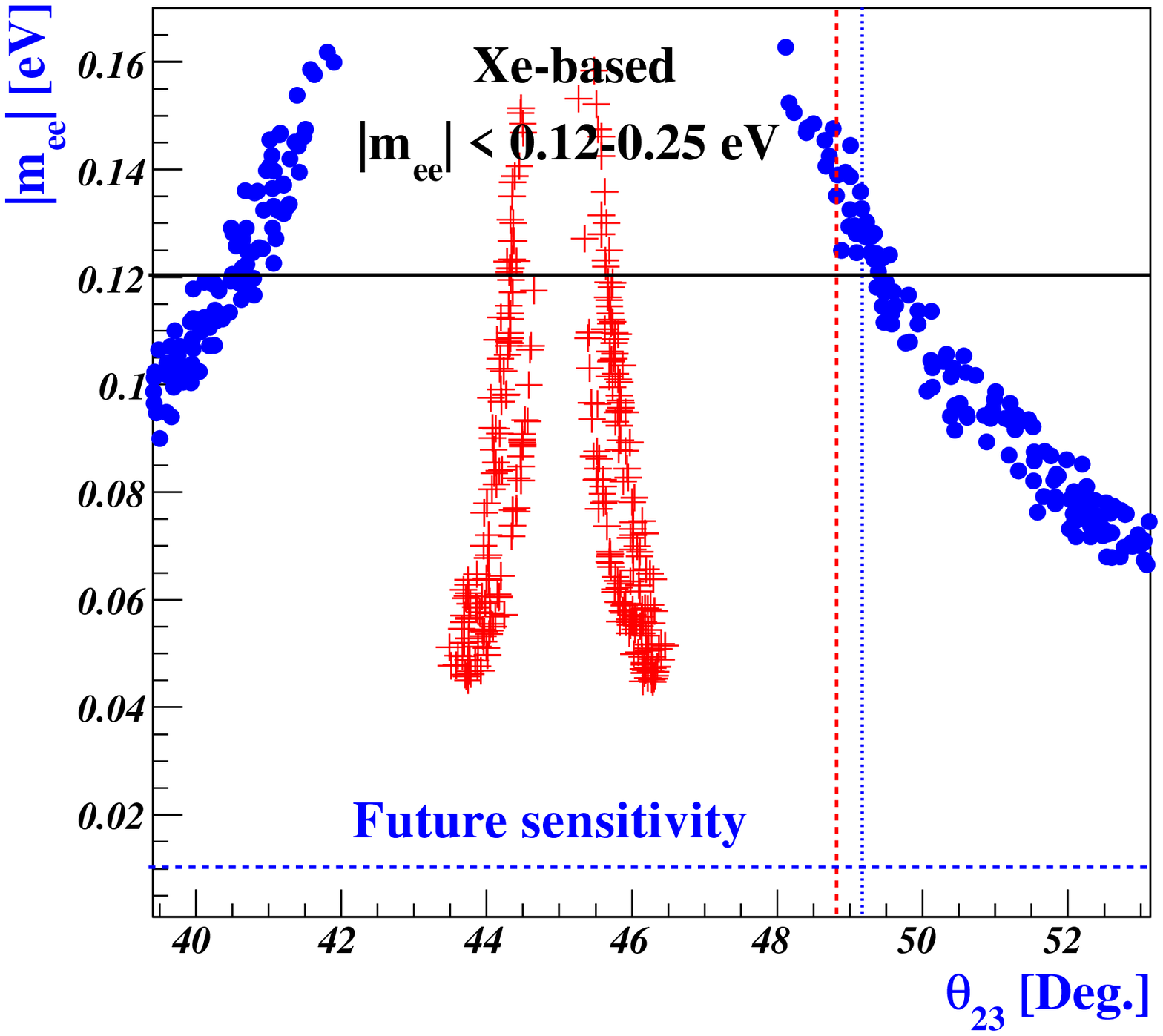,width=8.0cm,angle=0}
\end{minipage}
\caption{\label{FigA1}
 Left plot for predictions of $\delta_{CP}$ as a function of $\theta_{23}$, while right plot for model prediction of $|m_{ee}|\equiv|({\cal M}_\nu)_{ee}|$ in terms of $\theta_{23}$. Here the vertical dashed (dotted) lines show the best-fit values for NO (IO), and the blue dots and red crosses correspond to IO and NO, respectively. And the horizontal solid (dashed) lines show the Xe-based current bounds (near future reachable sensitivity) of $0\nu\beta\beta$ experiments.}
\end{figure}
Since the $0\nu\beta\beta$-decay is a probe of lepton number violation at low energy, its measurement could be the strongest evidence for lepton number violation at high energy. In other words, the discovery of $0\nu\beta\beta$-decay would suggest the Majorana character of the neutrinos and thus the existence of heavy Majorana neutrinos (via seesaw mechanism), which are a crucial ingredient for leptogenesis~\cite{Khlopov, review}.
In the model, the effective neutrino mass $|({\cal M}_{\nu})_{ee}|$ that characterizes the amplitude for $0\nu\beta\beta$-decay is given by
 \begin{eqnarray}
  |({\cal M}_{\nu})_{ee}|= m_{0}\left|\frac{3+\tilde{\kappa}\,e^{i\phi}}{1+\tilde{\kappa}\,e^{i\phi}}\right|~.
  \label{mee}
 \end{eqnarray}
This shows that in the model the rate of $0\nu\beta\beta$-decay depends on the parameters $m_{0}$, $\tilde{\kappa}$, and $\phi$ associated with the heavy Majorana neutrinos in Eq.~(\ref{MR1}).
Fig.~\ref{FigA1} indicates the importance of the precise measurements of the atmospheric mixing angle $\theta_{23}$ to distinguish between normal mass ordering and inverted one; here the blue dots and red crosses correspond to the IO and the NO, respectively. The IO is very predictive on $\delta_{CP}$ and $|({\cal M}_\nu)_{ee}|\equiv|m_{ee}|$, while the NO is less predictive on those. The left plot in Fig.~\ref{FigA1} shows the predictions on $\delta_{CP}$ in terms of the large uncertainty on $\theta_{23}$,
on the other hand, the right plot stands for the model predictions on $|({\cal M}_\nu)_{ee}|$ in terms of $\theta_{23}$.
Within the model, future precise measurements of $\theta_{23}$ should be able to distinguish between IO and NO. For NO, $\theta_{23}$ would be close to $44^\circ$ or $46^\circ$. For IO, $\theta_{23}$ would be in the range $[38^\circ,42^\circ] \cup [48^\circ,53^\circ]$, that is $3^\circ$ to $8^\circ$ away from maximality. In turn, such precise measurements of $\theta_{23}$ would restrict the possible range of $\delta_{CP}$ in the model. A value of $\theta_{23}$ slightly larger than maximal, i.e.\ $\theta_{23} \in [45^\circ,47^\circ]$, would imply an NO and $\delta_{CP} \in [90^\circ,270^\circ]$, while a value of $\theta_{23}$ slightly smaller than maximal, i.e.\ $\theta_{23} \in [43^\circ,45^\circ]$, would imply an NO and $\delta_{CP} \in [0,90^\circ] \cup [270^\circ,360^\circ]$. A value of $\theta_{23}$ considerably larger or smaller than maximal, i.e.\ $[38^\circ,42^\circ] \cup [48^\circ,53^\circ]$, would imply IO and $\delta_{CP}$ within few degrees of $70^\circ$, $110^\circ$, $250^\circ$, or $290^\circ$.

\begin{figure}[h]
\begin{minipage}[h]{7.5cm}
\epsfig{figure=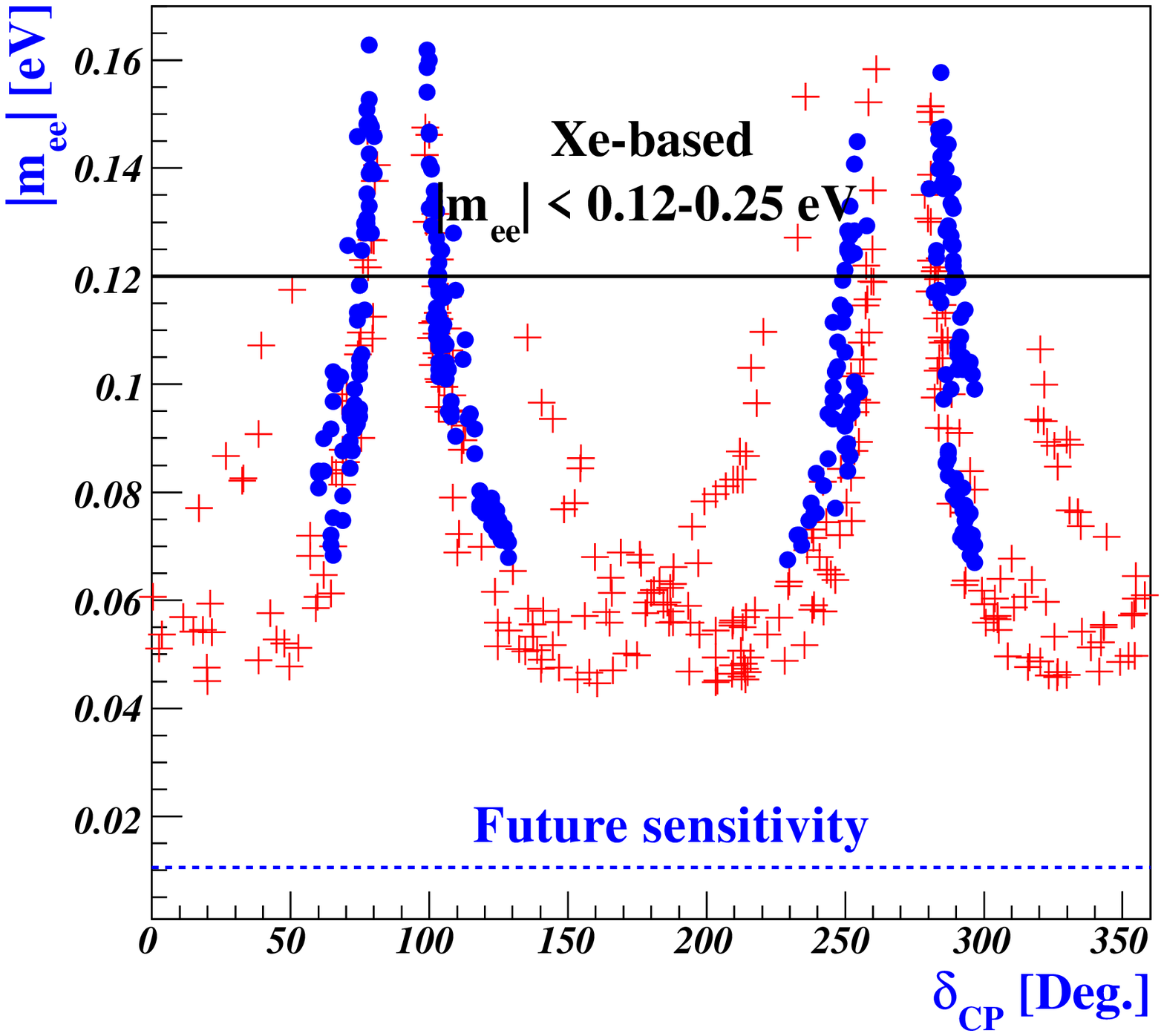,width=8.0cm,angle=0}
\end{minipage}
\hspace*{1.0cm}
\begin{minipage}[h]{7.5cm}
\epsfig{figure=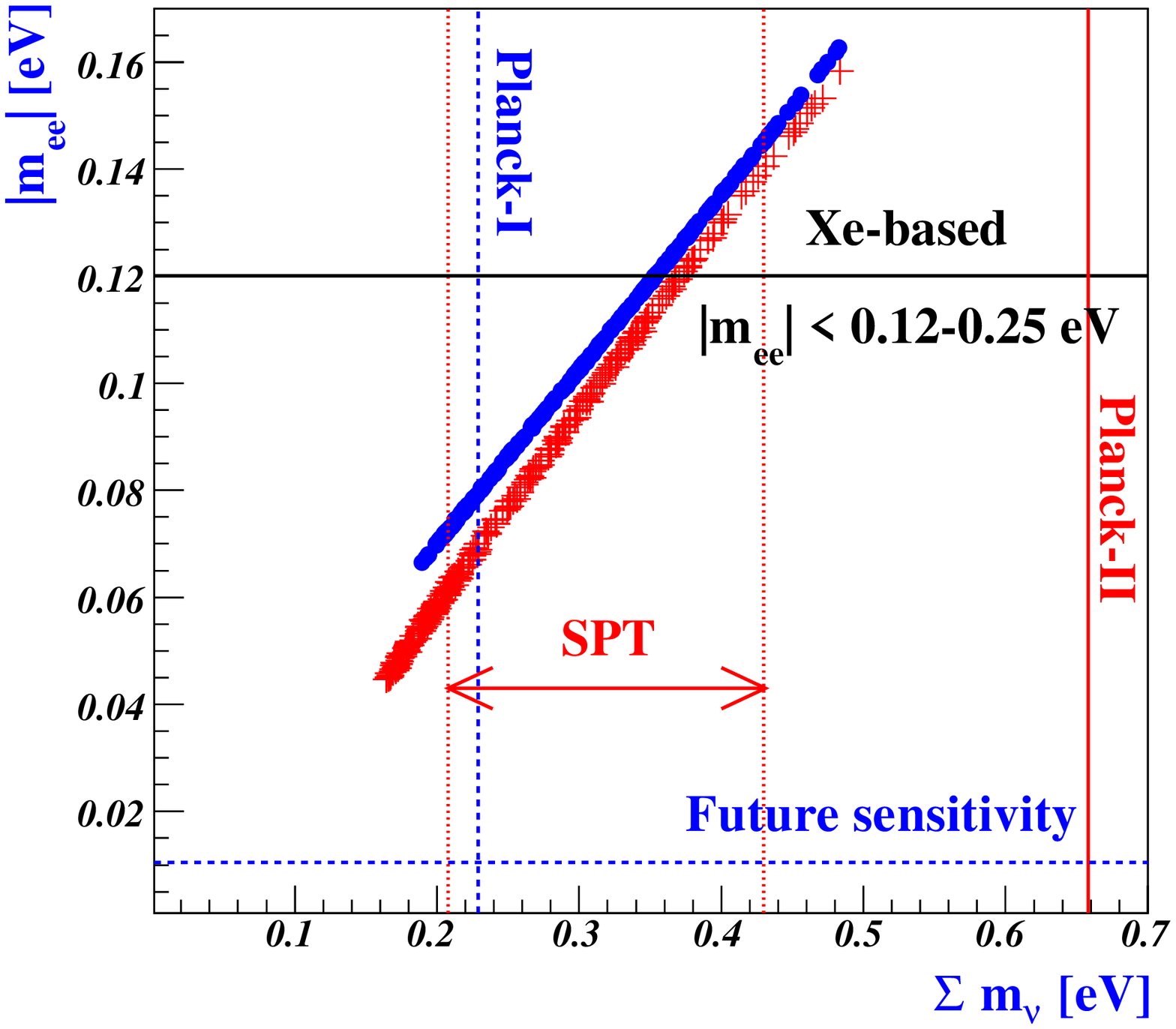,width=8.0cm,angle=0}
\end{minipage}
\caption{\label{FigA2}
  Plots for predictions of $|m_{ee}|\equiv|({\cal M}_\nu)_{ee}|$ in terms of $\delta_{CP}$ (left) and $\sum m_{\nu}$ (right). The horizontal solid (dashed) lines show the Xe-based current bounds (near future reachable sensitivity) of $0\nu\beta\beta$ experiments. In the right plot the vertical dashed (solid) lines indicate the cosmological Planck-I (Planck-II) upper bounds, while the two vertical dotted lines show the median value from the SPT in Ref.~\cite{SPTC}).}
\end{figure}
Recently, the T2K collaboration analyzes that the recent measurements of $\theta_{13}$ combined with the T2K data result in
 \begin{eqnarray}
  \theta_{23}=46.61^{+3.17}_{-2.18},\qquad\theta_{13}=9.10^{+0.47}_{-0.49},\qquad|\Delta m^{2}_{32}|=2.51\pm0.11
 \label{T2K}
 \end{eqnarray}
and exclude values of $\delta_{CP}$ between $25.2^{\circ}$ and $156.6^{\circ}$ with $90\%$ probability, which points highest posterior probability in the normal mass ordering~\cite{dePerio:2014zna}. Interestingly, as can be seen in Fig.~\ref{FigA1}, the recent analysis by the T2K collaboration, Eq.~(\ref{T2K}), favor the data points (red crosses) indicating the NO.
In the near future, KamLAND-Zen~\cite{Gando:2012zm}, EXO~\cite{Auger:2012ar}, and GERDA~\cite{Agostini:2013mzu} are expected~\footnote{AMoRE (Advanced Mo based Rare process Experiment) collaboration is now planning to search for $0\nu\beta\beta$-decay of $^{100}{\rm Mo}$ isotope, which reaches the sensitivity of the effective Majorana neutrino mass $|({\cal M}_\nu)_{ee}|\sim0.02-0.06$ eV~\cite{Bhang:2012gn}.} to probe the range $0.01~\text{eV}<|({\cal M}_{\nu})_{ee}|<0.1~\text{eV}$.
If these experiments measure a value of $|({\cal M}_{\nu})_{ee}|>0.01$ eV, the normal mass hierarchical spectrum would be strongly disfavored~\cite{Bilenky:2001rz}. Fig.~\ref{FigA2} directly shows that the model predictions lie on the testable region of those experiments. The correlations shown in the left plot in Fig.~\ref{FigA2} indicate that  in our model precise measurements of or improved upper bounds on $|({\cal M}_{\nu})_{ee}|$ from $0\nu\beta\beta$-decay experiments may be able to restrict the possible ranges of $\delta_{CP}$, and in some cases may even distinguish NO from IO.
In the right plot in Fig.~\ref{FigA2}, the more stringent Planck I limit cuts into our region of points and starts to disfavor a quasi-degenerate light neutrino mass spectrum. Interestingly, the data given in Eq.~(\ref{SPT}) from the SPT Collaboration~\cite{SPTC} favor our model as indicated in the left plot in Fig.~\ref{FigA2}.
Fig.~\ref{FigA2} explicitly shows that the current $0\nu\beta\beta$-decay experiments also cut into our region of points, and the near-future $0\nu\beta\beta$-decay experiments can test our model completely.
Remark that the tritium beta decay experiment KATRIN~\cite{KATRIN} may be not expected to reach into our model region. KATRIN will be sensitive to an effective electron neutrino mass  $m_{\beta}=\sqrt{\sum_{i}|U_{ei}|^{2}\,m^2_{\nu_i}}$~\cite{beta} down to about $0.2$ eV, while our model produces values in the range $0.050\lesssim m_{\nu_e}\lesssim0.160$ eV for NO and $0.051\lesssim m_{\nu_e}\lesssim0.171$ eV for IO.

\subsection{Quark sector}
\noindent In the quark sector, the superpotential $W_{q}$ driven by $\Phi_{T},\Phi_{S},\Theta,\Psi$, invariant under $SU(2)_L\times U(1)_{Y}\times A_{4}\times U(1)_{X}$, are given at leading order by
 \begin{eqnarray}
 W^{u}_{q} &=& y_{u}\,Q_{1}u^{c}\,H_{u}+y_{c}\,Q_{2}\,c^{c}\,H_{u}+y_{t}\,Q_{3}\,t^{c}\,H_{u}\,,\label{lagrangianU}\\
 W^{d}_{q} &=& y_{d}\,Q_{1}(D^c\Phi_{S})_{{\bf 1}}\,\frac{H_{d}}{\Lambda}+y_{s}\,Q_{2}(D^c\Phi_{S})_{{\bf 1}'}\,\frac{H_{d}}{\Lambda}+y_{b}\,Q_{3}(D^c\Phi_{S})_{{\bf 1}''}\,\frac{H_{d}}{\Lambda}\,.
 \label{lagrangianD}
 \end{eqnarray}
In the above superpotential, each quark sector has three independent Yukawa terms at the leading: apart from the Yukawa couplings, each up-type quark sector does not involve flavon fields, while the down-type quark sector involves the $A_{4}$-triplet flavon fields $\Phi_{T}$ and $\Phi_{S}$. The left-handed quark doublets $Q_1,Q_2,Q_3$ transform as ${\bf 1}, {\bf 1}''$, and ${\bf 1}'$, respectively; the right-handed quarks $u^{c}\sim{\bf 1}, c^{c}\sim{\bf 1}', t^{c}\sim {\bf 1}''$ and $D^{c}\equiv\{d^{c}, s^{c}, b^{c}\}\sim {\bf 3}$. Since the right-handed down-type quark transforms as ${\bf 3}$, in contrast with the up-type quark sector, the down-type quark sector can have nontrivial next-to-leading order terms as will be shown in Eq.~(\ref{lagrangianD1}).

According to the $U(1)_{X}$ quantum numbers assigned in Table~\ref{DrivingRef} and \ref{reps}, it is expected that the flavon field $A_{4}$-singlet $\Psi$ derives higher-dimensional operators, which are eventually visualized into the Yukawa couplings of up-type quarks as a function of flavon field $\Psi$, {\it i.e.} $y_{u,c}=y_{u,c}(\Psi)$, except for the top Yukawa coupling :
\begin{eqnarray}
y_{u}&=&\hat{y}_{u}\left(\frac{\Psi}{\Lambda}\right)^{8},\qquad\qquad y_{c}=\hat{y}_{c}\left(\frac{\Psi}{\Lambda}\right)^{4},\qquad\qquad y_{t}=\hat{y}_{t}\,
 \label{YukawaWqu}
\end{eqnarray}
and, similarly, the Yukawa couplings of
down-type quarks as a function of flavon field $\Psi$, {\it i.e.} $y_{d,s}=y_{d,s}(\Psi)$, except for the Yukawa coupling $y_{b}$ :
\begin{eqnarray}
y_{d}&=&\hat{y}_{d}\left(\frac{\Psi}{\Lambda}\right)^{3},\qquad\qquad y_{s}=\hat{y}_{s}\left(\frac{\Psi}{\Lambda}\right)^2,\qquad\qquad y_{b}=\hat{y}_{b}\,.
 \label{YukawaWq}
\end{eqnarray}
Recalling that all the hat Yukawa couplings are of order unity and complex numbers.

Similar to the lepton sector, even the flavon fields $A_{4}$-triplet $\Phi_{S,T}$ and $A_{4}$-singlets $\Theta$, $\Psi$ derive higher-dimensional operators, they are all forbidden or vanishing. Notice that the effects of non-trivial next-to-leading order operators will be discussed in Sec.~\ref{corrects}. There a few comments are in order: (i) next-to-next-to-leading order operators driven by $\Phi_{S}$ or $\Theta$, and higher dimensional operators including $(\Phi_{S}\Phi_{S})_{{\bf 1},{\bf 1}',{\bf 1}''{\bf 3}}$ are all forbidden by the $U(1)_{X}$, (ii) higher dimensional operators driven by $(\Phi_{T}\Phi_{T})_{{\bf 1}',{\bf 1}''}$ are all vanishing due to the VEV alignment $\langle\Phi_T\rangle\sim v_{T}(1,0,0)$, for example, $\hat{y}_{if}\left(\frac{\Psi}{\Lambda}\right)^nQ_{i}f^cH_{u}(\Phi_{T}\Phi_{T})_{{\bf 1}', {\bf 1}''}/\Lambda^2$ where $i=1,2,3$, $f=u,c,t$ and $n\geq1$ (integer), and $(i,f)\neq(1,u),(2,c),(3,t)$,
(iii) higher dimensional operators through the insertions of $(\Phi_{T}\Phi_{T})_{{\bf 1}}$ or $(\Phi_{T}\Phi_{T})_{{\bf 3}_s}$, $(\Phi_{S}\Phi_{T})_{{\bf 1},{\bf 1}',{\bf 1}''}$ have a VEV in the same direction as $\Phi_{T}$ due to the VEV alignment $\langle\Phi_{T}\rangle\sim v_{T}(1,0,0)$, all of which are absorbed into a redefinition of the leading terms, and (iv) higher dimensional operators via the insertion of $H_{u}H_{d}$ and $\Psi\tilde{\Psi}$ are all absorbed into a redefinition of the leading order terms.


After the breaking of the flavor and electroweak symmetries, with the VEV alignments as in Eq.~(\ref{vev}), in the weak eigenstate basis the up- and down-type quark mass terms in Eqs.~(\ref{lagrangianU}) and (\ref{lagrangianD}), and the charged current interactions between quarks, can be written in matrix form as
 \begin{eqnarray}
 -{\cal L}_{q} &=& \overline{q^{u}_{L}}\,\mathcal{M}_{u}\,q^{u}_{R}+\overline{q^{d}_{L}}\,\mathcal{M}_{d}\,q^{d}_{R}
  +\frac{g}{\sqrt{2}}W^{+}_{\mu} ~\overline{q^{u}_{L}}\,\gamma^{\mu}\,q^{d}_{L} + {\rm h.c.} ~.
 \label{lagrangianChA}
 \end{eqnarray}
Here $q^{u}=(u,c,t)$, $q^{d}=(d,s,b)$, and
 \begin{eqnarray}
 {\cal M}_{u}&=& {\left(\begin{array}{ccc}
 y_{u} &  0 &  0 \\
 0 &  y_{c} &  0   \\
 0 &  0  &  y_{t}
 \end{array}\right)}v_{u}= {\left(\begin{array}{ccc}
 \hat{y}_{u}\,\lambda^8 &  0 &  0 \\
 0 &  \hat{y}_{c}\,\lambda^4 &  0   \\
 0 &  0  &  \hat{y}_{t}
 \end{array}\right)}v_{u}\,, \label{Quark1}\\
 {\cal M}_{d}&=&{\left(\begin{array}{ccc}
 y_{d} &  y_{d} &  y_{d} \\
 y_{s} &  y_{s} &  y_{s}   \\
 y_{b} &  y_{b}  &  y_{b}
 \end{array}\right)}\frac{v_{S}}{\sqrt{2}\Lambda}v_{d}
 ={\left(\begin{array}{ccc}
 \hat{y}_{d}\,\lambda^3 &  \hat{y}_{d}\,\lambda^3 &  \hat{y}_{d}\,\lambda^3 \\
 \hat{y}_{s}\,\lambda^2 &  \hat{y}_{s}\,\lambda^2 &  \hat{y}_{s}\,\lambda^2   \\
 \hat{y}_{b} &  \hat{y}_{b}  &  \hat{y}_{b}
 \end{array}\right)}\frac{v_{S}}{\sqrt{2}\Lambda}v_{d}\,.
 \label{Quark2}
 \end{eqnarray}
Naively speaking, since the leading matrix ${\cal M}_d$ has 6 physical parameters, while observables are seven (CKM parameters: 4, down-type quark masses: 3), its alone may not generate the correct CKM matrix.
With Eqs.~(\ref{Quark1}) and (\ref{Quark2}) they directly show that the mass
spectra of quarks are strongly hierarchical, {\it i.e.}, the masses of the third
generation fermions are much heavier than those of the first and second generation quarks.

Due to the diagonal form in Eq.~(\ref{Quark1}), the contributions of the up-type quark sector to the CKM matrix are absent. The mass eigenvalues of the up-type quark can be made real and positive by the field redefinitions $q^{u}_{L}\rightarrow P^{u}_{L}\,q^{u}_{L}$ and $q^{u}_{R}\rightarrow P^{u}_{R}\,q^{u}_{R}$ (here, $P^{u}_{L(R)}$ is a diagonal matrix of phase factors):
 \begin{eqnarray}
  \widehat{\mathcal{M}}_{u}=P^{u}_{L}\,\mathcal{M}_{u}\,P^{u\ast}_{R}={\rm diag}(m_{u},m_{c},m_{t})~.
 \end{eqnarray}
The corresponding up-type quark masses are given as
 \begin{eqnarray}
 m_{t}&\equiv& |\hat{y}_{t}|\,v_{u}\,,\qquad
 m_{c}\equiv |y_{c}|\,v_{u}=\lambda^{4}\,v_{u}\,|\hat{y}_{c}|\,,\qquad
 m_{u}\equiv |y_{u}|\,v_{u}=\lambda^{8}\,v_{u}\,|\hat{y}_{u}|\,,
 \label{Top1}
 \end{eqnarray}
which are comparable with the results calculated from the measured values~\cite{PDG}
 \begin{eqnarray}
  \frac{m_{u}}{m_{t}} &\simeq& 1.4\times 10^{-5}\,,\qquad\quad
  \frac{m_{c}}{m_{t}}\simeq7.4\times10^{-3}\,.
\label{MuRatio}
 \end{eqnarray}
From the top Yukawa coupling and pole mass ($\hat{y}_{t}$ and $m_{t}$) and the neutral Higgs VEV ratio ($\tan\beta=v_{u}/v_{d}$),
by requiring $\hat{y}_{t}$ to be order of one, $1/\sqrt{10}\lesssim|\hat{y}_{t}|\lesssim\sqrt{10}$, we have following allowed range for $\tan\beta$:
 \begin{eqnarray}
  1.7\lesssim\tan\beta<10
  \label{tanbeta}
 \end{eqnarray}
where~\footnote{We take a lower bound of $\tan\beta$ preferred in the Minimal Supersymmetric Standard Model (MSSM). For $\tan\beta<1.7$ the top quark Yukawa coupling blows up before the momentum scale $\mu\approx2\times10^{16}$ GeV.} we have used $m_{t}=173.07\pm0.52\pm0.72$ GeV~\cite{PDG}.

On the other hand,  ${\cal M}_{d}$ in Eq.~(\ref{Quark2}) generates the down-type quark masses:
 \begin{eqnarray}
 \widehat{\mathcal{M}}_{d}=V^{d\dag}_{L}\,{\cal M}_{d}\,V^{d}_{R}
 ={\rm diag}(m_{d},m_{s},m_{b})\,,
 \label{Quark21}
 \end{eqnarray}
where $V^{d}_{L}$ and $V^{d}_{R}$ are the diagonalization matrices for ${\cal M}_{d}$. Then $V^{d}_{L}$ and $V^{d}_{R}$ can be determined by diagonalizing the matrices
${\cal M}_{d}{\cal M}^{\dag}_{d}$ and ${\cal M}^{\dag}_{d}{\cal M}_{d}$, respectively, indicated from
Eq.~(\ref{Quark21}). Especially, the mixing matrix $V^{d}_{L}$ becomes one of the matrices composing
the CKM mixing matrix.
The Hermitian matrix ${\cal M}_{d}{\cal M}^{\dag}_{d}$ is diagonalized by the mixing matrix $V^{d}_{L}$:
 \begin{eqnarray}
 V^{d\dag}_{L}{\cal M}_{d}{\cal M}^{\dag}_{d}V^{d}_{L}&=&
 v^2_{d}\frac{3}{2}\left(\frac{v_{S}}{\Lambda}\right)^2\,V^{d\dag}_{L}{\left(\begin{array}{ccc}
 \lambda^{6}|\hat{y}_{d}|^2 & \lambda^5\hat{y}_{d}\hat{y}^{\ast}_{s} & \lambda^3\hat{y}_{d}\hat{y}^\ast_{b}  \\
 \lambda^5\hat{y}^{\ast}_{d}\hat{y}_{s} & \lambda^4|\hat{y}_{s}|^2 & \lambda^2\hat{y}_{s}\hat{y}^\ast_{b}   \\
 \lambda^3\hat{y}^\ast_{d}\hat{y}_{b} & \lambda^2\hat{y}^\ast_{s}\hat{y}_{b}  & |\hat{y}_{b}|^2
 \end{array}\right)}V^{d}_{L}\nonumber\\
 &=&{\rm diag}(|m_{d}|^{2}, |m_s|^{2}, |m_{b}|^{2})\,.
 \label{MDMD0}
 \end{eqnarray}
Due to the strong hierarchal structure of the Hermitian matrix, one can fit the results calculated from the measured values~\cite{PDG} :
 \begin{eqnarray}
  \frac{m_{d}}{m_{b}} &\simeq&  1.2\times 10^{-3}\,,\qquad
  \frac{m_{s}}{m_{b}} \simeq 2.4\times10^{-2}\,.
 \label{massRatio}
 \end{eqnarray}
However, as mentioned before, one could not obtain the correct CKM mixing matrix (it seems difficult to reproduce the correct CKM matrix in the standard parameterization in Ref.~\cite{PDG}). Therefore, we should include nontrivial next-to-leading order corrections in order to obtain the correct CKM matrix.

\section{Higher order Corrections}
\label{corrects}
Higher-dimensional operators invariant under $A_{4}\times U(1)_X$ symmetry, suppressed by additional powers of the cutoff scale $\Lambda$, can be added to the leading terms in the superpotential. The mass and mixing matrices of fermions can be corrected by both a shift of the vacuum configuration and nontrivial next-to-leading operators contributing to the Yukawa superpotential $W_{f}$.
We have shown in the previous section that the next-to-leading order corrections in the charged lepton and up-type quark Yukawa superpotentials are either vanishing or absorbed into a redefinition of the leading order terms. Here, we show that the next leading corrections in the Dirac neutrino, Majorana neutrino and down-type quark sectors are well under control.

\subsection{Corrections to the Yukawa superpotential}
\noindent In addition to the leading order Yukawa superpotential $W_{f}$, we should also consider those higher dimensional operators that could be induced by the flavon field $\Phi_{T}$ which is not charged under the $U(1)_{X}$.

\subsubsection{Corrections to the lepton sector}
\label{correLepton}
\noindent At the next leading order in the Majorana neutrino sector those operators triggered by the field $\Phi_{T}$ are written as
\begin{eqnarray}
(N^cN^c\Theta\Phi_{T})_{{\bf 1}}/\Lambda,\qquad(N^cN^c\Phi_{S}\Phi_{T})_{{\bf 1}}/\Lambda\,.
\end{eqnarray}
Here the first term, after symmetry breaking, is absorbed into the leading order terms in the superpotential~(\ref{lagrangian}) and the corresponding Yukawa couplings are redefined. On the other hand, the second term could be non-trivial and it can be clearly expressed as
\begin{eqnarray}
\Delta W_{\nu}&=&\frac{\hat{y}^{R}_{1}}{\Lambda}(N^cN^c)_{{\bf 1}}(\Phi_{S}\Phi_{T})_{{\bf 1}}+\frac{\hat{y}^{R}_{2}}{\Lambda}(N^cN^c)_{{\bf 1}'}(\Phi_{S}\Phi_{T})_{{\bf 1}''}+\frac{\hat{y}^{R}_{3}}{\Lambda}(N^cN^c)_{{\bf 1}''}(\Phi_{S}\Phi_{T})_{{\bf 1}'}\nonumber\\
&+&\frac{\hat{y}^{R}_{s}}{\Lambda}(N^cN^c)_{{\bf 3}_s}(\Phi_{S}\Phi_{T})_{{\bf 3}_s}+\frac{\hat{y}^{R}_{a}}{\Lambda}(N^cN^c)_{{\bf 3}_s}(\Phi_{S}\Phi_{T})_{{\bf 3}_s}\,.
\label{NewMR}
\end{eqnarray}
Indeed at order $1/\Lambda$, after symmetry breaking, there is a new structure contributing to $M_{R}$, whose contribution is written as
 \begin{eqnarray}
 \Delta M_{R}&=&\frac{v_{T}}{\Lambda\sqrt{6}}{\left(\begin{array}{ccc}
 \tilde{\kappa}_{1}+\frac{4}{3}\tilde{\kappa}_{s} & \tilde{\kappa}_{2}+\frac{1}{3}\tilde{\kappa}_{s}-\frac{1}{\sqrt{3}}\tilde{\kappa}_{a} &  \tilde{\kappa}_{3}+\frac{1}{3}\tilde{\kappa}_{s}+\frac{1}{\sqrt{3}}\tilde{\kappa}_{a} \\
 \tilde{\kappa}_{2}+\frac{1}{3}\tilde{\kappa}_{s}-\frac{1}{\sqrt{3}}\tilde{\kappa}_{a} &  \tilde{\kappa}_{3}+\frac{2}{3}\tilde{\kappa}_{s}-\frac{2}{\sqrt{3}}\tilde{\kappa}_{a} &  \tilde{\kappa}_{1}-\frac{2}{3}\tilde{\kappa}_{s}\\
 \tilde{\kappa}_{3}+\frac{1}{3}\tilde{\kappa}_{s}+\frac{1}{\sqrt{3}}\tilde{\kappa}_{a} &  \tilde{\kappa}_{1}-\frac{2}{3}\tilde{\kappa}_{s} &  \tilde{\kappa}_{2}-\frac{2}{3}\tilde{\kappa}_{s}+\frac{2}{\sqrt{3}}\tilde{\kappa}_{a}
 \end{array}\right)}M~,
 \label{MRCorr}
 \end{eqnarray}
where $\tilde{\kappa}_{i}\equiv\sqrt{\frac{3}{2}}\frac{v_{S}}{M}\hat{y}^{R}_{i}$ with $i=1,2,3,s,a$. Even though these corrections to the leading order picture seem to non-trivial, these can be kept small, below the percent level due to $v_{T}/\sqrt{6}\Lambda\simeq0.02$ with $v_{T}/\Lambda\simeq0.05$, Eq.~(\ref{scaleM}), and $\tilde{\kappa}_i\simeq\tilde{\kappa}$ with Eq.~(\ref{MR2}). Therefore, the mass and mixing matrices of neutrino at leading order can not be crucially changed.

\subsubsection{Corrections to the quark sector}
\noindent The non-trivial next leading order operators induced by the $\Phi_{T}$ field in the down-type quark sector are written as
 \begin{eqnarray}
 \Delta W^{d}_{q} &=& x_{d}\,Q_{1}(D^c\Phi_{T})_{{\bf 1}}\,\frac{\Theta}{\Lambda^2}\,H_{d}+x_{s}\,Q_{2}(D^c\Phi_{T})_{{\bf 1}'}\,\frac{\Theta}{\Lambda^2}\,H_{d}+x_{b}\,Q_{3}(D^c\Phi_{T})_{{\bf 1}''}\,\frac{\Theta}{\Lambda^2}\,H_{d}\,\nonumber\\
 &+&x^{as}_{d}\,Q_{1}(D^c\Phi_{T}\Phi_{S})_{{\bf 1}}\,\frac{H_{d}}{\Lambda^2}+x^{as}_{s}\,Q_{2}(D^c\Phi_{T}\Phi_{S})_{{\bf 1}'}\,\frac{H_{d}}{\Lambda^2}+x^{as}_{b}\,Q_{3}(D^c\Phi_{T}\Phi_{S})_{{\bf 1}''}\,\frac{H_{d}}{\Lambda^2}\,.
 \label{lagrangianD1}
 \end{eqnarray}
Here, the next-to-leading order terms associated with the field $\Theta$ play crucial roles for the CKM mixing angles to be correctly fitted, while the contributions associated with the field $\Phi_{S}$ including the coefficients $x^{s}_{f}$ (which are from symmetric operators) and $x^{a}_{f}$ (which are from anti-symmetric operators) do cancel each other out at leading contribution due to the character of symmetry and anti-symmetry (the first contributions to the CKM appear as $\lambda^4$). Moreover, these next-to-leading order terms are correlated with the mass scale of neutrino in Eq.~(\ref{Numass1}) and the $\mu$-term in Eq.~(\ref{muterm}) through the flavon field $\Phi_{T}$.

In the above superpotential~(\ref{lagrangianD1}), the Yukawa couplings of
down-type quarks are expressed as a function of flavon field $\Psi$, {\it i.e.} $\mathbf{x}_{d,s}=\mathbf{x}_{d,s}(\Psi)$ with $\mathbf{x}=x,\,x^{as}$ :
\begin{eqnarray}
\mathbf{x}_{d}&=&\hat{\mathbf{x}}_{d}\left(\frac{\Psi}{\Lambda}\right)^{3},\qquad\qquad \mathbf{x}_{s}=\hat{\mathbf{x}}_{s}\left(\frac{\Psi}{\Lambda}\right)^2,\qquad\qquad \mathbf{x}_{b}=\hat{\mathbf{x}}_{b}\,.
 \label{YukawaWq}
\end{eqnarray}
With the help of Eq.~(\ref{YukawaWq}) the corrections of down-type quark matrix $\Delta{\cal M}_{d}$ can be expressed as
 \begin{eqnarray}
 \Delta{\cal M}_{d}&=&{\left(\begin{array}{ccc}
 x_{d} & 0 & 0 \\
 0 & x_{s} & 0   \\
 0 & 0 & x_{b}
 \end{array}\right)}\frac{v_{T}v_{\Theta}}{2\Lambda^2}v_{d}+{\small\left(\begin{array}{ccc}
 \frac{2}{\sqrt{3}}x^{s}_{d} & x^{+}_{d} & x^{-}_{d} \\
 \frac{2}{\sqrt{3}}x^{s}_{s} & x^{+}_{s} & x^{-}_{s} \\
 \frac{2}{\sqrt{3}}x^{s}_{b} & x^{+}_{b} & x^{-}_{b}
 \end{array}\right)}\frac{v_{T}v_{S}}{2\Lambda^2}v_{d}\label{DQuarkCorr0}\\
 &=&
 {\left(\begin{array}{ccc}
 \lambda^3\,(\frac{\hat{x}_{d}}{\kappa}+\frac{2\hat{x}^{s}_{d}}{\sqrt{3}}) &  \lambda^3\,\hat{x}^{+}_{d} &  \lambda^3\,\hat{x}^{-}_{d} \\
 \lambda^2\,\frac{2\hat{x}^{s}_{s}}{\sqrt{3}} &  \lambda^2\,(\frac{\hat{x}_{s}}{\kappa}+\hat{x}^{+}_{s}) &  \lambda^2\,\hat{x}^{-}_{s} \\
 \frac{2\hat{x}^{s}_{b}}{\sqrt{3}} &  \hat{x}^{+}_{b}   &  \frac{\hat{x}_{b}}{\kappa}+\hat{x}^{-}_{b}
 \end{array}\right)}\frac{v_{T}v_{S}}{2\Lambda^2}v_{d},
 \label{DQuarkCorr}
 \end{eqnarray}
where $\hat{x}^{\pm}_{f}=-\frac{\hat{x}^{s}_{f}}{\sqrt{3}}\pm i\hat{x}^{a}_{f}$. Recalling that all the hat Yukawa couplings are of order unity and complex numbers.
Each row of the leading matrix in Eq.~(\ref{Quark2}) has the same entries, while for the next leading order matrix in the second matrix in Eq.~(\ref{DQuarkCorr0}) the first term in each row cancels out the second term plus third one, therefore in the production $({\cal M}_d+\Delta{\cal M}_d)({\cal M}^{\dag}_{d}+\Delta{\cal M}^{\dag}_d)$ the mismatch between the leading matrix in Eq.~(\ref{Quark2}) and the second matrix for the next leading matrices in Eq.~(\ref{DQuarkCorr0}) cancel each other out, and the mismatch between the first matrix in $\Delta{\cal M}_d$ and the second one can contribute to the CKM matrix but its effects is below the few percent level. However, a mismatch between the first matrix, $\Delta{\cal M}_d$, and the leading matrix, Eq.~(\ref{Quark2}), can reproduce the masses of down-type quarks, $|V_{ub}|$ and $\delta^{q}_{CP}$, once $\theta^{q}_{23}=A\lambda^2$ and $\theta^q_{12}=\lambda$ are determined. We will show this non-trivial effects and analyze its physical effects in Sec.~\ref{corrMassCKM}.

\subsection{Corrections to the vacuum alignment}
\label{VEVcorrects}
\noindent Now we consider higher dimensional operators induced by $\Phi_{T},\Phi_{S},\Theta, \Psi$ invariant under $A_{4}\times U(1)_X$ in the driving superpotential $W_{v}$, which are suppressed by additional powers of the cut-off scale $\Lambda$. They can lead to small deviations from the leading order vacuum alignments.

The next leading order superpotential $\delta W_{v}$, which is linear in the driving fields and invariant under $A_{4}\times U(1)_{X}$, is given by
 \begin{eqnarray}
 \delta W_{v}&=& \frac{1}{\Lambda}\Big\{a_{1}(\Phi_{T}\Phi_{T})_{{\bf 3s}}(\Phi_{T}\Phi^{T}_{0})_{{\bf 3a}}+a_{2}(\Phi_{T}\Phi_{T})_{{\bf 1}}(\Phi_{T}\Phi^{T}_{0})_{{\bf 1}}+a_{3}(\Phi_{T}\Phi_{T})_{{\bf 1}'}(\Phi_{T}\Phi^{T}_{0})_{{\bf 1}''}\nonumber\\
 &+&a_{4}(\Phi_{T}\Phi_{T})_{{\bf 1}''}(\Phi_{T}\Phi^{T}_{0})_{{\bf 1}'}+a_{5}\Psi\tilde{\Psi}(\Phi_{T}\Phi^{T}_{0})_{{\bf 1}}\Big\}\nonumber\\
 &+&\frac{1}{\Lambda}\Big\{b_{1}(\Phi_{S}\Phi_{S})_{{\bf 3s}}(\Phi_{T}\Phi^{S}_{0})_{{\bf 3a}}
 +b_{2}(\Phi_{S}\Phi_{S})_{{\bf 3s}}(\Phi_{T}\Phi^{S}_{0})_{{\bf 3s}}+b_{3}(\Phi_{S}\Phi_{S})_{{\bf 1}}(\Phi_{T}\Phi^{S}_{0})_{{\bf 1}}\nonumber\\
 &+&b_{4}(\Phi_{S}\Phi_{S})_{{\bf 1}'}(\Phi_{T}\Phi^{S}_{0})_{{\bf 1}''}+b_{5}(\Phi_{S}\Phi_{S})_{{\bf 1}''}(\Phi_{T}\Phi^{S}_{0})_{{\bf 1}'}+b_{6}\Phi^{S}_{0}(\Phi_{S}\Phi_{T})_{{\bf 3a}}\Theta\nonumber\\
 &+&b_{7}\Phi^{S}_{0}(\Phi_{S}\Phi_{T})_{{\bf 3s}}\Theta+b_{8}\Phi^{S}_{0}(\Phi_{S}\Phi_{T})_{{\bf 3a}}\tilde{\Theta}+b_{9}\Phi^{S}_{0}(\Phi_{S}\Phi_{T})_{{\bf 3s}}\tilde{\Theta}
 \nonumber\\
 &+&b_{10}(\Phi^{S}_{0}\Phi_{T})_{{\bf 1}}\Theta\Theta
 +b_{11}(\Phi^{S}_{0}\Phi_{T})_{{\bf 1}}\Theta\tilde{\Theta}+b_{12}(\Phi^{S}_{0}\Phi_{T})_{{\bf 1}}\tilde{\Theta}\tilde{\Theta} \Big\}\nonumber\\
 &+&\frac{\Theta_{0}}{\Lambda}\left\{c_{1}(\Phi_{S}\Phi_{S})_{{\bf 3s}}\Phi_{T}+c_{2}(\Phi_{S}\Phi_{T})_{{\bf 1}}\tilde{\Theta}\right\}+\frac{\Psi_{0}}{\Lambda}d_{1}(\Phi_{T}\Phi_{T})_{{\bf 3s}}\Phi_{T}\,.
 \label{Npotential}
 \end{eqnarray}
By keeping only the first order in the expansion, we obtain the minimization equations. The details are in Appendix~\ref{highCorrect}.
The corrections to the VEVs, Eqs.~(\ref{vevdirection1},\ref{vevdirection2},\ref{vevdirection3}), are of relative order $1/\Lambda$ and affect the flavon fields $\Phi_{S}$, $\tilde{\Theta}$ and $\Psi$, and the vacuum configuration is modified into
 \begin{eqnarray}
 &&\langle\Phi_{T}\rangle\rightarrow\frac{1}{\sqrt{2}}(v_{T}+\delta v_{T_1},0,0),\qquad\qquad\langle\Theta\rangle\rightarrow \frac{v_{\Theta}}{\sqrt{2}},\qquad\qquad\langle\tilde{\Theta}\rangle\rightarrow \delta \tilde{\Theta}~,\nonumber\\
 &&\langle\Phi_{S}\rangle\rightarrow\frac{1}{\sqrt{2}}(v_{S}+\delta v_{S_{1}}, v_{S}+\delta v_{S_{2}},  v_{S}+\delta v_{S_{3}}),\qquad\langle\Psi\rangle\rightarrow \frac{v_{\Psi}}{\sqrt{2}}+\delta v_{\Psi}\,.
 \label{NewVEV}
 \end{eqnarray}
If there are no fine-tuning among the dimensionless parameters ($a_{1}...a_{5}$, $b_{1}...b_{12}$, $c_{1}, c_{2}$, $d_{1}$), when $v_{T}/\Lambda\sim{\cal O}(0.1)$ it is expected that
 \begin{eqnarray}
  &&|\delta v_{\Psi}|\sim{\cal O}(0.01)\,v_{T}\,,\nonumber\\
  &&|\delta\tilde{\Theta}|\sim
  |\delta v_{S_{1}}|\sim |\delta v_{S_{2}}|\sim |\delta v_{S_{3}}|\sim{\cal O}(0.1)\,v_{S}\,.
 \label{sizeDevi}
 \end{eqnarray}
From Appendix~\ref{highCorrect}, given the expected range for $v_{T}/\Lambda$, we see that the shifts $|\delta\tilde{\Theta}|/v_{S},|\delta v_{S_i}|/v_S$ can be kept small enough, below the percent level without any fine-tuning.
The next leading order terms in the driving superpotential lead to small deviations from the leading order vacuum alignments. And the mass and mixing matrices are corrected by the shift of the vacuum configuration.

\subsubsection{Corrections to the Majorana neutrino sector}
The corrected vacuum alignments in Eq.~(\ref{NewVEV}) modify the leading order Majorana neutrino mass term into  $M'_{R}=M_{R}+\delta M_{R}$, while the Dirac neutrino mass term is not affected due to the redefinition of $\langle\Phi_{T}\rangle\rightarrow(v'_{T},0,0)$: with the redefinition of $M\rightarrow M=y_{\Theta}\frac{v_{\Theta}}{\sqrt{2}}+y_{\tilde{\Theta}}\delta\tilde{\Theta}$ the corrected heavy neutrino mass term reads
 \begin{eqnarray}
 \delta M_{R}&=&M\,e^{i\phi}{\left(\begin{array}{ccc}
 \frac{2\sqrt{2}}{3}\epsilon_{1} &  -\frac{\sqrt{2}}{3}\epsilon_{2}  &  -\frac{\sqrt{2}}{3}\epsilon_{3} \\
 -\frac{\sqrt{2}}{3}\epsilon_{2} &  \frac{2\sqrt{2}}{3}\epsilon_{3}  &  -\frac{\sqrt{2}}{3}\epsilon_{1}\\
 -\frac{\sqrt{2}}{3}\epsilon_{3} &  -\frac{\sqrt{2}}{3}\epsilon_{1} & \frac{2\sqrt{2}}{3}\epsilon_{2}
 \end{array}\right)}~,
 \label{DMR2}
 \end{eqnarray}
where $\epsilon_i=\frac{\delta v_{Si}}{v_S}\,\tilde{\kappa}$ with $i=1,2,3$. Because of Eq.~(\ref{sizeDevi}) it is expected that the magnitude of $\epsilon_i$ is of order 0.1 or can be controlled, below the percent level. Then the mixing angles and masses of the light neutrinos may not be crucially modified by the next-leading order results.

\subsubsection{Corrections to the down-type quark sector}
And, also the new vacuum in Eq.~(\ref{NewVEV}) modifies the leading order mass matrix of the down-type quark into ${\cal M}'_{d}={\cal M}_{d}+\delta {\cal M}_{d}$
 \begin{eqnarray}
 \delta {\cal M}_{d}&=&{\left(\begin{array}{ccc}
 y_{d}\,\frac{\delta v_{S1}}{v_{S}} & y_{d}\,\frac{\delta v_{S3}}{v_{S}} & y_{d}\,\frac{\delta v_{S2}}{v_{S}}  \\
 y_{s}\,\frac{\delta v_{S2}}{v_{S}} & y_{s}\,\frac{\delta v_{S1}}{v_{S}} & y_{s}\,\frac{\delta v_{S3}}{v_{S}} \\
 y_{b}\,\frac{\delta v_{S3}}{v_{S}} & y_{b}\,\frac{\delta v_{S2}}{v_{S}} & y_{b}\,\frac{\delta v_{S1}}{v_{S}}
 \end{array}\right)}\frac{v_{S}}{\Lambda\sqrt{2}}\,v_{d}\,.
 \label{VevCorrect}
 \end{eqnarray}
The corrections from the vacuum alignments in Eq.~(\ref{VevCorrect}) are absorbed into the leading order terms and can be redefined. In order to show that this correction does not crucially affect the generation of small mixing angles in the CKM matrix, we explicitly express the Hermitian matrix ${\cal M}'_{d}{\cal M}'^{\dag}_{d}$, which is diagonalized by the mixing matrix $V^{d}_{L}$:
 \begin{eqnarray}
 {\cal M}'_{d}{\cal M}'^{\dag}_{d}&\simeq&
 v^2_{d}\frac{3}{2}\left(\frac{v_{S}}{\Lambda}\right)^2\,{\left(\begin{array}{ccc}
 \lambda^{6}|\hat{y}_{d}|^2(1+\varepsilon) & \lambda^5\hat{y}_{d}\,\hat{y}^{\ast}_{s}(1+\varepsilon) & \lambda^3\hat{y}_{d}\,\hat{y}^\ast_{b}(1+\varepsilon)  \\
 \lambda^5\hat{y}^{\ast}_{d}\,\hat{y}_{s}(1+\varepsilon) & \lambda^4|\hat{y}_{s}|^2(1+\varepsilon) & \lambda^2\hat{y}_{s}\,\hat{y}^\ast_{b}(1+\varepsilon)   \\
 \lambda^3\hat{y}^\ast_{d}\,\hat{y}_{b}(1+\varepsilon) & \lambda^2\hat{y}^\ast_{s}\,\hat{y}_{b}(1+\varepsilon)  & |\hat{y}_{b}|^2(1+\varepsilon)
 \end{array}\right)}\,,
 \label{MDMD1}
 \end{eqnarray}
with $\varepsilon=2(\frac{\delta v_{S1}}{v_{S}}+\frac{\delta v_{S2}}{v_{S}}+\frac{\delta v_{S3}}{v_{S}})/3$. It is easy to find that this matrix could not lead to the correct CKM mixing angles.
So, in this work we will not consider the next to leading order contributions of vacuum alignments which may not crucially change the leading order results of $W_{f}$.

\subsection{Corrected masses and the CKM matrix}
\label{corrMassCKM}
\noindent The light neutrino mass matrix can be modified by both the non-trivial operators, Eq.~(\ref{NewMR}), and by the shift of the vacuum alignment, Eq.~(\ref{NewVEV}). The remaining results modify ${\cal M}_{\nu}$ in Eq.~(\ref{mass matrix}) into ${\cal M}'_{\nu}={\cal M}_{\nu}+\Delta {\cal M}_{\nu}$
 \begin{eqnarray}
  \Delta {\cal M}_{\nu}=m_{D}M^{-1}_{R}\,\Delta M_{R}\,M^{-1}_{R}m^{T}_{D}+m_{D}M^{-1}_{R}\,\delta M_{R}\,M^{-1}_{R}m^{T}_{D}+{\cal O}\left(\epsilon^2_i, \frac{v^2_{T}}{\Lambda^2}\right)\,.
 \label{}
 \end{eqnarray}
As expected from Sec.~\ref{correLepton} and~\ref{VEVcorrects}, the corrections from these non-leading terms can be kept small enough, below the percent level. Therefore, it is expected that corrections from the leading order results can be obtained for all measurable quantities at approximately the same level.

As seen in Eq.~(\ref{MDMD1}), including the corrections from the shift of the vacuum configuration of down-type quark, they can be all absorbed into a redefinition of the overall factor. So, here considering the corrections from the nontrivial next leading operators in Yukawa superpotential, Eq.~(\ref{lagrangianD1}), we obtain the Hermitian matrix $\tilde{{\cal M}}_{d}\tilde{{\cal M}}^{\dag}_{d}$ :
 \begin{eqnarray}
 \tilde{{\cal M}}_{d}\tilde{{\cal M}}^{\dag}_{d}&=&
 v^2_{d}\frac{3}{2}\left(\frac{v_{S}}{\Lambda}\right)^2{\left(\begin{array}{ccc}
 \lambda^{6}|\hat{y}_{d}|^2(1+\varepsilon_{dd}) & \lambda^5\hat{y}_{d}\hat{y}^{\ast}_{s}(1+\varepsilon_{ds}) & \lambda^3\hat{y}_{d}\hat{y}^\ast_{b}(1+\varepsilon_{db})  \\
 \lambda^5\hat{y}^{\ast}_{d}\hat{y}_{s}(1+\varepsilon^\ast_{ds}) & \lambda^4|\hat{y}_{s}|^2(1+\varepsilon_{ss}) & \lambda^2\hat{y}_{s}\hat{y}^\ast_{b}(1+\varepsilon_{sb})   \\
 \lambda^3\hat{y}^\ast_{d}\hat{y}_{b}(1+\varepsilon^\ast_{db}) & \lambda^2\hat{y}^\ast_{s}\hat{y}_{b}(1+\varepsilon^\ast_{sb})  & |\hat{y}_{b}|^2(1+\varepsilon_{bb})
 \end{array}\right)}\nonumber\\
 &+&{\cal O}\left(\frac{v^2_{T}}{\Lambda^2},\frac{1}{\kappa}\frac{v^2_{T}}{\Lambda^2}\right)
 \label{MDMD}
 \end{eqnarray}
where
 \begin{eqnarray}
  \varepsilon_{\alpha\alpha}&=&\frac{1}{3\sqrt{2}\kappa}\frac{v_{T}}{\Lambda}\left(\frac{\hat{x}_{\alpha}}{\hat{y}_{\alpha}}+\frac{\hat{x}^{\ast}_{\alpha}}{\hat{y}^{\ast}_{\alpha}}\right)+\left(\frac{1}{3\sqrt{2}\kappa}\frac{v_{T}}{\Lambda}\right)^2\frac{|\hat{x}_{\alpha}|^2}{|\hat{y}_{\alpha}|^2}\,,\quad\varepsilon_{\alpha\beta}=\frac{1}{3\sqrt{2}\kappa}\frac{v_{T}}{\Lambda}\left(\frac{\hat{x}_{\alpha}}{\hat{y}_{\alpha}}+\frac{\hat{x}^{\ast}_{\beta}}{\hat{y}^\ast_{\beta}}\right)\,.\nonumber
 \end{eqnarray}
Here $\tilde{{\cal M}}_{d}\equiv{\cal M}_d+\Delta{\cal M}_d$ and the Hermitian matrix is diagonalized as $V^{d\dag}_{L}\tilde{{\cal M}}_{d}\tilde{{\cal M}}^{\dag}_{d}V^{d}_{L}={\rm diag}(|m_{d}|^{2}, |m_s|^{2}, |m_{b}|^{2})$ by the mixing matrix $V^{d}_{L}$.
Recalling that $\kappa\equiv v_{S}/v_{\Theta}$. Notice here that the parameters $\varepsilon_{\alpha\alpha},\varepsilon_{\alpha\beta}$ are only associated with the next leading operators driven by the $\Theta$ field of $\Delta W^{d}_{q}$ in the Yukawa superpotential~(\ref{lagrangianD1}), while the contributions associated with the $\Phi_S$ field do cancel out each other and do not play a part.
Due to the strong hierarchical structure of the Hermitian matrix, we can obtain the mixing matrix $V^{d}_{L}$ of the down-type quarks: under the constraint of unitarity up to ${\cal O}(\lambda^{3})$, it can be written as
 \begin{eqnarray}
 V^d_{L}&=&{\left(\begin{array}{ccc}
 1-\frac{1}{2}\lambda^2\,\Gamma^2 &  \lambda\,\Gamma\,e^{i\phi^{d}_{3}} &  \lambda^3\,B\,e^{i\phi^{d}_{2}} \\
 -\lambda\,\Gamma\,e^{-i\phi^{d}_{3}} &  1-\frac{1}{2}\lambda^2\,\Gamma^2 &  \lambda^2\,A\,e^{i\phi^{d}_{1}} \\
 \lambda^3(A\,\Gamma\,e^{-i(\phi^{d}_{1}+\phi^{d}_{3})}-B\,e^{-i\phi^{d}_{2}}) &  -\lambda^2\,A\,e^{-i\phi^{d}_{1}} &  1
 \end{array}\right)}P_{d}+{\cal O}(\lambda^4)
 \label{VDMixing}
 \end{eqnarray}
with the phases
 \begin{eqnarray}
  \phi^{d}_{1}&=&\frac{1}{2}\arg\{\hat{y}_s\,\hat{y}^{\ast}_b(1+\varepsilon_{sb})\},\quad
  \phi^{d}_{2}=\frac{1}{2}\arg\left\{\frac{\hat{y}_d(1+\varepsilon_{db})}{\hat{y}_s(1+\varepsilon_{sb})}\right\},\quad
  \phi^{d}_{3}=\frac{1}{2}\arg\left(\Sigma\right)-\frac{\phi^{d}_{2}}{2},\nonumber
 \end{eqnarray}
and the associated parameters
 \begin{eqnarray}
  A&=&\frac{|\hat{y}_{s}(1+\varepsilon_{sb})|}{|\hat{y}_{b}(1+\varepsilon_{bb})|},\qquad\qquad\qquad
  B=\frac{|\hat{y}_d(1+\varepsilon_{db})|}{|\hat{y}_b(1+\varepsilon_{bb})|},\nonumber\\
  \Gamma&=&\frac{\left|\Sigma\right|(1+\varepsilon_{bb})}{|\hat{y}_{s}|^2\left\{\left(\frac{1}{3\kappa}\frac{v_{T}}{\Lambda}\right)^2\,\Gamma_1+\left(\frac{1}{3\kappa}\frac{v_{T}}{\Lambda}\right)^3\,\Gamma_2+\left(\frac{1}{3\kappa}\frac{v_{T}}{\Lambda}\right)^4\,\Gamma_3\right\}}.\nonumber
 \end{eqnarray}
Here $\Sigma=\hat{y}_d\,\hat{y}^{\ast}_s(1+\varepsilon_{ds})\,e^{i\phi^{d}_{1}}-A\,\hat{y}_d\,\hat{y}^{\ast}_b(1+\varepsilon_{db})\,e^{-i\phi^{d}_{1}}$, $\Gamma_1=\frac{\hat{x}_s}{\hat{y}_{s}}\frac{\hat{x}^{\ast}_b}{\hat{y}^{\ast}_{b}}+\frac{\hat{x}^{\ast}_s}{\hat{y}^{\ast}_{s}}\frac{\hat{x}_b}{\hat{y}_{b}}$, $\Gamma_2=\left(\frac{\hat{x}_b}{\hat{y}_{b}}+\frac{\hat{x}^\ast_b}{\hat{y}^\ast_{b}}\right)\frac{|\hat{x}_s|^2}{|\hat{y}_{s}|^2}+\left(\frac{\hat{x}_s}{\hat{y}_{s}}+\frac{\hat{x}^\ast_s}{\hat{y}^\ast_{s}}\right)\frac{|\hat{x}_b|^2}{|\hat{y}_{b}|^2}$, and $\Gamma_3=\frac{|\hat{x}_s|^2}{|\hat{y}_{s}|^2}\frac{|\hat{x}_b|^2}{|\hat{y}_{b}|^2}$.
In Eq.~(\ref{VDMixing}) the diagonal phase matrix $P_{d}$ can be rotated away by redefinition of quark fields.
Then from the charged current interactions of quark sector, we can obtain the CKM matrix
 \begin{eqnarray}
  V_{\rm CKM}= V^{u\dag}_{L}V^{d}_{L}= V^{d}_{L}~.
 \label{mixing relation}
 \end{eqnarray}
It is very crucial to note that the next-leading order terms denoted as $\varepsilon_{\alpha\alpha},\varepsilon_{\alpha\beta}$ lead to the correct CKM matrix.
From Eqs.~(\ref{VDMixing}) and (\ref{mixing relation}), if we set
 \begin{eqnarray}
  \frac{|\hat{y}_d(1+\varepsilon_{db})|}{|\hat{y}_s(1+\varepsilon_{sb})|} = \sqrt{\rho^2+\eta^2}\,, \qquad\qquad\Gamma=1\,,
 \label{ckmPara}
 \end{eqnarray}
and by redefining the quark fields with the transformation $c\rightarrow c\,e^{i\phi^{d}_{3}}$, $s\rightarrow s\,e^{i\phi^{d}_{3}}$, $b\rightarrow b\,e^{i(\phi^{d}_{1}+\phi^{d}_{3})}$ and $t\rightarrow t\,e^{-i(\phi^{d}_{1}+\phi^{d}_{3})}$,
then we obtain the CKM matrix in the Wolfenstein parametrization~\cite{Wolfenstein:1983yz} given by
 \begin{eqnarray}
  V_{\rm CKM}={\left(\begin{array}{ccc}
   1-\lambda^2/2 & \lambda & A\lambda^3\,(\rho+i\eta) \\
   -\lambda & 1-\lambda^2/2 & A\lambda^2 \\
   A\,\lambda^3(1-\rho+i\eta) & -A\,\lambda^2 & 1 \\
   \end{array}\right)}+{\cal O}(\lambda^{4})~,
 \label{QM}
 \end{eqnarray}
with the CKM CP phase $\delta^{q}_{CP}=\phi^{d}_{1}+\phi^{d}_{3}-\phi^{d}_{2}$, or equivalently $\delta^{q}_{CP}=\tan^{-1}(\rho/\eta)$.
From the global fits to the quark mixing matrix reported in Ref.~\cite{ckmfitter}, the best-fit values of the parameters $\lambda$, $A$, $\bar{\rho}$,
$\bar{\eta}$ with $3\sigma$ errors are
 \begin{eqnarray}
  \lambda &=& \sin\theta_{C}=0.22457^{+0.00200}_{-0.00027}\,,\qquad\qquad\, A=0.823^{+0.025}_{-0.049}~, \nonumber\\
  \bar{\rho} &=& 0.129^{+0.075}_{-0.027}\,,\qquad\qquad\qquad\qquad\qquad\bar{\eta}=0.348^{+0.037}_{-0.044}~,
 \end{eqnarray}
where $\bar{\rho}=\rho(1-\lambda^{2}/2)$ and $\bar{\eta}=\eta(1-\lambda^{2}/2)$.
The effects caused by CP violation are always proportional to the Jarlskog invariant~\cite{Jarlskog:1985ht} in the quark sector is given by
 \begin{eqnarray}
  J^{q}_{CP}=-{\rm Im}[V_{ud}V_{tb}V^{\ast}_{ub}V^{\ast}_{td}]\simeq A^{2} \lambda^{6}\eta~.
 \label{Jcp1}
 \end{eqnarray}
whose value is $3.02^{+0.42}_{-0.36}\times10^{-5}$ at $3\sigma$ level~\cite{ckmfitter}.
Numerically, it reads $J^{q}_{CP}\simeq0.2\times\lambda^{6}$.
And, the corresponding mass eigenvalues are given in a good approximation as
 \begin{eqnarray}
  m_{b}&\simeq&\sqrt{\frac{3}{2}}\,\frac{v_{S}}{\Lambda}\,v_{d}\,|\hat{y}_{b}|\sqrt{1+\varepsilon_{bb}}\,,\nonumber\\
  m_{s}&\simeq&\lambda^2\,\sqrt{\frac{3}{2}}\,\frac{v_{S}}{\Lambda}\,v_{d}\sqrt{|\Sigma|}\,,\nonumber\\
  m_{d}&\simeq&\lambda^3\,\sqrt{\frac{3}{2}}\,\frac{v_{S}}{\Lambda}\,v_{d}|\hat{y}_{d}|\,\left(\frac{1}{3\sqrt{2}\kappa}\frac{v_{T}}{\Lambda}\right)\left\{\Gamma_a+\left(\frac{1}{3\kappa}\frac{v_{T}}{\Lambda}\right)\,\Gamma_b\right\}^{\frac{1}{2}},
 \label{dMasses}
 \end{eqnarray}
where
$\Gamma_a=\frac{\hat{x}_d}{\hat{y}_{d}}\frac{\hat{x}^{\ast}_b}{\hat{y}^{\ast}_{b}}+\frac{\hat{x}^{\ast}_d}{\hat{y}^{\ast}_{d}}\frac{\hat{x}_b}{\hat{y}_{b}}$, and $\Gamma_b=\left(\frac{\hat{x}_d}{\hat{y}_{d}}+\frac{\hat{x}^\ast_d}{\hat{y}^\ast_{d}}\right)\frac{|\hat{x}_b|^2}{|\hat{y}_{b}|^2}+\left(\frac{\hat{x}_b}{\hat{y}_{b}}+\frac{\hat{x}^\ast_b}{\hat{y}^\ast_{b}}\right)\frac{|\hat{x}_d|^2}{|\hat{y}_{d}|^2}$.
Considering the expected value for the VEVs for $v_{S},v_{T}$ and $v_{d}$ with Eqs.~(\ref{MassRangge},\ref{tanbeta}), these results can be in a good agreement with the empirical down-type quark masses calculated from the measured values~\cite{PDG}.

\section{a light Axion}
 \label{U1-Axion}

The QCD Lagrangian has a CP-violating term
 \begin{eqnarray}
   {\cal L}_{\vartheta} &=& \vartheta_{\rm eff}\frac{\alpha_s}{8\pi}G^{a\mu\nu}\tilde{G}^{a}_{\mu\nu}
\label{QCDlag}
 \end{eqnarray}
where $-\pi\leq\vartheta_{\rm eff}\leq\pi$ is the effective $\vartheta$ parameter defined, in the basis where quark masses are real and positive, diagonal, and $\gamma_5$-free, as
 \begin{eqnarray}
  \vartheta_{\rm eff}=\vartheta+\arg\left\{\det(\mathcal{M}_{u})\det(\mathcal{M}_{d})\right\}\,.
\label{QCDlag1}
 \end{eqnarray}
Here the angle $\vartheta$ is given above the electroweak scale, which is the coefficient of $\vartheta\,g^{2}_{s}\,G^{a\mu\nu}\tilde{G}^{a}_{\mu\nu}/32\pi^{2}$ where $G^a$ is the color field strength tensor and its dual $\tilde{G}^{a}_{\mu\nu}=\frac{1}{2}\varepsilon_{\mu\nu\rho\sigma}G^{a\mu\nu}$, coming from the strong interaction. And, the second term comes from a chiral transformation of weak interaction for diagonalization of the quark mass matrices by $\psi_q\rightarrow e^{-i\gamma_{5}\arg[\det m_{q}]/2}\psi_q$, directly indicating the CKM CP phase $\delta_{CP}$ in Eq.~(\ref{Jcp1}), which is of order unity. However, experimental bounds on CP violation in strong interactions are very tight, the strongest ones coming from the limits on the electric dipole moment of the neutron $d_{n}<0.29\times10^{-25}~e$~\cite{Beringer:1900zz} which implies $|\vartheta_{\rm eff}|<0.56\times10^{-10}$. $\vartheta_{\rm eff}$ should be very small to make a theory consistent with experimental bounds.
A huge cancelation between $\vartheta$ and $\arg\left\{\det(\mathcal{M}_{u})\det(\mathcal{M}_{d})\right\}$ suggests that there should be a physical process.

Until now, the most elegant solution to the strong CP problem is the PQ mechanism, which yields a light pseudo-Nambu-Goldstone boson, called the axion~\cite{Peccei-Quinn, axion}.
There are two prototype models by what couples to $U(1)_{\rm PQ}$: (i) the Kim-Shifman-Vainshtein-Zakharov (KSVZ) model~\cite{KSVZ}, where only new heavy quarks charged under $U(1)_{\rm PQ}$ are introduced, and (ii) the Dine-Fischler-Srednicki-Zhitnitsky (DFSZ) model~\cite{DFSZ}, where only known quarks exist and Higgs doublets carry PQ charges. And there are good reviews Refs.~\cite{Kim:1986ax,Cheng:1987gp,Peccei} on the axion.

Now, based on the model described by the superpotential~(\ref{potential}), (\ref{lagrangian}), (\ref{lagrangianU}) and (\ref{lagrangianD}) we wish to discuss an automatic theory for strong CP invariance introducing the so-called ``flavored-PQ symmetry" $U(1)_{X}$ (which is introduced for describing the SM fermion mass hierarchies) with non-Abelian $A_{4}$ symmetry in the superpotential as in Table~\ref{DrivingRef} and \ref{reps}. The flavored PQ symmetry $U(1)_{X}$ guarantees the absence of bare mass terms.
The model incorporates the SM gauge singlet flavon fields ${\cal F}_A=\Phi_{S},\Theta,\Psi,\tilde{\Psi}$ with the following interactions invariant under the $U(1)_{X}\times A_{4}$ and the resulting chiral symmetry, {\it i.e.}, the kinetic and Yukawa terms, and the scalar potential $V_{\rm SUSY}$ in SUSY limit~\footnote{In our superpotential, the superfields $\Phi_S, \Theta$ and $\Psi(\tilde{\Psi})$ are gauge singlets and have $-2p$ and $-q(q)$ $X$-charges, respectively.
Given soft SUSY-breaking potential, the radial components of the $X$-fields $|\Phi_S|$, $|\Theta|$ $|\Psi|$ and $|\tilde{\Psi}|$ are stabilized. The $X$-fields contain the axion, saxion (the scalar partner of the axion), and axino (the fermionic superpartner of the axion).} are of the form
 \begin{eqnarray}
  {\cal L} &=& \partial_{\mu}{\cal F}^{\dag}_A\,\partial^{\mu}{\cal F}_A+{\cal L}_{Y}-V_{\rm SUSY}+{\cal L}_{\vartheta}\,,
  \label{AxionLag}
 \end{eqnarray}
in which the $V_{\rm SUSY}$ term is replaced by $V_{total}$, Eq.~(\ref{Vtotal}), when SUSY breaking effects are considered.
The kinetic term is written as
 \begin{eqnarray} \partial_{\mu}\Phi^{\dag}_{S}\partial^{\mu}\Phi_{S}+\partial_{\mu}\Theta^{\dag}\partial^{\mu}\Theta+\partial_{\mu}\Psi^{\dag}\partial^{\mu}\Psi+\partial_{\mu}\tilde{\Psi}^{\dag}\partial^{\mu}\tilde{\Psi}\,.
 \label{AxionLag1}
 \end{eqnarray}
The relevant Yukawa interaction term with chiral fermions $\psi$ charged under the flavored PQ symmetry $U(1)_X$ symmetry is given as
 \begin{eqnarray}
  {\cal L}_{Y} &=&
  -\frac{1}{2}y_{\Theta}\Theta(\overline{N^{c}_{R}}N_{R})_{{\bf 1}}-\frac{y_{R}}{2} (\overline{N^{c}_{R}}N_{R})_{{\bf 3}_{s}}\Phi_{S}-\overline{\psi}_{L}Y_\psi(\Psi,\Phi_{S},\Theta)\,\psi_{R}H_{u,d}+\text{h.c.}\,.
  \label{AxionLag2}
 \end{eqnarray}
And the relevant $F$-term scalar potential term is given as
 \begin{eqnarray}
  V_{\rm SUSY}&=&\left|\frac{2g_{1}}{\sqrt{3}}\left(\Phi_{S1}\Phi_{S1}-\Phi_{S2}\Phi_{S3}\right)+g_{2}\Phi_{S1}\tilde{\Theta}\right|^{2}\nonumber\\
  &+&\left|\frac{2g_{1}}{\sqrt{3}}\left(\Phi_{S2}\Phi_{S2}-\Phi_{S1}\Phi_{S3}\right)+g_{2}\Phi_{S3}\tilde{\Theta}\right|^{2}\nonumber\\
  &+&\left|\frac{2g_{1}}{\sqrt{3}}\left(\Phi_{S3}\Phi_{S3}-\Phi_{S1}\Phi_{S2}\right)+g_{2}\Phi_{S2}\tilde{\Theta}\right|^{2}\nonumber\\
  &+&\left|g_{3}\left(\Phi_{S1}\Phi_{S1}+2\Phi_{S2}\Phi_{S3}\right)+g_{4}\Theta^{2}+g_{5}\Theta\tilde{\Theta}+g_{6}\tilde{\Theta}^{2}\right|^{2}+\left|g_{7}\Psi\tilde{\Psi}+\mu^{2}_{\Psi}\right|^2+...
  \label{AxionLag3}
 \end{eqnarray}
Here dots represent the other scalar potential $\{...\}=\sum_{i}\left|\frac{\partial W_{v}}{\partial\varphi_{i}}\right|^{2}$ with $\varphi_{i}=\{\Phi^{T}_{0},\Phi_{T},\Phi_{S},\Theta,\tilde{\Theta},\Psi,\tilde{\Psi}\}$, and all of those are irrelevant for our discussion (c.f. Eq.~(\ref{drivingF-term})).

After getting VEVs $\langle\Theta\rangle,\langle\Phi_{S}\rangle\neq0$ (which generates the heavy neutrino masses given by Eq.~(\ref{MR1})) and $\langle\Psi\rangle\neq0$, the flavored PQ symmetry $U(1)_{X}$ is spontaneously broken at a scale much higher than the electroweak scale and is realized by the existence of the NG mode $A$ that couples to ordinary quarks at the tree level through the Yukawa couplings as in Eq.~(\ref{AxionLag2}) (see also Eqs.~(\ref{lagrangianU}) and (\ref{lagrangianD})), and the resulting NG boson becomes the axion~\footnote{The VEV configurations in Eqs.~(\ref{vevdirection1},\ref{vevdirection2},\ref{vevdirection3}) break the $U(1)_X$ spontaneously and the superpotential dependent on the driving field $\Theta_0$ in Eq.~(\ref{potential}) becomes, for simplicity, if we let $\Phi_{S1}=\Phi_{S2}=\Phi_{S3}$,
 $W_{\Theta_0} = \Theta_{0}\left(g_{3}\,\Phi_{S}\Phi_{S}+g_{4}\,\Theta\Theta+6\kappa\, g_3\left\{v_{\Theta}\Phi_{Si}-v_S\Theta\right\}+g_{5}\,(\Theta+2\frac{v_S}{\kappa})\tilde{\Theta}+g_{6}\,\tilde{\Theta}\tilde{\Theta}\right)$ after shifting by $v_{\Theta},v_S$. This shows clearly that the linear combination $(v_{\Theta}\Theta+v_{S}\Phi_{Si})/\sqrt{v^2_{\Theta}+v^2_{S}}$ is a massless superfield.}. Through triangle anomalies, the axion mixes with mesons (leading to a non-zero mass), and thus couples to photons, nucleons, and leptons. The explicit breaking of the $U(1)_X$ by the chiral anomaly effect further breaks it down to $Z_{N}$ discrete symmetry, where $N$ is the color anomaly number. At the QCD phase transition, the $Z_{N}$ symmetry is spontaneously broken, and which gives rise to a domain wall problem~\cite{Sikivie:1982qv}.
Such domain wall problem can be overcome because the model has two anomalous axial $U(1)$ symmetries which are generated by the charges $X_1$ and $X_2$, $U(1)_{X}\equiv U(1)_{X_1}\times U(1)_{X_2}$.

The scalar fields $\Phi_{S},\Theta$ and $\Psi(\tilde{\Psi})$ have $X$-charges $X_{1}=-2p$ and $X_{2}=-q(q)$, respectively, that is
 \begin{eqnarray}
 \Phi_{S_i}\rightarrow e^{i\xi_1 X_1}\Phi_{S_i},\quad\Theta\rightarrow e^{i\xi_1 X_1}\Theta\,;\quad\Psi\rightarrow e^{i\xi_2 X_2}\Psi,\quad\tilde{\Psi}\rightarrow e^{-i\xi_2 X_2}\tilde{\Psi}
 \label{X_sclar}
 \end{eqnarray}
where $\xi_k$ ($k=1,2$) are constants. So, the potential $V_{\rm SUSY}$ has $U(1)_X$ global symmetry.
In order to extract NG bosons resulting from spontaneous breaking of $U(1)_{X}$ symmetry, we set the decomposition of complex scalar fields as follows~\footnote{Note that the massless modes are not contained in the fields $\tilde{\Theta},\Phi_{T},\Phi^{T}_{0},\Phi^{S}_{0},\Theta_{0},\Psi_{0}$.}
 \begin{eqnarray}
  \Phi_{Si}=\frac{e^{i\frac{\phi_{S}}{v_{S}}}}{\sqrt{2}}\left(v_{S}+h_{S}\right)\,,\quad\Theta=\frac{e^{i\frac{\phi_{\theta}}{v_{\Theta}}}}{\sqrt{2}}\left(v_{\Theta}+h_{\Theta}\right)\,,\quad\Psi=\frac{e^{i\frac{\phi_{\Psi}}{v_{\Psi}}}}{\sqrt{2}}\left(v_{\Psi}+h_{\Psi}\right)\,,
  \label{NGboson}
 \end{eqnarray}
in which we have assumed $\Phi_{S1}=\Phi_{S2}=\Phi_{S3}\equiv\Phi_{Si}$. And the NG modes $A_1$, $A_2$ are expressed as
 \begin{eqnarray}
  A_1=\frac{v_{S}\,\phi_{S}+v_{\Theta}\,\phi_{\theta}}{\sqrt{v^{2}_{S}+v^{2}_{\Theta}}}\,,\qquad A_{2}=\phi_{\Psi}
 \end{eqnarray}
with the angular fields $\phi_{S}$, $\phi_{\theta}$ and $\phi_{\Psi}$.
With Eqs.~(\ref{AxionLag1}) and (\ref{NGboson}), the derivative couplings of $A$ arise from the kinetic terms
 \begin{eqnarray}
  \partial_{\mu}{\cal F}^{\ast}_{k}\,\partial^{\mu}{\cal F}_{k}
  &=&\frac{1}{2}\left(\partial_{\mu}A_1\right)^2\left(1+\frac{h_{\cal F}}{v_{\cal F}}\right)^2+\frac{1}{2}\left(\partial_{\mu}A_2\right)^2\left(1+\frac{h_{\Psi}}{v_{\Psi}}\right)^2+\frac{1}{2}\left(\partial_{\mu}h_{\cal F}\right)^2+\frac{1}{2}\left(\partial_{\mu}h_{\Psi}\right)^2\nonumber\\
  &+&...
  \label{derivative}
 \end{eqnarray}
where $v_{\cal F}=v_{\Theta}(1+\kappa^2)^{1/2}$ and $h_{\cal F}=(\kappa h_{S}+h_{\Theta})/(1+\kappa^2)^{1/2}$, and the dots stand for the orthogonal components $h^{\bot}_{\cal F}$ and $A^{\bot}_{1}$. Recalling that $\kappa\equiv v_{S}/v_{\Theta}$. Clearly, the derivative interactions of $A_k$ ($k=1,2$) are suppressed by the VEVs $v_{\cal F}$ and $v_{\Psi}$. From Eq.~(\ref{derivative}), performing $v_{\cal F}, v_{\Psi}\rightarrow\infty$, the NG modes $A_{1,2}$, whose interactions are determined by symmetry, are distinguished from the radial modes, like $h_{\cal F}, h_{\Psi}$, which are model-dependent (SUSY breaking mechanism) and invariant under the symmetry.

The model has two anomalous $U(1)$ symmetries, $U(1)_{X_1}\times U(1)_{X_2}$, with respective anomalies $N_1$ and $N_2$, both of which are the coefficients of the $U(1)_{X_k}-SU(3)_C-SU(3)_C$ anomaly, so there are two would-be axions $A_{1}$ and $A_{2}$, with the transformation of the phase fields $A_{1}\rightarrow A_{1}+\frac{v_{\cal F}X_1}{N_1}\,\xi_{1}$ and $A_{2}\rightarrow A_{2}+\frac{v_{\Psi}X_2}{N_2}\,\xi_{2}$, respectively~\cite{TwoU}.
Their charges $X_1$ and $X_2$ are linearly independent. And the color anomaly coefficients are obtained by letting $2\sum_{\psi_i}X_{k\psi_i}\,{\rm Tr}(t^at^b)=N_k\delta^{ab}$, where the $t^a$ are the generators of the representation of $SU(3)$ to which $\psi$ belongs and the sum runs over all Dirac fermion $\psi$ with $X$-charge.
Since the two $U(1)$s are broken by two types of field attaining VEVs, a new PQ symmetry $U(1)_{\tilde{X}}$ which is a linear combination of the two $U(1)$s has anomaly, while another $U(1)$ is anomaly-free (it is the broken $U(1)_{f}$ symmetry by $\langle\Theta\rangle,\langle\Phi_S\rangle\neq0$ responsible for lepton number violation).
Under $U(1)_{\tilde{X}}\times U(1)_f$ the fields are transformed as
 \begin{eqnarray}
  &&{\cal F}_{1}=\frac{v_{\cal F}\,e^{i\frac{A_1}{v_{\cal F}}}}{\sqrt{2}}\left(1+\frac{h_{\cal F}}{v_{\cal F}}\right)\,;\qquad{\cal F}_{1}\rightarrow e^{iX_{1}\,\xi_1}{\cal F}_{1}\,,\quad\text{with}~~\xi_1=N_2\,\alpha\,,\nonumber\\
  &&{\cal F}_{2}=\frac{v_{\Psi}\,e^{i\frac{A_2}{v_{\Psi}}}}{\sqrt{2}}\left(1+\frac{h_{\Psi}}{v_{\Psi}}\right)\,;\qquad {\cal F}_{2}\rightarrow e^{iX_{2}\,\xi_2}{\cal F}_{2}\,,\quad\text{with}~~\xi_2=-N_1\,\alpha\,.
  \label{Ahnaxion01}
 \end{eqnarray}
One linear combination of the phase fields $A_{1}$ and $A_{2}$ becomes the axion ($\equiv A$), and the other orthogonal combination corresponds to the Goldstone boson ($\equiv G$):
 \begin{eqnarray}
  {\left(\begin{array}{c}
 A \\
 G
 \end{array}\right)}={\left(\begin{array}{cc}
 \cos\vartheta &  \sin\vartheta \\
 -\sin\vartheta &  \cos\vartheta
 \end{array}\right)}{\left(\begin{array}{c}
 A_1 \\
 A_2
 \end{array}\right)}
 \end{eqnarray}
Here the $G$ is the ``true" Goldstone boson of the spontaneously broken $U(1)_{f}$. And since the Goldstone boson interactions arise only through the derivative couplings as Eq.~(\ref{derivative}), we can have the nonlinearly realized global symmetry below the symmetry breaking scale
 \begin{eqnarray}
  U(1)_f:\qquad G\rightarrow G+\Upsilon \text{(constant)}\,.
 \end{eqnarray}
Then the angle is obtained as $\cos\vartheta=-\frac{\tilde{X}_2\,v_{\Psi}}{\sqrt{\left(\tilde{X}_1\,v_{\cal F}\right)^2+\left(-\tilde{X}_2\,v_{\Psi}\right)^2}}$ and $\sin\vartheta=\frac{\tilde{X}_1\,v_{\cal F}}{\sqrt{\left(\tilde{X}_1\,v_{\cal F}\right)^2+\left(-\tilde{X}_2\,v_{\Psi}\right)^2}}$ with $\tilde{X}_1\equiv N_2\,X_1$ and $\tilde{X}_2\equiv-N_1\,X_2$. Therefore,
the axion $A$ and the Goldstone boson $G$ can be expressed as
 \begin{eqnarray}
 A=\frac{-A_1\tilde{X}_2\,v_{\Psi}+A_2\tilde{X}_1\,v_{\cal F}}{\sqrt{\left(\tilde{X}_1\,v_{\cal F}\right)^2+\left(-\tilde{X}_2\,v_{\Psi}\right)^2}}\,,\qquad G=\frac{-A_1\tilde{X}_1\,v_{\cal F}-A_2\tilde{X}_2\,v_{\Psi}}{\sqrt{\left(\tilde{X}_1\,v_{\cal F}\right)^2+\left(-\tilde{X}_2\,v_{\Psi}\right)^2}}\,.
  \label{AxioAhn}
 \end{eqnarray}

Meanwhile, the $X$-current for $U(1)_{\tilde{X}}$ with the condition~(\ref{Ahnaxion01}) is given by
 \begin{eqnarray}
  J^{\tilde{X}}_{\mu}=i\tilde{X}_{1}{\cal F}^{\dag}_{1}\overleftrightarrow{\partial}_{\mu}{\cal F}_{1}-i\tilde{X}_{2}{\cal F}^{\dag}_{2}\overleftrightarrow{\partial}_{\mu}{\cal F}_{2}+\frac{1}{2}\sum_{\psi} \tilde{X}_{\psi}\bar{\psi}\gamma_{\mu}\gamma_{5}\psi
\label{AxialC}
 \end{eqnarray}
where $\psi=$ all $X$-charged Dirac fermions and $\tilde{X}_{\psi}\equiv\tilde{X}_{1\psi}-\tilde{X}_{2\psi}$, which is conserved, $\partial^{\mu}J^{\tilde{X}}_{\mu}=0$, up to the triangle anomaly. This current creates a massless particle, the axion.
The $X$-current in Eq.~(\ref{AxialC}) is now decoupled in the limit $v_{\cal F},v_{\Psi}\rightarrow\infty$ as
 \begin{eqnarray}
  J^{\tilde{X}}_{\mu}&=&\tilde{X}_1\,v_{\cal F}\,\partial_{\mu}A_1+(-\tilde{X}_2\,v_{\Psi})\,\partial_{\mu}A_2+\frac{1}{2}\sum_{\psi} \tilde{X}_{\psi}\bar{\psi}\gamma_{\mu}\gamma_{5}\psi\nonumber\\
  &=&\frac{\partial_{\mu}A}{\sqrt{\left(\frac{1}{2v_{\cal F}\tilde{X}_1}\right)^2+\left(-\frac{1}{2v_{\Psi}\tilde{X}_2}\right)^2}}+\frac{1}{2}\sum_{\psi} \tilde{X}_{\psi}\bar{\psi}\gamma_{\mu}\gamma_{5}\psi\,,
  \label{Jx}
 \end{eqnarray}
which corresponds to the charge flow satisfying the current conservation equation if the symmetry is exact. Since the $J^{\tilde{X}}_{\mu}$ does not couple to the Goldstone boson $G$ in Eq.~(\ref{AxioAhn}), requiring $J^{\tilde{X}}_{\mu}$ not to create $G$ from the vacuum $\langle0|J^{\tilde{X}}_{\mu}|G\rangle=0$, it follows
 \begin{eqnarray}
  \left(\tilde{X}_1\,v_{\cal F}\right)^2=\left(\tilde{X}_2\,v_{\Psi}\right)^2\,.
  \label{AhnMass}
 \end{eqnarray}
This indicates that, if one of symmetry breaking scales is determined, the other one is automatically fixed.
The NG boson $A$ (which will be the axion) possess the decay constant, $f_{A}$, defined by
 \begin{eqnarray}
  \langle0|J^{\tilde{X}}_{\mu}(x)|A(p)\rangle=ip_{\mu}\,f_{A}\,e^{-ip\cdot x}\,.
  \label{fA}
 \end{eqnarray}
From Eqs.~(\ref{Jx}) and (\ref{fA}), we obtain the spontaneous symmetry breaking scale
 \begin{eqnarray}
  f_{A}= \left\{\left(\frac{1}{2v_{\cal F}\tilde{X}_1}\right)^2+\left(\frac{1}{-2v_{\Psi}\tilde{X}_2}\right)^2\right\}^{-\frac{1}{2}}\,,
  \label{fA1}
 \end{eqnarray}
which will be more reduced to $f_{A}= \sqrt{2}\,N_2|X_1|v_{\cal F}=\sqrt{2}\,N_1|X_2|v_{\Psi}$ by using Eq.~(\ref{AhnMass}).
Under the $U(1)_{\tilde{X}}$ transformation, the axion field $A$ translates with the axion decay constant $F_{A}$
 \begin{eqnarray}
  A\rightarrow A+F_{A}\,\alpha\,\qquad\text{with}\,\,F_{A}\equiv f_{A}/N\,,
 \end{eqnarray}
where $\alpha\equiv\sum_i\alpha_{i}$ and $N=2N_1N_2$. Note here that if $N$ were large, then $F_A$ can be lowered significantly compared to the symmetry breaking scale.

However, the current $J^{\tilde{X}}_{\mu}$ is anomalous, that is, it is violated at one loop by the triangle anomaly $\partial^{\mu}J^{\tilde{X}}_{\mu}=N\frac{g^{2}_{s}}{32\pi^2}G^{a}_{\mu\nu}\tilde{G}^{a\mu\nu}$~\cite{anomaly}.
Then the corresponding Lagrangian has the form
 \begin{eqnarray}
  {\cal L}_{\rm eff}\ni\frac{g^{2}_{s}}{32\pi^2}\left(\vartheta_{\rm eff}+\frac{A_1}{f_{a1}}N_1+\frac{A_2}{f_{a2}}N_2\right)G^{a}_{\mu\nu}\tilde{G}^{a\mu\nu}=\frac{g^{2}_{s}}{32\pi^2}\left(\vartheta_{\rm eff}+\frac{A}{F_{A}}\right)G^{a}_{\mu\nu}\tilde{G}^{a\mu\nu}
 \end{eqnarray}
where $f_{a1}\equiv X_1v_{\cal F}$ and $f_{a2}\equiv X_2v_{\Psi}$.
Since $\vartheta_{\rm eff}$ is an angle of mod $2\pi$, after chiral rotations on Dirac fermion charged under $U(1)_{X_1}\times U(1)_{X_2}$, the Lagrangian should be invariant under
 \begin{eqnarray}
  \frac{A_1}{f_{a1}}\rightarrow\frac{A_1}{f_{a1}}+\frac{2\pi}{N_1}n_1\,,\qquad\frac{A_2}{f_{a2}}\rightarrow\frac{A_2}{f_{a2}}+\frac{2\pi}{N_2}n_2\,,
 \end{eqnarray}
where $n_{1,2}$ are non-negative integers. So,
it is clear to see the following by replacing $n_{i}$ with $N_{\rm DW}N_i$: if $N_{1}$ and $N_{2}$ are relative prime (so, the domain wall number $N_{\rm DW}=1$), there can be no $Z_{N_{\rm DW}}$ discrete symmetry and therefore no domain wall problem. Our model ($N_1=3$, $N_2=17$) corresponds to the case.

The heavy neutrinos and SM fermions get the flavored PQ symmetry $U(1)_X$ breaking mass terms and the effective Yukawa couplings, respectively, and the remaining massless (at this level) modes $A_1$ of the scalar $\Phi_{S}$ (or $\Theta$) and $A_{2}$ of the scalar $\Psi$ appear as phases:
 \begin{eqnarray}
  -{\cal L}_{Y} \rightarrow
  \frac{e^{i\frac{A_{1}}{v_{\cal F}}}}{2} \overline{N^{c}_{R}}M_RN_{R}
  +\overline{Q}_{L}\,Y_U\,U_{R}H_{u}+e^{i\frac{A_{1}}{v_{\cal F}}}\,\overline{Q}_{L}\,Y_D\,D_{R}H_{d}+\overline{\ell}_{L}\,Y_L\,\ell_{R}H_{d}+\text{h.c.}\,.
  \label{XYukawa}
 \end{eqnarray}
Here $U_{R}=(u_R,c_{R},t_{R})^{T}$, $D_{R}=(d_{R},s_{R},b_{R})^{T}$, and the Yukawa matrices $Y_{U},Y_{L}$ and $Y_{D}$ are expressed as
 \begin{eqnarray}
 Y_{U}&=&{\left(\begin{array}{ccc}
 y_{u}\,e^{8i\frac{A_{2}}{v_{\Psi}}} & 0 & 0  \\
 0 & y_{c}\,e^{4i\frac{A_{2}}{v_{\Psi}}} & 0 \\
 0 & 0 & y_{t}
 \end{array}\right)},\,\nonumber\\
 Y_{L}&=&{\left(\begin{array}{ccc}
 y_{e}\,e^{8i\frac{A_{2}}{v_{\Psi}}} & 0 & 0  \\
 0 & y_{\mu}\,e^{4i\frac{A_{2}}{v_{\Psi}}} & 0 \\
 0 & 0 & y_{\tau}\,e^{2i\frac{A_{2}}{v_{\Psi}}}
 \end{array}\right)}\,,\nonumber\\
 Y_{D}
 &=&{\left(\begin{array}{ccc}
 e^{3i\frac{A_{2}}{v_{\Psi}}} &  0 &  0 \\
 0 &  e^{2i\frac{A_{2}}{v_{\Psi}}} &  0   \\
 0 &  0  &  1
 \end{array}\right)}{\left(\begin{array}{ccc}
 \tilde{y}_{d} &  y_{d} &  y_{d} \\
 y_{s} & \tilde{y}_{s} &  y_{s}   \\
 y_{b} &  y_{b}   &  \tilde{y}_{b}
 \end{array}\right)}\frac{v_{S}}{\Lambda}\,,
 \end{eqnarray}
where $\tilde{y}_{f}=y_{f}+x_{f}\,\frac{1}{\kappa}\frac{v_{T}}{\Lambda}$ with $f=d,s,b$.
Note that all of Yukawa couplings above are dependent of the phases.
The Yukawa Lagrangian of the fermions in Eq.~(\ref{XYukawa}) have the $\tilde{X}$-symmetry with the transformation parameter $\alpha$ under
 \begin{eqnarray}
 U(1)_{\tilde{X}}: &&N_{R}\rightarrow e^{-i\frac{\tilde{X}_1}{2}\alpha}N_{R},\quad\qquad D_{R}\rightarrow e^{-i\tilde{X}_1\alpha}D_{R},~~\quad\qquad u_{R}\rightarrow e^{-5i\tilde{X}_2\alpha}u_{R},\nonumber\\
 &&c_{R}\rightarrow  e^{-2i\tilde{X}_2\alpha}c_{R},~\quad\qquad
Q_{L_1}\rightarrow e^{3i\tilde{X}_2\alpha}Q_{L_1},\quad\qquad Q_{L_2}\rightarrow e^{2i\tilde{X}_2\alpha}Q_{L_2},\nonumber\\
  &&e_{R}\rightarrow  e^{-i(\frac{\tilde{X}_1}{2}+8\tilde{X}_2)\alpha}e_{R},~\quad \mu_{R}\rightarrow  e^{-i(\frac{\tilde{X}_1}{2}+4\tilde{X}_2)\alpha}\mu_{R},~\quad \tau_{R}\rightarrow  e^{-i(\frac{\tilde{X}_1}{2}+2\tilde{X}_2)\alpha}\tau_{R},\nonumber\\
  &&\ell_{L}\rightarrow e^{-i\frac{\tilde{X}_1}{2}\alpha}\ell_{L},\qquad\qquad\qquad\qquad\text{ others}=\text{invariant},
 \end{eqnarray}
where we took, without loss of generality, the quantum number $r$ to be zero.
At energies below the electroweak scale, all quarks and leptons obtain masses.
From Eqs.~(\ref{lagrangianA}) and (\ref{lagrangianChA}) (see also Eq.~(\ref{XYukawa})) the fermion mass matrix is defined as $-{\cal L}_{M}=\bar{\psi}_{L}{\cal M}_{\psi}\psi_{R}+\text{h.c.}$. The axion coupling matrices to the up-type quarks, charged leptons, and down-type quarks, respectively, are diagonalized through bi-unitary transformations: $V^{\dag}_L{\cal M}_\psi V_R=\widehat{{\cal M}}_\psi$ (diagonal), $\psi^{0}_{L}=V^\dag_L\,\psi_{L}$ ($\psi^{0}_{L}$: mass eigenstates) and $\psi^{0}_{R}=V^{\dag}_{R}\,\psi_{R}$ ($\psi^{0}_{R}$: mass eigenstates). These transformation include in particular the chiral transformation necessary to make ${\cal M}_{u}$ and ${\cal M}_{d}$ real and positive. This induce a contribution to the QCD vacuum angle, {\it i.e.} $\vartheta\rightarrow\vartheta_{\rm eff}=\vartheta+\arg\left\{\det(\mathcal{M}_{u})\det(\mathcal{M}_{d})\right\}$ as in Eq.~(\ref{QCDlag1}). Note here that under the chiral rotation of the quark field given by Eq.~(\ref{chiralR}) the effective QCD angle $\vartheta_{\rm eff}$ is invariant.
The physical structure of the Lagrangian given by Eqs.~(\ref{XYukawa}) and (\ref{chiralR}) may be examined if we diagonalize the mass matrices for fermions. After diagonalization, between $1$ GeV and $246$ GeV the axion-fermion Lagrangian are expressed as
 \begin{eqnarray}
  -{\cal L}^{a-q} &\simeq&\frac{A_1}{f_{a1}}\left\{X_{1d}\,m_d\,\bar{d}i\gamma_{5}d+X_{1s}\,m_s\,\bar{s}i\gamma_{5}s+X_{1b}\,m_b\,\bar{b}i\gamma_{5}b\right\}\nonumber\\
  &+&\frac{A_2}{f_{a2}}\big\{X_{u}\,m_u\,\bar{u}i\gamma_{5}u
  +X_{c}\,m_c\,\bar{c}i\gamma_{5}c+X_{2d}\,m_d\,\bar{d}i\gamma_{5}d+X_{2s}\,m_s\,\bar{s}i\gamma_{5}s\big\}\nonumber\\
  &+&m_u\,\bar{u}u+m_c\,\bar{c}c+m_t\,\bar{t}t+m_d\,\bar{d}d+m_s\,\bar{s}s+m_b\,\bar{b}b-\bar{q}i\gamma_{\mu}D^{\mu}q,\label{AxionLag12}\\
  -{\cal L}^{a-\ell} &\simeq&\frac{A_2}{f_{a2}}\left\{X_{e}\,m_e\,\bar{e}\,i\gamma_{5}\,e+X_{\mu}\,m_\mu\,\bar{\mu}\,i\gamma_{5}\,\mu+X_{\tau}\,m_\tau\,\bar{\tau}\,i\gamma_{5}\,\tau\right\}\nonumber\\
  &+&m_e\,\bar{e}e+m_\mu\,\bar{\mu}\mu+m_\tau\,\bar{\tau}\tau-\bar{\ell}i\gamma_{\mu}D^{\mu}\ell
  \label{AxionLag13}
 \end{eqnarray}
in which $q=u,c,t,d,s,b$, $\ell=e,\mu,\tau$ represent mass eigenstates, and $D_{\mu}$ are the covariant derivatives for the $SU(3)\times SU(2)\times U(1)$ gauge interactions of the SM.  The axion couplings are model dependent with the elements of the matrices, so the $X$-charges of the fermions are given as $X_u=8X_2$, $X_c=4X_2$, $X_e=8X_2$, $X_\mu=4X_2$,  $X_\tau=2X_2$, $X_{1d}=X_{1s}=X_{1b}=X_1$, $X_{2d}=3X_2$ and $X_{2s}=2X_2$. Recalling that $X_1=-2p$ and $X_2=-q$.
The above axion-SM fermion interactions are applicable above $1$ GeV such as in $J/\Psi$ and $\Upsilon$ decays. It is clear that the hadronic axion does not couple to leptons at tree level, whereas the new Goldstone bosons, $A_1$ and/or $A_2$, interact with both quarks and leptons. Such couplings, however, are suppressed by factors $v/f_{a1}$ or $v/f_{a2}$. Consequently, both the hadronic axion and the new Goldstone bosons are invisible.
Below the QCD scale (1 GeV$\approx4\pi f_{\pi}$), the axion-hadron interactions are meaningful rather than the axion-quark couplings: the chiral symmetry is broken and $\pi, K$ and $\eta$ are produced as pseudo-Goldstone bosons. Then the axion coupling to quarks is changed as will be seen in the following subsection.

\subsection{Axion interactions with quarks, leptons, gluons, and photons}
Now, through a chiral rotation on $\psi$, we can dispose of the $\vartheta_{\rm eff}$ angle in Eq.~(\ref{QCDlag}). Let us chiral-rotate the $f$-th $\psi$ in the Fujikawa measure of the path integral that a rotation of
 \begin{eqnarray}
  \psi_f\rightarrow \text{exp}\left(i\frac{\alpha_{f}\gamma_5}{2}\right)\psi_f\qquad\text{with}\,\,\alpha_{f}\equiv\rho \tilde{X}_{\psi_f}=\rho(\tilde{X}_{1\psi_f}-\tilde{X}_{2\psi_f})
\label{chiralR}
 \end{eqnarray}
on Dirac spinors contribute
 \begin{eqnarray}
  {\cal L}\rightarrow{\cal L}+\frac{g^{2}_{s}}{16\pi^2}\sum_{\psi_f}\rho \tilde{X}_{\psi_f} G^{a}_{\mu\nu}\tilde{G}^{b\mu\nu}\,{\rm Tr}(t^at^b)={\cal L}+\frac{g^{2}_{s}}{32\pi^2}\,\rho N\,G^{a}_{\mu\nu}\tilde{G}^{a\mu\nu}
 \label{angleT}
 \end{eqnarray}
to the Lagrangian, where the $N$ is the axion color anomaly of the $U(1)_{\tilde{X}}$ symmetry. And the second term in Eq.~(\ref{angleT}) is obtained by letting $2\sum_{\psi_f}\tilde{X}_{1\psi_f}\,{\rm Tr}(t^at^b)-2\sum_{\psi_f}\tilde{X}_{2\psi_f}\,{\rm Tr}(t^at^b)=N\delta^{ab}$, where the sum runs over all $\psi$ with $\tilde{X}$-charge.

Through a rotation Eq.~(\ref{chiralR}), {\it i.e.} $\psi_f\rightarrow \text{exp}\{i\frac{\tilde{X}_{\psi}}{N}\frac{A}{F_{A}}\frac{\gamma_5}{2}\}\psi_f$, we obtain
the vanishing anomaly terms by adding the QCD vacuum given in Lagrangian~(\ref{QCDlag}) to the above Lagrangian
 \begin{eqnarray}
   {\cal L}_{\vartheta} &=& \left(\vartheta_{\rm eff}+\frac{A_1}{F_{a_1}}+\frac{A_2}{F_{a_2}}\right)\frac{\alpha_s}{8\pi}G^{a\mu\nu}\tilde{G}^{a}_{\mu\nu}\equiv\left(\vartheta_{\rm eff}+\frac{A}{F_{A}}\right)\frac{\alpha_s}{8\pi}G^{\mu\nu a}\tilde{G}^{a}_{\mu\nu}\,.
 \label{angleT1}
 \end{eqnarray}
Here $F_{a_i}=f_{a_i}/N_i$ with $i=1,2$.
At low energies $A$ will get a VEV, $\langle A\rangle=-F_{A}\vartheta_{\rm eff}$, eliminating the constant $\vartheta_{\rm eff}$ term. The axion then is the excitation of the $A$ field, $a=A-\langle A\rangle$.
Since the SM fields $\psi$ have $U(1)_{\rm EM}$ charges, the axion coupling to photon will be added to the Lagrangian through a rotation Eq.~(\ref{chiralR}), which survive to the QCD scale:
 \begin{eqnarray}
  {\cal L}\rightarrow{\cal L}+e^{2}\frac{2\rho\sum_\psi \tilde{X}_{\psi}(Q^{\rm em}_i)^2}{32\pi^2}F_{\mu\nu}\tilde{F}^{\mu\nu}={\cal L}+\frac{e^{2}}{32\pi^2}\left(\frac{E}{N}\right)\frac{A}{F_{A}}F_{\mu\nu}\tilde{F}^{\mu\nu}
 \label{EManomaly}
 \end{eqnarray}
with the axion electromagnetic anomaly $E=2\sum_{\psi} \tilde{X}_{1\psi_f}(Q^{\rm em}_{f})^2-2\sum_{\psi} \tilde{X}_{2\psi_f}(Q^{\rm em}_{f})^2$ for here $\psi=$ all $\tilde{X}$-charged Dirac fermions, where $F_{\mu\nu}$ is the electromagnetic field strength and its dual $\tilde{F}^{\mu\nu}$. Note that since the field $A$ is not a constant, this term is not a total derivative, and so can not be neglected.

In order to remove the axion fields from the Yukawa interactions in Eqs.~(\ref{AxionLag12}) and (\ref{AxionLag13}), instead of using Eq.~(\ref{chiralR}) we transform the quark and lepton fields by the chiral rotations
 \begin{eqnarray}
&&D_{R}\rightarrow e^{i\frac{X_1A_1}{f_{a1}}}D_{R},~~\quad\qquad\qquad u_{R}\rightarrow e^{i\frac{5X_2A_2}{f_{a2}}}u_{R},\qquad\qquad c_{R}\rightarrow  e^{i\frac{2X_2A_2}{f_{a2}}}c_{R},\nonumber\\
 &&~~~
Q_{L_1}\rightarrow e^{i\frac{3X_2A_2}{f_{a2}}}Q_{L_1},~\qquad\quad Q_{L_2}\rightarrow e^{i\frac{2X_2A_2}{f_{a2}}}Q_{L_2},~~\qquad\quad\ell_{L}\rightarrow e^{i\frac{X_1}{2}\frac{A_1}{f_{a1}}}\ell_{L},\nonumber\\
  &&e_{R}\rightarrow  e^{i(\frac{X_1A_1}{2f_{a1}}+\frac{8X_2A_2}{f_{a2}})}e_{R},~\quad \mu_{R}\rightarrow  e^{i(\frac{X_1A_1}{2f_{a1}}+\frac{4X_2A_2}{f_{a2}})}\mu_{R},~\quad \tau_{R}\rightarrow  e^{i(\frac{X_1A_1}{2f_{a1}}+\frac{2X_2A_2}{f_{a2}})}\tau_{R}.
 \end{eqnarray}
Then derivative interactions from the kinetic terms for the fermions are generated
 \begin{eqnarray}
  -{\cal L}^{a-q} &\simeq&\frac{\partial_{\mu}A_1}{2f_{a1}}\left\{X_{1d}\,\bar{d}\gamma^{\mu}\gamma_{5}d+X_{1s}\,\bar{s}\gamma^{\mu}\gamma_{5}s+X_{1b}\,\bar{b}\gamma^{\mu}\gamma_{5}b\right\}\nonumber\\
  &+&\frac{\partial_{\mu}A_2}{2f_{a2}}\big\{X_{u}\,\bar{u}\gamma^{\mu}\gamma_{5}u
  +X_{c}\,\bar{c}\gamma^{\mu}\gamma_{5}c+X_{2d}\,\bar{d}\gamma^{\mu}\gamma_{5}d+X_{2s}\,\bar{s}\gamma^{\mu}\gamma_{5}s\big\}\nonumber\\
  &+&m_u\,\bar{u}u+m_c\,\bar{c}c+m_t\,\bar{t}t+m_d\,\bar{d}d+m_s\,\bar{s}s+m_b\,\bar{b}b-\bar{q}i\gamma_{\mu}D^{\mu}q,\label{AxionLag14}\\
  -{\cal L}^{a-\ell} &\simeq&\frac{\partial_{\mu}A_2}{2f_{a2}}\left\{X_{e}\,\bar{e}\,\gamma^{\mu}\gamma_{5}\,e+X_{\mu}\,\bar{\mu}\,\gamma^{\mu}\gamma_{5}\,\mu+X_{\tau}\,\bar{\tau}\,\gamma^{\mu}\gamma_{5}\,\tau\right\}\nonumber\\
  &+&m_e\,\bar{e}e+m_\mu\,\bar{\mu}\mu+m_\tau\,\bar{\tau}\tau-\bar{\ell}i\gamma_{\mu}D^{\mu}\ell,
  \label{AxionLag15}
 \end{eqnarray}
both of which are equivalent to Eqs.~(\ref{AxionLag12},\ref{AxionLag13}). And the derivative interactions can also be simplified, and in turn which can be expressed in terms of the axion $A$ as
 \begin{eqnarray}
  \frac{1}{2}\sum_\psi \left(\frac{\partial_{\mu}A_1}{f_{a1}}X_{1\psi}+\frac{\partial_{\mu}A_2}{f_{a2}}X_{2\psi}\right)\bar{\psi}\gamma^{\mu}\gamma_{5}\psi
  =\frac{\partial_{\mu}A}{f_A}\sum_{\psi}\tilde{X}_{\psi}\bar{\psi}\gamma^{\mu}\gamma_{5}\psi\,.
 \end{eqnarray}

At energies far below $f_{A}$, after integrating out the $X$-charge carrying heavy degree of freedoms, in terms of the physical axion field $``a"$ (which is the excitation with the vacuum expectation removed) we can obtain the following effective Lagrangian ${\cal L}$~\footnote{Refs.~\cite{CASPEr} has recently considered several interesting effects arising from and detection schemes based on some of these effects for the axion couplings to quarks, leptons and gluons.}  including the SM Lagrangian ${\cal L}_{\rm SM}$:
 \begin{eqnarray}
  {\cal L} &\ni&\frac{1}{2}(\partial_{\mu}a)^2-\frac{\partial_{\mu}a}{f_{A}}\sum_\psi \tilde{X}_{\psi}\bar{\psi}\gamma^{\mu}\gamma_{5}\psi
  +\frac{g^{2}_{s}}{32\pi^2}\frac{a}{F_A}G^{a}_{\mu\nu}\tilde{G}^{a\mu\nu}+\frac{e^{2}}{32\pi^2}\left(\frac{E}{N}\right)\frac{a}{F_A}F_{\mu\nu}\tilde{F}^{\mu\nu}.
  \label{AxionLag01}
 \end{eqnarray}

\subsection{Axion mass and Axion-photon coupling}

Now, below the $SU(2)\times U(1)$ breaking scale where all quarks and leptons obtain masses, the $X$-current given in Eq.~(\ref{Jx}) is constructed from the axion, quark and lepton transformations under the $X$-symmetry. The reason that the axion gets a mass is that the $X$-current has the color anomaly. Then we neglect the lepton current for the axion mass.

We integrate out the heavy quarks ($c, b, t$) to obtain the effective couplings just above QCD scale.
Now there are three light quarks ($u,d,s$). In order to obtain the axion mass and derive the axion coupling to photons, we eliminate the coupling of axions to gluons through rotation of the light quark fields
 \begin{eqnarray}
  q\rightarrow \text{exp}\left(-i\alpha_q\frac{\gamma_5}{2}\right)q\qquad\text{with}\,\,q=u,d,s\,.
\label{chiralR2}
 \end{eqnarray}
With the above chiral-rotation, such that $a/F_A-\sum_q\alpha_q=0$, the quark-axion sector of the Lagrangian (\ref{AxionLag01}) reads
 \begin{eqnarray}
  {\cal L}_{A} &=&i\bar{q}\gamma_{\mu}D_{\mu}q+\frac{1}{2}(\partial^{\mu}a)^2-\frac{\partial^{\mu}a}{f_A}\sum_q\left(\tilde{X}_q+\alpha_q\right)\bar{q}\gamma^{\mu}\gamma_5q\nonumber\\
  &-&\left(\sum_{q=u,d,s} m_{q}\bar{q}_Le^{i\alpha_q}q_R+\text{h.c.}\right)+\frac{e^{2}}{32\pi^2}\left(\frac{E}{N}\frac{a}{F_A}-6\sum_{q}\alpha_q(Q^{\rm em}_q)^2\right)F_{\mu\nu}\tilde{F}^{\mu\nu}\,.
  \label{AxionLag02}
 \end{eqnarray}
As can be seen here, the CP violating $\vartheta_{\rm eff}$ term at the minimum is canceled out, which provides a dynamical solution to the CP problem~\cite{Peccei-Quinn}, but there is a phase in $m_q$.
Clearly, we have some freedom in choosing the phase~\footnote{In the case that $m_{u}, m_{d}$ and $m_{s}$ are equal, it is natural to choose these phase to be the same, {\it i.e.} $\alpha_{u}=\alpha_{d}=\alpha_{s}\equiv\alpha/3$~\cite{Dine:2000cj}.}:
since the QCD vacuum is a flavor singlet, {\it i.e.} $\langle\bar{u}u\rangle=\langle\bar{d}d\rangle=\langle\bar{s}s\rangle$, the $\alpha_{q}$ is determined by the flavor singlet condition, that is, $\alpha_{u}m_u=\alpha_{d}m_d=\alpha_{s}m_s$. From $a/F_A-\sum_q\alpha_q=0$ we obtain
 \begin{eqnarray}
  \alpha_u&=&\frac{a}{F_A}\frac{1}{1+z+w}\,,\qquad
  \alpha_d=\frac{a}{F_A}\frac{z}{1+z+w}\,,\qquad
  \alpha_s=\frac{a}{F_A}\frac{w}{1+z+w}\,,
 \end{eqnarray}
where $z=m_{u}\langle\bar{u}u\rangle/m_{d}\langle\bar{d}d\rangle=m_{u}/m_{d}$ and $w=m_{u}\langle\bar{u}u\rangle/m_{s}\langle\bar{s}s\rangle=m_{u}/m_{s}$ in the $SU(3)_{\rm flavor}$ symmetric vacuum.
Considering $u$, $d$ and $s$ quarks, the chiral symmetry breaking effect due to the mixing between axion and light mesons is
 \begin{eqnarray}
  \sum_{q}\alpha_q(Q^{\rm em}_q)^2=\frac{4+z+w}{9(1+z+w)}\frac{a}{F_A}\,.
 \end{eqnarray}
And the value of $E/N$ is determined by the $X$-charge carrying quarks and leptons
 \begin{eqnarray}
  \frac{E}{N}&=&\frac{2\cdot[(\tilde{X}_e+\tilde{X}_\mu+\tilde{X}_\tau)(-1)^2+3(\tilde{X}_u+\tilde{X}_c)\left(\frac{2}{3}\right)^2+3(\tilde{X}_d+\tilde{X}_s+\tilde{X}_b)\left(-\frac{1}{3}\right)^2]}{2(X_{1d}+X_{1s}+X_{1b})(X_{u}+X_{c}+X_{2d}+X_{2s})}
 \end{eqnarray}
which corresponds to $112/51$, where $N_1=3$, $N_2=17$ for the given $X$-charges $X_1=X_2=1$. Here the axion color anomaly $N$ and electromagnetic anomaly $E$ are given below Eq.~(\ref{angleT}) and Eq.~(\ref{EManomaly}), respectively.

And, at below the QCD scale where the quarks have hadronized into mesons, which will result in mixing between axions and NG mesons of the broken chiral $SU(3)_L\times SU(3)_R$, the kinetic terms vanish
 \begin{eqnarray}
  {\cal L}_{A} &=&-
 \left(\sum_{q=u,d,s} m_{q}\bar{q}_Le^{i\alpha_q}q_R+\text{h.c.}\right)+\frac{e^{2}}{32\pi^2}\left(\frac{E}{N}-\frac{2}{3}\,\frac{4+z+w}{1+z+w}\right)\frac{a}{F_A}F_{\mu\nu}\tilde{F}^{\mu\nu}\,.
 \end{eqnarray}
From the effective Lagrangian~(\ref{XYukawa}) or Eq.~(\ref{AxionLag12}) the interaction for the light quarks preserves the $X$-symmetry, while it does not preserve the chiral symmetry. So, we may include the effects of the Yukawa interactions in the effective Lagrangian by adding a term which explicitly breaks the symmetry. Let us consider the form of the chiral Lagrangian
 \begin{eqnarray}
  {\cal L}_{\rm eff} &=& -\frac{f^{2}_{\pi}}{4}{\rm Tr}\left[D_{\mu}\Sigma^{\dag}D^{\mu}\Sigma\right]-\frac{1}{2}\mu f^{2}_{\pi}{\rm Tr}\left[\Sigma{\cal A} M_q+(\Sigma{\cal A}M_q)^{\dag}\right]
  \label{ChLagran}
 \end{eqnarray}
where
$\Sigma\equiv{\rm exp}\left[2i\pi^aT^a/f_{\pi}\right]$ ($a=1,...,8$) is the meson field, $T^a$ are the generators of $SU(3)$, $D_{\mu}$ is the appropriate covariant derivatives which introduce the electroweak interactions, $f_{\pi}=93$ MeV, $\mu$ is an undetermined constant, which is related to explicit chiral symmetry breaking, $M_q={\rm diag}(m_{u},m_{d},m_{s})$ is the light quark mass matrix, and ${\cal A}={\rm diag}(e^{i\alpha_{u}}, e^{i\alpha_{d}}, e^{i\alpha_{s}})$ is the axion phase rotation. The first term in the above Lagrangian~(\ref{ChLagran}) is invariant under global transformation $\Sigma\rightarrow g_{L}\Sigma g^{\dag}_{R}$ where $g_{L}=I$ (unit matrix) and $g_{R}={\rm diag}(e^{i\alpha_{1}}, e^{i\alpha_{2}}, e^{i\alpha_{3}})$, while the second term is not invariant. Thus, the axion and mesons will acquire masses from the second term in the Lagrangian~(\ref{ChLagran}). Note that the invariance of the above Lagrangian~(\ref{ChLagran}) under $U(1)_{\tilde{X}}$ requires that $\Sigma$ transform as
 \begin{eqnarray}
 \Sigma\rightarrow\Sigma{\left(\begin{array}{ccc}
 e^{-i\alpha\,\tilde{X}_{u}} &  0 &  0 \\
 0 &  e^{-i\alpha\,\tilde{X}_{d}} &  0 \\
 0 &  0 &  e^{-i\alpha\,\tilde{X}_{s}}
 \end{array}\right)}\,;\qquad A\rightarrow A+F_A\alpha\,.
 \end{eqnarray}
Even the $A$ field is generated at the high energy, it develops a VEV below QCD scale. Expanding $\Sigma$ and considering the constant term corresponding to ground state energy, the $A$ potential is given as
 \begin{eqnarray}
 V(A)
 &=&-\mu f^{2}_{\pi}\Big\{m_{u}\cos\frac{1}{1+z+w}\left(\frac{A}{F_A}+\vartheta_{\rm eff}\right)\nonumber\\
 &+&m_{d}\cos\frac{z}{1+z+w}\left(\frac{A}{F_A}+\vartheta_{\rm eff}\right)
 +m_{s}\cos\frac{w}{1+z+w}\left(\frac{A}{F_A}+\vartheta_{\rm eff}\right)\Big\}\,,
 \end{eqnarray}
which is minimized when $\langle A\rangle=-\vartheta_{\rm eff}F_A$.
Then the axion mass is proportional to the curvature of the effective potential induced by the anomaly.
Expanding $V(A)$ at the minimum gives the axion mass
 \begin{eqnarray}
 m^2_{a}=\left\langle\frac{\partial^2V(A)}{\partial a^2}\right\rangle_{\langle A\rangle=-\vartheta_{\rm eff}F_A}=\frac{ f^{2}_{\pi}}{F^2_{A}}\frac{\mu m_u}{1+z+w}.
  \label{axiMass1}
 \end{eqnarray}
\begin{figure}[t]
\includegraphics[width=11.0cm]{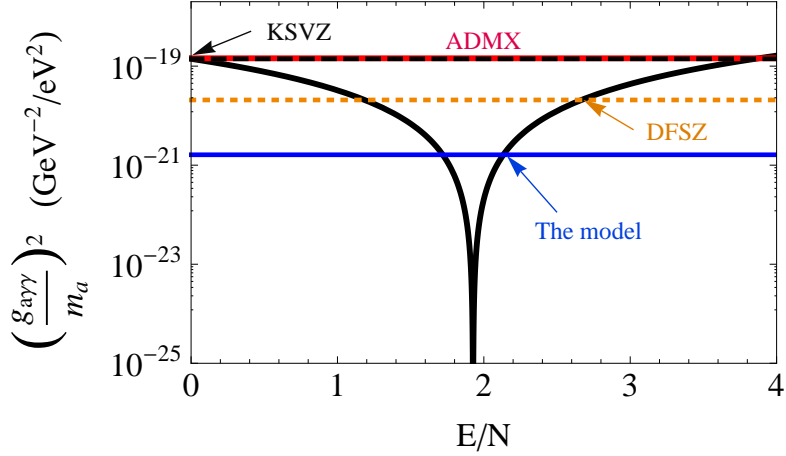}
\caption{\label{Fig2} Plot of $(g_{a\gamma\gamma}/m_{a})^2$ versus $E/N$ for $z=0.56$. The solid-red line represents the experimental upper bound $(g_{a\gamma\gamma}/m_{a})^2\leq1.44\times10^{-19}\,{\rm GeV}^{-2}\,{\rm eV}^{-2}$ from ADMX~\cite{Asztalos:2003px}. Here the dashed-black, dotted-brown, and solid-blue lines stand for $(g_{a\gamma\gamma}/m_{a})^2=1.404\times10^{-19}\,{\rm GeV}^{-2}\,{\rm eV}^{-2}$ for $E/N=0$, $2.074\times10^{-20}\,{\rm GeV}^{-2}\,{\rm eV}^{-2}$ for $E/N=8/3$, and $2.754\times10^{-21}\,{\rm GeV}^{-2}\,{\rm eV}^{-2}$ for $E/N=112/51$, respectively.}
\end{figure}
The physical axion/ meson states and the mixing parameters may be determined from the axion/ meson mass matrix which can be obtained by expanding the symmetry breaking part in Lagrangian~(\ref{ChLagran}) and taking the terms quadratic in the fields (see Eq.~(\ref{neut-A})). The axion mass in terms of the pion mass is obtained as
 \begin{eqnarray}
 m^{2}_{a}F^{2}_{A}=m^{2}_{\pi^0}f^{2}_{\pi}F(z,w)\,,
\label{axiMass2}
 \end{eqnarray}
where $m^2_{\pi^0}$ is the $\pi^{0}\pi^{0}$ entry of ${\cal M}^2$ in Eq.~(\ref{neut-C}), and
 \begin{eqnarray}
 F(z,w)=\frac{z}{(1+z)(1+z+w)}\,, \qquad F_A=\left\{\left(\frac{1}{F_{a1}}\right)^2+\left(\frac{1}{F_{a2}}\right)^2\right\}^{-\frac{1}{2}}\,.
 \end{eqnarray}
It is clear that the axion mass vanishes in the limit $m_u$ or $m_{d}\rightarrow0$. The axion mass derived in Eq.~(\ref{axiMass2}) is equivalent to Eq.~(\ref{axiMass1}).
In order to estimate the axion mass, first we determine the parameters $\mu m_{u}$ and $w$ as a function of $z$ from the physical masses of the NG bosons. In Eq.~(\ref{neut-A}) they can be extracted as $\mu m_{u}= (108.3{\rm MeV})^2\,z, w=0.315\,z$.
Then we can estimate the axion mass
 \begin{eqnarray}
 m_{a}\simeq2.53\times10^{-5}{\rm eV}\left(\frac{10^{12}{\rm GeV}}{3\sqrt{2}\,F_{A}}\right)\,,
 \end{eqnarray}
where the Weinberg value for $z\equiv m_{u}/m_{d}=0.56$~\cite{WB} and Eq.~(\ref{AhnMass}) are used.
After integrating out the heavy $\pi^{0}$ and $\eta$ at low energies, there is an effective low energy Lagrangian with an axion-photon coupling $g_{a\gamma\gamma}$:
 \begin{eqnarray}
{\cal L}_{a\gamma\gamma}= \frac{1}{4}g_{a\gamma\gamma}\,a_{\rm phys}\,F^{\mu\nu}\tilde{F}_{\mu\nu}=-g_{a\gamma\gamma}\,a_{\rm phys}\,\vec{E}\cdot\vec{B}\,,
 \end{eqnarray}
where $\vec{E}$ and $\vec{B}$ are the electromagnetic field components.
And the axion-photon coupling can be expressed in terms of the axion mass, pion mass, pion decay constant, $z$ and $w$:
 \begin{eqnarray}
 g_{a\gamma\gamma}=\frac{\alpha_{\rm em}}{2\pi}\frac{m_a}{f_{\pi}m_{\pi^0}}\frac{1}{\sqrt{F(z,w)}}\left(\frac{E}{N}-\frac{2}{3}\,\frac{4+z+w}{1+z+w}\right)\,.
 \end{eqnarray}
\begin{figure}[h]
\includegraphics[width=11.0cm]{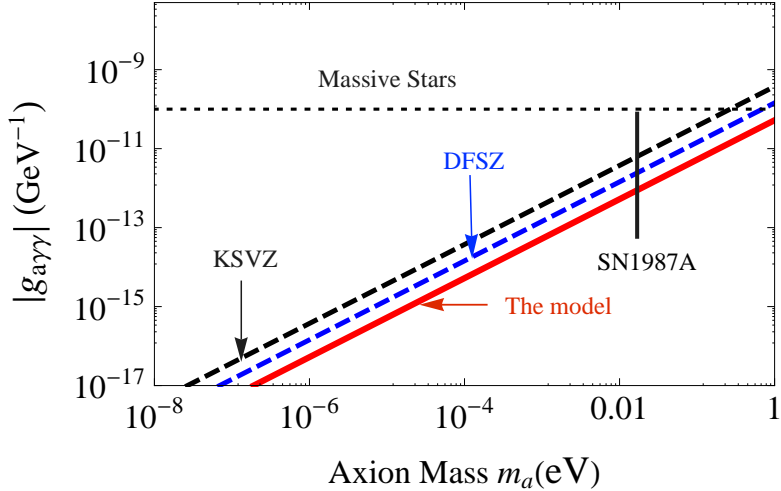}
\caption{\label{Fig3} Plot of $|g_{a\gamma\gamma}|$ versus $m_{a}$ for KSVZ (black dashed line), DFSZ (blue dashed line) and our model (red solid line) in terms of $E/N=$ $0$, $8/3$ and $112/51$, respectively. Here the horizontal dotted line stands for the upper bound $|g_{a\gamma\gamma}|\lesssim1\times10^{-10}$ GeV$^{-1}$ which is from globular-cluster stars~\cite{PDG}. And the black bar corresponding to $m_a\lesssim16$ meV is the constraint derived from the measured duration of the neutrino signal of the supernova SN1987A~\cite{PDG}. Especially, in the model, for $f_{A}=10^{12}$ GeV we obtain $m_{a}=2.53\times10^{-5}$ eV and $|g_{a\gamma\gamma}|=1.33\times10^{-15}$ GeV$^{-1}$.}
\end{figure}
The axion coupling to photon $g_{a\gamma\gamma}$ divided by the axion mass $m_{a}$ is dependent on $E/N$. Fig.~\ref{Fig2} shows the $E/N$ dependence of $(g_{a\gamma\gamma}/m_{a})^2$ so that the experimental limit is independent of the axion mass $m_{a}$: the value of $(g_{a\gamma\gamma}/m_{a})^2$ of our model is 1 or 2 orders of magnitude lower than that of the conventional axion model, {\it i.e.} KSVZ or DFSZ model. For the Weinberg value $z=0.56$, the anomaly value $E/N=112/51$ predicts $(g_{a\gamma\gamma}/m_{a})^2=2.754\times10^{-21}\,{\rm GeV}^{-2}\,{\rm eV}^{-2}$ which is lower than the ADMX (Axion Dark Matter eXperiment) bound~\cite{Asztalos:2003px}.
Fig.~\ref{Fig3} shows the plot for the axion-photon coupling $|g_{a\gamma\gamma}|$ as a function of the axion mass $m_{a}$ in terms of anomaly values $E/N=0, 8/3, 112/51$ which correspond to the KSVZ, DFSZ and our model, respectively. The model will be testable in the very near future through experiment such as that at the Center for Axion and Precision Physics research (CAPP)~\cite{CAPP}.

\section{Conclusion}

We have suggested a $\mu-\tau$ power law under which certain elements associated with the muon and tau flavors in the lepton mass matrices are distinguished, such that relatively large 13 mixing angle and bi-large mixing ones could be derived. According to this, we have proposed a neat and economical model for both the fermion mass hierarchy problem of the standard model and a solution of the strong CP problem, in a way that no domain wall problem occurs, based on $A_{4}\times U(1)_{X}$ symmetry in a supersymmetric framework. Here the global $U(1)_X$ symmetry that can explain the above problems is referred to as ``flavored Peccei-Quinn symmetry". In the model, a direct coupling of the SM gauge singlet flavon fields responsible for spontaneous symmetry breaking to ordinary quarks and leptons, both of which carry $X$-charges, comes to pass through Yukawa interactions. All the VEVs (scaled by the cutoff scale $\Lambda$) breaking the symmetries are connected each other. So, the other VEVs scales are automatically determined, once a VEV scale is fixed through low energy phenomenology. In the model, the scale of Peccei-Quinn symmetry breaking is shown to be roughly located around $10^{12}$ GeV section through its connection to the fermion masses.

On phenomenology, we have examined leptonic $CP$ violation and neutrinoless double beta ($0\nu\beta\beta$) decay : Figs.~\ref{FigA1} and \ref{FigA2} show the main results. Future precise measurement on the atmospheric mixing angle $\theta_{23}$ is of importance in order to distinguish between normal mass ordering (NO) and inverted one (IO) in the model. The value of $\theta_{23}$ would lie on $|\theta_{23}-45^{\circ}|\sim1^{\circ}$ for NO, while $|\theta_{23}-45^{\circ}|\sim3^{\circ}-8^{\circ}$ for IO. Moreover, the model predictions have showed that the IO is more predictive on Dirac CP phase $\delta_{CP}\sim70^{\circ}, 110^{\circ}, 250^{\circ}, 290^{\circ}$ than the NO $\delta_{CP}\in[90^{\circ}, 270^{\circ}]$ for $\theta_{23}\sim46^{\circ}$ and $\delta_{CP}\in[0^{\circ}, 90^{\circ}]$ and $[270^{\circ}, 360^{\circ}]$ for $\theta_{23}\sim44^{\circ}$, and the effective neutrino mass proportional to the $0\nu\beta\beta$ decay $|m_{ee}|\sim0.044-0.16$ eV for NO and $0.066-0.171$ eV for IO. And also we have showed that the model naturally describes the fermion mass and mixing hierarchies of the standard model, which are in good agreement with the present data. Interestingly, we have showed model predictions on the axion mass $m_{a}\simeq2.53\times10^{-5}$ eV and the axion coupling to photon $g_{a\gamma\gamma}\simeq1.33\times10^{-15}~{\rm GeV}^{-1}$. In turn, the square of the ratio between them is shown to be 1 or 2 orders of magnitude lower than that of the conventional axion model, {\it i.e.} KSVZ or DFSZ model.  The model can be testable in the very near future through on-going experiments for neutrino oscillation, neutrinoless double beta decay and axion.

\newpage

\appendix
\section{The $A_{4}$ Group}
 \label{A4group}
The  group $A_{4}$ is the symmetry group of the
tetrahedron, isomorphic to the finite group of the even permutations of four
objects. The group $A_{4}$ has two generators, denoted $S$ and $T$, satisfying the relations $S^{2}=T^{3}=(ST)^{3}=1$. In the three-dimensional complex representation, $S$ and $T$ are given by
 \begin{eqnarray}
 S=\frac{1}{3} \, {\left(\begin{array}{ccc}
 -1 &  2 &  2 \\
 2 &  -1 & 2 \\
 2 &  2 &  -1
 \end{array}\right)}~,\qquad T={\left(\begin{array}{ccc}
 1 &  0 &  0 \\
 0 &  \omega &  0 \\
 0 &  0 &  \omega^2
 \end{array}\right)}~.
 \label{generator}
 \end{eqnarray}
$A_{4}$ has four irreducible representations: one triplet ${\bf
3}$ and three singlets ${\bf 1}, {\bf 1}', {\bf 1}''$.
An $A_4$ singlet $a$ is invariant under the action of $S$ ($Sa=a$), while the action of $T$ produces $Ta=a$ for ${\bf 1}$, $Ta=\omega a$ for ${\bf 1}'$, and $Ta=\omega^2 a$ for ${\bf 1}''$, where $\omega=e^{i2\pi/3}=-1/2+i\sqrt{3}/2$ is a complex cubic-root of unity.
Products of two $A_4$ representations decompose into irreducible representations according to the following multiplication rules: ${\bf 3}\otimes{\bf 3}={\bf 3}_{s}\oplus{\bf
3}_{a}\oplus{\bf 1}\oplus{\bf 1}'\oplus{\bf 1}''$, ${\bf
1}'\otimes{\bf 1}''={\bf 1}$, ${\bf 1}'\otimes{\bf 1}'={\bf 1}''$
and ${\bf 1}''\otimes{\bf 1}''={\bf 1}'$. Explicitly, if $(a_{1},
a_{2}, a_{3})$ and $(b_{1}, b_{2}, b_{3})$ denote two $A_4$ triplets, then we have Eq.~(\ref{A4reps}).

To make the presentation of our model physically more transparent, we define the $T$-flavor quantum number $T_f$ through the eigenvalues of the operator $T$, for which $T^3=1$. In detail, we say that a field $f$ has $T$-flavor $T_f=0$, +1, or -1 when it is an eigenfield of the $T$ operator with eigenvalue $1$, $\omega$, $\omega^2$, respectively (in short, with eigenvalue $\omega^{T_f}$ for $T$-flavor $T_f$, considering the cyclical properties of the cubic root of unity $\omega$). The $T$-flavor is an additive quantum number modulo 3. We also define the $S$-flavor-parity through the eigenvalues of the operator $S$, which are +1 and -1 since $S^2=1$, and we speak of $S$-flavor-even and $S$-flavor-odd fields.
For $A_4$-singlets, which are all $S$-flavor-even, the $\mathbf{1}$ representation has no $T$-flavor ($T_f=0$), the $\mathbf{1}'$ representation has $T$-flavor $T_f=+1$, and the $\mathbf{1}''$ representation has $T$-flavor $T_f=-1$. Since for $A_4$-triplets, the operators $S$ and $T$ do not commute, $A_4$-triplet fields cannot simultaneously have a definite $T$-flavor and a definite $S$-flavor-parity.

The real representation, in which $S$ is diagonal, is obtained through the unitary transformation
\begin{align}
A \to A'=U_{\omega}\,A\,U^{\dag}_{\omega},
\end{align}
where $A$ is any $A_4$ matrix in the real representation and
\begin{align}
U_{\omega}=\frac{1}{\sqrt{3}}{\left(\begin{array}{ccc}
 1 &  1 &  1 \\
 1 & \omega & \omega^{2} \\
 1 & \omega^{2} & \omega
 \end{array}\right)}.
 \label{eq:Uomega}
\end{align}
We have
 \begin{eqnarray}
 S'={\left(\begin{array}{ccc}
 1 &  0 &  0 \\
 0 &  -1 & 0 \\
 0 &  0 &  -1
 \end{array}\right)}~,\qquad T'={\left(\begin{array}{ccc}
 0 &  1 &  0 \\
 0 &  0 &  1 \\
 1 &  0 &  0
 \end{array}\right)}~.
 \label{generator2}
 \end{eqnarray}
For reference, an $A_4$ triplet field with $T$-flavor eigenfields $(a_1,a_2,a_3)$ in the complex representation can be expressed in terms of components $(a_{R1},a_{R2},a_{R3})$ as
\begin{align}
 a_{1R} = \frac{a_{1}+a_{c2}+a_{3}}{\sqrt{3}} , \quad
 a_{2R} = \frac{a_{1}+\omega\,a_{2}+\omega^2 a_{3}}{\sqrt{3}} , \quad
 a_{3R} = \frac{a_{1}+\omega^2 a_{2}+\omega\,a_{3}}{\sqrt{3}} .
 \label{eq:Ua1}
\end{align}
Inversely,
 \begin{align}
 a_{1}  = \frac{a_{1R}+a_{2R}+a_{3R}}{\sqrt{3}} , \quad
 a_{2}  = \frac{a_{1R}+\omega^2 a_{2R}+\omega \,a_{3R}}{\sqrt{3}} , \quad
 a_{3}  = \frac{a_{1R}+\omega \,a_{2R}+\omega^2 a_{3R}}{\sqrt{3}} .
 \label{eq:Ua2}
 \end{align}
Now, in the $S$ diagonal basis the product rules of two triplets $(a_{R1},a_{R2},a_{R3})$ and $(b_{R1},b_{R2},b_{R3})$ according to ${\bf 3}\otimes{\bf 3}={\bf 3}_{s}\oplus{\bf
3}_{a}\oplus{\bf 1}\oplus{\bf 1}'\oplus{\bf 1}''$ are as follows
 \begin{eqnarray}
  (a_R\otimes b_R)_{{\bf 3}_{\rm s}} &=& (a_{2R}\,b_{3R}+a_{3R}\,b_{2R}, \,a_{3R}\,b_{1R}+a_{1R}\,b_{3R}, \,a_{1R}\,b_{2R}+a_{2R}\,b_{1R})~,\nonumber\\
  (a_R\otimes b_R)_{{\bf 3}_{\rm a}} &=& (a_{2R}\,b_{3R}-a_{3R}\,b_{2R}, \,a_{3R}\,b_{1R}-a_{1R}\,b_{3R}, \,a_{1R}\,b_{2R}-a_{2R}\,b_{1R})~,\nonumber\\
  (a_R\otimes b_R)_{{\bf 1}} &=& a_{1R}\,b_{1R}+a_{2R}\,b_{2R}+a_{3R}\,b_{3R}~,\nonumber\\
  (a_R\otimes b_R)_{{\bf 1}'} &=& a_{1R}\,b_{1R}+\omega^{2} a_{2R}\,b_{2R}+\omega\, a_{3R}\,b_{3R}~,\nonumber\\
  (a_R\otimes b_R)_{{\bf 1}''} &=& a_{1R}\,b_{1R}+\omega\,a_{2R}\,b_{2R}+\omega^{2} a_{3R}\,b_{3R}~.
 \end{eqnarray}

\section{}
\subsection{Vacuum configuration for the driving fields}
 \label{driving}
From the vanishing of the F-terms associated to the flavons, the vacuum configuration of the driving fields $\Phi^{T}_{0},\Phi^{S}_{0},\Theta_{0},\Psi_{0}$ are determined by
 \begin{eqnarray}
  \frac{\partial W_{v}}{\partial\Phi_{T1}}&=&\frac{2\tilde{g}}{\sqrt{3}}\left(2\Phi_{T1}\Phi^{T}_{01}-\Phi_{T2}\Phi^{T}_{03}-\Phi_{T3}\Phi^{T}_{02}\right)+\tilde{\mu}\Phi^{T}_{01}=0\,,\nonumber\\
  \frac{\partial W_{v}}{\partial\Phi_{T2}}&=&\frac{2\tilde{g}}{\sqrt{3}}\left(2\Phi_{T2}\Phi^{T}_{02}-\Phi_{T3}\Phi^{T}_{01}-\Phi_{T1}\Phi^{T}_{03}\right)+\tilde{\mu}\Phi^{T}_{03}=0\,,\nonumber\\
  \frac{\partial W_{v}}{\partial\Phi_{T3}}&=&\frac{2\tilde{g}}{\sqrt{3}}\left(2\Phi_{T3}\Phi^{T}_{03}-\Phi_{T2}\Phi^{T}_{01}-\Phi_{T1}\Phi^{T}_{02}\right)+\tilde{\mu}\Phi^{T}_{02}=0\,,
 \end{eqnarray}
 \begin{eqnarray}
  \frac{\partial W_{v}}{\partial\Phi_{S1}}&=&\frac{2g_{1}}{\sqrt{3}}\left(2\Phi_{S1}\Phi^{S}_{01}-\Phi_{S2}\Phi^{S}_{03}-\Phi_{S3}\Phi^{S}_{02}\right)+g_{2}\Phi^{S}_{01}\tilde{\Theta}+2g_{3}\Phi_{S1}\Theta_{0}=0\,,\nonumber\\
  \frac{\partial W_{v}}{\partial\Phi_{S2}}&=&\frac{2g_{1}}{\sqrt{3}}\left(2\Phi_{S2}\Phi^{S}_{02}-\Phi_{S3}\Phi^{S}_{01}-\Phi_{S1}\Phi^{S}_{03}\right)+g_{2}\Phi^{S}_{03}\tilde{\Theta}+2g_{3}\Phi_{S3}\Theta_{0}=0\,,\nonumber\\
  \frac{\partial W_{v}}{\partial\Phi_{S3}}&=&\frac{2g_{1}}{\sqrt{3}}\left(2\Phi_{S3}\Phi^{S}_{03}-\Phi_{S1}\Phi^{S}_{02}-\Phi_{S2}\Phi^{S}_{01}\right)+g_{2}\Phi^{S}_{02}\tilde{\Theta}+2g_{3}\Phi_{S2}\Theta_{0}=0\,,
 \end{eqnarray}
 \begin{eqnarray}
  \frac{\partial W_{v}}{\partial\Theta}&=&\Theta_{0}\left(2g_{4}\Theta+g_{5}\tilde{\Theta}\right)=0\,,\nonumber\\
  \frac{\partial W_{v}}{\partial\tilde{\Theta}}&=&\Theta_{0}\left(g_{5}\Theta+2g_{6}\tilde{\Theta}\right)+g_{2}\left(\Phi_{S1}\Phi^{S}_{01}+\Phi_{S2}\Phi^{S}_{03}+\Phi_{S3}\Phi^{S}_{02}\right)=0\,,\nonumber\\
  \frac{\partial W_{v}}{\partial\Psi}&=&g_{7}\Psi_{0}\tilde{\Psi}=0\,,\nonumber\\
  \frac{\partial W_{v}}{\partial\tilde{\Psi}}&=&g_{7}\Psi_{0}\Psi=0\,.
  \label{drivingF-term}
 \end{eqnarray}
From this set of ten equations, we obtain
 \begin{eqnarray}
\langle\Phi^{T}_{0}\rangle=(0,0,0)\,,\qquad\langle\Phi^{S}_{0}\rangle=(0,0,0)\,,\qquad\langle\Theta_{0}\rangle=0\,,\qquad\langle\Psi_{0}\rangle=0\,,
 \label{Drdirection1}
 \end{eqnarray}
which are valid to all orders.

\subsection{Correction to the vacuum configuration}
 \label{highCorrect}
By keeping only the first order in the expansion, the minimization equations become
 \begin{eqnarray}
  \frac{2\,\tilde{g}\,\delta v_{T1}}{\sqrt{3}}+\frac{a_2\,v^{2}_{T}}{\Lambda}+\frac{a_5\,v^{2}_{\Psi}}{\Lambda}=0\,,\qquad\delta v_{T2}=0\,,\qquad\delta v_{T3}=0\,,
 \end{eqnarray}
 \begin{eqnarray}
  &&\frac{2\sqrt{3}\,g_{1}}{3}\left(2\,\delta v_{S1}-\delta v_{S2}-\delta v_{S3}\right)+g_{2}\,\delta\tilde{\Theta}+p_{1}\,v_{S}=0\,,\nonumber\\
  &&\frac{2\sqrt{3}\,g_{1}}{3}\left(2\,\delta v_{S_{2}}-\delta v_{S_{1}}-\delta v_{S_{3}}\right)+g_{2}\,\delta\tilde{\Theta}+p_{2}\,v_{S}=0\,,\nonumber\\
  &&\frac{2\sqrt{3}\,g_{1}}{3}\left(2\,\delta v_{S_{3}}-\delta v_{S_{1}}-\delta v_{S_{2}}\right)+g_{2}\,\delta\tilde{\Theta}+p_{3}\,v_{S}=0\,,
 \end{eqnarray}
 \begin{eqnarray}
  2\,g_{3}\left(\delta v_{S_{1}}+\delta v_{S_{2}}+\delta v_{S_{3}}\right)+\left(2\,g_{4}\,\delta\Theta+g_{5}\,\delta\tilde{\Theta}\right)\sqrt{-\frac{3g_{3}}{g_{4}}}=0\,,
 \end{eqnarray}
 \begin{eqnarray}
  2\,g_{7}\,v_{\Psi}\,\delta v_{\Psi}+d_1\frac{2}{\sqrt{3}}\frac{v^{3}_{T}}{\Lambda}=0\,,
 \end{eqnarray}
where $p_{1}=\frac{v_{T}}{\Lambda}\left\{3\,b_3+2\,b_{7}\sqrt{-\frac{g_{3}}{g_{4}}}-\frac{3\,b_{10}\,g_{3}}{g_{4}}\right\}$, $p_{2}=\frac{v_{T}}{\Lambda}\left\{3\,b_5-\left(i\sqrt{3}\,b_{6}+b_{7}\right)\sqrt{-\frac{g_{3}}{g_{4}}}\right\}$ and $p_{3}=\frac{v_{T}}{\Lambda}\left\{3\,b_4+\left(i\sqrt{3}\,b_{6}-b_{7}\right)\sqrt{-\frac{g_{3}}{g_{4}}}\right\}$.
These equations can be solved by
 \begin{eqnarray}
 \delta v_{T1}&=&-\frac{a_{2}v^{2}_{T}+a_{5}v^{2}_{\Psi}}{\Lambda}\frac{\sqrt{3}}{2g}\,,\qquad \qquad\qquad\qquad\qquad\delta v_{T2}=\delta v_{T3}=0\,,\nonumber\\
  \delta\tilde{\Theta}&=&-\frac{p_{1}+p_{2}+p_{3}}{3g_{2}}\,v_{S}\,,\qquad\qquad\qquad\qquad\qquad\qquad \delta\Theta=0\,,\nonumber\\
  \delta v_{S_{1}}&=&(g'_{2}+g'_{5})\frac{p_{1}+p_{2}+p_{3}}{9g_{2}}\,v_{S}-\frac{\sqrt{3}}{6g_{1}}p_{1}\,v_{S}\,,\nonumber\\
  \delta v_{S_{2}}&=&(g'_{2}+g'_{5})\frac{p_{1}+p_{2}+p_{3}}{9g_{2}}\,v_{S}-\frac{\sqrt{3}}{6g_{1}}p_{2}\,v_{S}\,,\nonumber\\
  \delta v_{S_{3}}&=&(g'_{2}+g'_{5})\frac{p_{1}+p_{2}+p_{3}}{9g_{2}}\,v_{S}-\frac{\sqrt{3}}{6g_{1}}p_{3}\,v_{S}\,,\nonumber\\
  \delta v_{\Psi}&=&\frac{d_1}{g_7\sqrt{3}}\frac{v_{T}}{v_{\Psi}}\frac{v^2_{T}}{\Lambda}\,,
 \end{eqnarray}
in which $g'_{2}=\frac{\sqrt{3}}{2}\frac{g_{2}}{g_{1}}$ and $g'_{5}=g_{5}\sqrt{\frac{-3}{4g_{3}\,g_{4}}}$.

\section{Mixing between Axion and meson}
 \label{ECL}

The mass terms reads
 \begin{eqnarray}
  {\cal L}_{\rm mass} &=& \mu m_{u}\Big\{\frac{f^{2}_{\pi}}{2(1+z+w)F^{2}_{A}}a^{2}+\frac{1+z}{2z}\pi^{2}_{0}+\frac{w+4z+zw}{6zw}\eta^{2}\nonumber\\
  &-&\frac{1-z}{2\sqrt{3}z}\pi_{0}\eta+\left(\frac{z+w}{zw}\right)\bar{K}^{0}K^{0}+\frac{1+w}{w}K^{+}\bar{K}^{-}+\frac{1+z}{z}\pi^{+}\pi^{-}\Big\}\,.
  \label{neut-A}
 \end{eqnarray}
As for the axion-photon coupling, both the $\pi^{0}$ and $\eta$ couple to photons through triangle anomalies.
However, from Eq.~(\ref{neut-A}) we see that there are no mixings with the axion and the heavy $\pi^0$ and $\eta$.
We explicitly show the mass squared terms in Eq.~(\ref{neut-A}) and the boson-photon-photon couplings $G_{a\gamma\gamma}, G_{\pi\gamma\gamma}$, and $G_{\eta\gamma\gamma}$ for the axion, $\pi^{0}$ and $\eta$, respectively:
 \begin{eqnarray}
  \frac{1}{2}{\left(\begin{array}{ccc}
 a &  \pi^{0} &  \eta
 \end{array}\right)}{\cal M}^2{\left(\begin{array}{c}
 a\\
 \pi^{0} \\
 \eta
 \end{array}\right)}+\frac{1}{4}{\left(\begin{array}{ccc}
 a &  \pi^{0} &  \eta
 \end{array}\right)}{\left(\begin{array}{c}
 G_{a\gamma\gamma}\\
 G_{\pi\gamma\gamma} \\
 G_{\eta\gamma\gamma}
 \end{array}\right)}F\tilde{F}
  \label{neut-B}
 \end{eqnarray}
where
 \begin{eqnarray}
 {\cal M}^2={\left(\begin{array}{ccc}
 \mu m_{u}\frac{f^{2}_{\pi}}{F^{2}_{A}(1+z+w)} & 0 & 0 \\
 0 &  \mu m_{u}\frac{1+z}{z} &  \mu m_{u}\frac{z-1}{\sqrt{3}z} \\
 0 &  \mu m_{u}\frac{z-1}{\sqrt{3}z} &  \mu m_{u}\frac{w+4z+zw}{3zw}
 \end{array}\right)}\,.
  \label{neut-C}
 \end{eqnarray}
Diagonalization of the mass squared matrix ${\cal M}^2$ in a basis $a-\pi^{0}-\eta$ basis, one can find the physical masses for the axion $a$, $\pi^{0}$, and $\eta$. And, the physical masses for $\pi^0$ and $K^0$ mesons as well as the electromagnetic contributions to the physical $\pi^{\pm}$ and $K^{\pm}$ mesons are expressed as
 \begin{eqnarray}
 (m^{2}_{\pi^0})_{\rm phys}&=&2\mu m_{u}\left(\frac{z+w+zw-\sqrt{(z+w+zw)^2-3zw(1+z+w)}}{3zw}\right)\,,\nonumber\\
 \nonumber\\
 (m^{2}_{K^0})_{\rm phys}&=& \mu m_{u}\left(\frac{1}{z}+\frac{1}{w}\right)\,,~~\qquad(m^{2}_{K^\pm}-m^{2}_{\pi^\pm})_{\rm phys}=\mu m_{u}\left(\frac{1}{w}-\frac{1}{z}\right)\,.
\label{physMeson}
 \end{eqnarray}

\acknowledgments{ I thank prof. E. J. Chun and my good friend Chang Sub for useful comments.
}



\begin{thebibliography}{99}
\def\plb#1#2#3{Phys.\ Lett.\       {\bf B#1}, (#3) #2}
\def\npb#1#2#3{Nucl.\ Phys.\       {\bf B#1}, (#3) #2}
\def\prd#1#2#3{Phys.\ Rev.\        {\bf D#1}, (#3) #2}
\def\prl#1#2#3{Phys.\ Rev.\ Lett.\ {\bf #1},  (#3) #2}
\def\mpl#1#2#3{Mod.\ Phys.\ Lett.\ {\bf A#1}, (#3) #2}
\def\rep#1#2#3{Phys.\ Rep.\        {\bf #1},  (#3) #2}
\def\sci#1#2#3{Science             {\bf #1},  (#3) #2}
\def\astro#1#2#3{Astrophys.\ J.\   {\bf #1},  (#3) #2}
\def\epj#1#2#3{Eur.\ Phys.\ J.  {\bf C#1},  (#3) #2}
\def\jhep#1#2#3{JHEP               {\bf #1},  (#3) #2}
\def\jpg#1#2#3{J.\ Phys.\        {\bf G#1},  (#3) #2}
\def\ijmp#1#2#3{Int.\ J.\ Mod.\ Phys.\ {\bf #1},  (#3) #2}
\def\ptp#1#2#3{Prog.\ Theor.\ Phys.\ {\bf #1},  (#3) #2}

\bibitem{Minkowski:1977sc}
  P.~Minkowski,
  Phys.\ Lett.\ B {\bf 67}, 421 (1977).  

\bibitem{Froggatt:1978nt}
  C.~D.~Froggatt and H.~B.~Nielsen,
  Nucl.\ Phys.\ B {\bf 147}, 277 (1979).  

\bibitem{Xing:2014sja}
  Z.~z.~Xing,
  Int.\ J.\ Mod.\ Phys.\ A {\bf 29}, 1430067 (2014)  [arXiv:1411.2713 [hep-ph]].  

\bibitem{Peccei-Quinn}
  R.~D.~Peccei and H.~R.~Quinn,
  Phys.\ Rev.\ Lett.\  {\bf 38}, 1440 (1977).  

\bibitem{axion}
R.~D.~Peccei and H.~R.~Quinn,
 Phys.\ Rev.\ D {\bf 16}, 1791 (1977).  
S.~Weinberg,
  Phys.\ Rev.\ Lett.\  {\bf 40}, 223 (1978);  
F.~Wilczek,
Phys.\ Rev.\ Lett.\  {\bf 40}, 279 (1978).  

\bibitem{Beringer:1900zz}
  J.~Beringer {\it et al.}  [Particle Data Group Collaboration],
  Phys.\ Rev.\ D {\bf 86}, 010001 (2012).  

\bibitem{nonAbelian}
K.~S.~Babu, E.~Ma and J.~W.~F.~Valle,
  Phys.\ Lett.\ B {\bf 552}, 207 (2003)
  [arXiv:hep-ph/0206292];
M.~Hirsch, J.~C.~Romao, S.~Skadhauge, J.~W.~F.~Valle and A.~Villanova del Moral,
  arXiv:hep-ph/0312244;
  Phys.\ Rev.\  D {\bf 69}, 093006 (2004)
  [arXiv:hep-ph/0312265];
A.~Zee,
  Phys.\ Lett.\ B {\bf 630}, 58 (2005)
  [arXiv:hep-ph/0508278];
X.~G.~He, Y.~Y.~Keum and R.~R.~Volkas,
  JHEP {\bf 0604}, 039 (2006)
  [arXiv:hep-ph/0601001];
C.~Hagedorn, M.~Lindner and R.~N.~Mohapatra,
  JHEP {\bf 0606}, 042 (2006)
  [arXiv:hep-ph/0602244];
L.~Lavoura and H.~Kuhbock,
  Mod.\ Phys.\ Lett.\  A {\bf 22}, 181 (2007)
  [arXiv:hep-ph/0610050];
S.~F.~King and M.~Malinsky,
  Phys.\ Lett.\  B {\bf 645}, 351 (2007)
  [arXiv:hep-ph/0610250];
S.~Morisi, M.~Picariello and E.~Torrente-Lujan,
  Phys.\ Rev.\  D {\bf 75}, 075015 (2007)
  [arXiv:hep-ph/0702034];
M.~Honda and M.~Tanimoto,
  Prog.\ Theor.\ Phys.\  {\bf 119}, 583 (2008)
  [arXiv:0801.0181 [hep-ph]];
C.~S.~Lam,
  Phys.\ Rev.\ Lett.\  {\bf 101}, 121602 (2008)
  [arXiv:0804.2622 [hep-ph]];
M.~C.~Chen and S.~F.~King,
  JHEP {\bf 0906}, 072 (2009)  [arXiv:0903.0125 [hep-ph]].  

\bibitem{Altarelli:2010gt}
  G.~Altarelli and F.~Feruglio,
  Rev.\ Mod.\ Phys.\  {\bf 82}, 2701 (2010)  [arXiv:1002.0211 [hep-ph]].  

\bibitem{Ahn:2012cg}
  Y.~H.~Ahn, S.~Baek and P.~Gondolo,
  Phys.\ Rev.\ D {\bf 86}, 053004 (2012)  [arXiv:1207.1229 [hep-ph]];  
  Y.~H.~Ahn and S.~K.~Kang,
  Phys.\ Rev.\ D {\bf 86}, 093003 (2012);  
  Y.~H.~Ahn, S.~K.~Kang and C.~S.~Kim,
  Phys.\ Rev.\ D {\bf 87}, no. 11, 113012 (2013)  [arXiv:1304.0921 [hep-ph]];  
  M.~D.~Campos, A.~E.~Carcamo Hernandez, S.~Kovalenko, I.~Schmidt and E.~Schumacher,
  Phys.\ Rev.\ D {\bf 90}, 016006 (2014)  [arXiv:1403.2525 [hep-ph]];  
  A.~E.~Carcamo Hernandez, I.~de Medeiros Varzielas, S.~G.~Kovalenko, H.~Pas and I.~Schmidt,
  Phys.\ Rev.\ D {\bf 88}, no. 7, 076014 (2013)  [arXiv:1307.6499 [hep-ph]].


\bibitem{extraU(1)}
  E.~J.~Chun and A.~Lukas,
  Phys.\ Lett.\ B {\bf 387}, 99 (1996)  [hep-ph/9605377];
  K.~S.~Babu, I.~Gogoladze and K.~Wang,
  Nucl.\ Phys.\ B {\bf 660}, 322 (2003)  [hep-ph/0212245].  

\bibitem{Krauss:1988zc}
  L.~M.~Krauss and F.~Wilczek,
  Phys.\ Rev.\ Lett.\  {\bf 62}, 1221 (1989).

\bibitem{Ma:2001dn}
  E.~Ma and G.~Rajasekaran,
  Phys.\ Rev.\  D {\bf 64}, 113012 (2001)
  [arXiv:hep-ph/0106291].

\bibitem{mutau}
 T. Fukuyama and H. Nishiura, arXiv:hep-ph/9702253;
 R. N. Mohapatra and S. Nussinov, Phys. Rev. {\bf D 60}, 013002 (1999);
 E. Ma and M. Raidal, Phys. Rev. Lett. {\bf 87}, 011802 (2001);
 C. S. Lam, [arXiv:hep-ph/0104116]; T. Kitabayashi and M. Yasue, Phys.Rev. D {\bf 67} 015006 (2003);
 W. Grimus and L. Lavoura, arXiv:hep-ph/0305046; 0309050;W. Grimus and L. Lavoura, Phys.\ Lett.\ B {\bf 572}, 189 (2003);
 Y. Koide, Phys.Rev. D {\bf 69}, 093001 (2004); A. Ghosal, hep-ph/0304090;
 W.~Grimus and L.~Lavoura, J.\ Phys.\ G {\bf 30}, 73 (2004);
 R.~N.~Mohapatra and W.~Rodejohann,
  Phys.\ Rev.\ D {\bf 72}, 053001 (2005) [hep-ph/0507312];
 Y.~H. Ahn, Sin Kyu Kang, C. S. Kim, Jake Lee, arXiv:hep-ph/0602160;
 Y.~H.~Ahn,  S.~K.~Kang, C.~S.~Kim and J.~Lee, Phys.\ Rev.\  D {\bf 75}, 013012 (2007).

\bibitem{Barr:1982uj}
  S.~M.~Barr, X.~C.~Gao and D.~Reiss,
  Phys.\ Rev.\ D {\bf 26}, 2176 (1982);
K.~Choi and J.~E.~Kim,
  Phys.\ Rev.\ Lett.\  {\bf 55}, 2637 (1985);
S.~M.~Barr, K.~Choi and J.~E.~Kim,
  Nucl.\ Phys.\ B {\bf 283}, 591 (1987).  

\bibitem{Sikivie:1982qv}
  P.~Sikivie,
  Phys.\ Rev.\ Lett.\  {\bf 48}, 1156 (1982).

\bibitem{PDG}
  J.~Beringer {\it et al.}  [Particle Data Group Collaboration],
  Phys.\ Rev.\ D {\bf 86}, 010001 (2012).  

\bibitem{Forero:2014bxa}
  D.~V.~Forero, M.~Tortola and J.~W.~F.~Valle,
  arXiv:1405.7540 [hep-ph].  

\bibitem{An:2012eh}
  F.~P.~An {\it et al.}  [DAYA-BAY Collaboration],
  Phys.\ Rev.\ Lett.\  {\bf 108}, 171803 (2012)  [arXiv:1203.1669 [hep-ex]].

\bibitem{Ahn:2012nd}
  J.~K.~Ahn {\it et al.}  [RENO Collaboration],
  Phys.\ Rev.\ Lett.\  {\bf 108}, 191802 (2012)  [arXiv:1204.0626 [hep-ex]].

\bibitem{Other}
 K.~Abe {\it et al.} [T2K Collaboration],
 Phys.\ Rev.\ Lett.\ \ {\bf 107}, 041801  (2011)  [arXiv:1106.2822 [hep-ex]];
 see also: T. Nakaya [for the T2K Collaboration], talk at the Neutrino 2012 conference, http://neu2012.kek.jp/;
 P.~Adamson {\it et al.} [MINOS Collaboration],
 Phys.\ Rev.\ Lett.\ \ {\bf 107}, 181802  (2011)  [arXiv:1108.0015 [hep-ex]];
 Y.~Abe {\it et al.}  [Double Chooz Collaboration],
 Phys.\ Rev.\ D {\bf 86}, 052008 (2012)  [arXiv:1207.6632 [hep-ex]].

\bibitem{Capozzi:2013csa}
  F.~Capozzi, G.~L.~Fogli, E.~Lisi, A.~Marrone, D.~Montanino and A.~Palazzo,
  Phys.\ Rev.\ D {\bf 89}, 093018 (2014)
  [arXiv:1312.2878 [hep-ph]].

\bibitem{GonzalezGarcia:2012sz}
  M.~C.~Gonzalez-Garcia, M.~Maltoni, J.~Salvado and T.~Schwetz,
JHEP {\bf 1212}, 123 (2012)  [arXiv:1209.3023 [hep-ph]], and 2014 update at www.nu-fit.org.  


\bibitem{Harrison:2002er}
  P.~F.~Harrison, D.~H.~Perkins and W.~G.~Scott,
  Phys.\ Lett.\  B {\bf 530}, 167 (2002)
  [arXiv:hep-ph/0202074].

\bibitem{Democra}
  J.~T.~Goldman and G.~J.~Stephenson,
  Phys.\ Rev.\ D {\bf 24}, 236 (1981);  
Y.~Koide,
  Phys.\ Rev.\ Lett.\  {\bf 47}, 1241 (1981);  
  Phys.\ Rev.\ D {\bf 28}, 252 (1983);  
  L.~Lavoura,
  Phys.\ Lett.\ B {\bf 228}, 245 (1989).  

\bibitem{Harari:1978yi}
  H.~Harari, H.~Haut and J.~Weyers,
  Phys.\ Lett.\ B {\bf 78}, 459 (1978).  


\bibitem{Ahn:2014zja}
  Y.~H.~Ahn and P.~Gondolo,
  Phys.\ Rev.\ D {\bf 91}, no. 1, 013007 (2015)  [arXiv:1402.0150 [hep-ph]].  

\bibitem{Altarelli:2005yx}
  G.~Altarelli and F.~Feruglio,
  Nucl.\ Phys.\  B {\bf 741}, 215 (2006)
  [arXiv:hep-ph/0512103].

\bibitem{Altarelli:2005yp}
  G.~Altarelli and F.~Feruglio, Nucl.\ Phys.\  B {\bf 720}, 64 (2005) [arXiv:hep-ph/0504165].

\bibitem{Luty:2005sn}
  M.~A.~Luty,
  hep-th/0509029.  

\bibitem{Ahn:2011i}
  Y.~H.~Ahn, H.~-Y.~Cheng and S.~Oh,
  Phys.\ Rev.\ D {\bf 84}, 113007 (2011)  [arXiv:1107.4549 [hep-ph]].  

\bibitem{sumrule}
J.~Barry and W.~Rodejohann,
  Nucl.\ Phys.\ B {\bf 842}, 33 (2011)  [arXiv:1007.5217 [hep-ph]];  
  L.~Dorame, D.~Meloni, S.~Morisi, E.~Peinado and J.~W.~F.~Valle,
  Nucl.\ Phys.\ B {\bf 861}, 259 (2012)  [arXiv:1111.5614 [hep-ph]];  
  S.~F.~King, A.~Merle and A.~J.~Stuart,
  JHEP {\bf 1312}, 005 (2013)  [arXiv:1307.2901 [hep-ph]].  

\bibitem{Khlopov}
  M.~Y.~.Khlopov and A.~D.~Linde,
  Phys.\ Lett.\ B {\bf 138}, 265 (1984);  
F.Balestra, G.Piragino, D.B.Pontecorvo, M.G.Sapozhnikov,
I.V.Falomkin, M.Yu.Khlopov,
Yadernaya Fizika (1984) V. 39, PP. 990-997. [English translation:
Sov.J.Nucl.Phys. (1984) V.39, PP.626-631];
M.Yu. Khlopov, Yu.L.Levitan, E.V.Sedelnikov and I.M.Sobol,
Yadernaya Fizika (1994) V. 57, PP. 1466-1470
[English translation: Phys.Atom.Nucl. (1994) V.57, PP.1393-1397].

\bibitem{review}
  M.~Fukugita and T.~Yanagida, Phys.\ Lett.\  B {\bf 174}, 45 (1986);
  G.~F.~Giudice {\it et al.}, Nucl.\ Phys.\ B {\bf 685}, 89 (2004) [arXiv:hep-ph/0310123];
  W.~Buchmuller, P.~Di Bari and M.~Plumacher, Annals Phys.\  {\bf 315}, 305 (2005)  [arXiv:hep-ph/0401240];
  A.~Pilaftsis and T.~E.~J.~Underwood, Phys.\ Rev.\  D {\bf 72}, 113001 (2005)  [arXiv:hep-ph/0506107].

\bibitem{Schwingenheuer:2012zs}
  B.~Schwingenheuer,
  Annalen Phys.\  {\bf 525}, 269 (2013)  [arXiv:1210.7432 [hep-ex]];  
L.~J.~Kaufman,
  arXiv:1305.3306 [nucl-ex].  

\bibitem{Ade:2013zuv}
  P.~A.~R.~Ade {\it et al.}  [Planck Collaboration],
arXiv:1303.5076 [astro-ph.CO].  

\bibitem{SPTC}
Z. Hou {\it et al}. 2014 ApJ {\bf 782}, 74   [arXiv:1212.6267].

\bibitem{Antusch:2005gp}
  S.~Antusch, J.~Kersten, M.~Lindner, M.~Ratz and M.~A.~Schmidt,
  JHEP {\bf 0503}, 024 (2005)  [hep-ph/0501272].  

\bibitem{dePerio:2014zna}
  P.~de Perio [ for the T2K Collaboration],
  arXiv:1405.3871 [hep-ex].  

\bibitem{Gando:2012zm}
  A.~Gando {\it et al.}  [KamLAND-Zen Collaboration],
Phys.\ Rev.\ Lett.\  {\bf 110}, no. 6, 062502 (2013)  [arXiv:1211.3863 [hep-ex]].  

\bibitem{Auger:2012ar}
  M.~Auger {\it et al.}  [EXO Collaboration],
Phys.\ Rev.\ Lett.\  {\bf 109}, 032505 (2012)  [arXiv:1205.5608 [hep-ex]].  

\bibitem{Agostini:2013mzu}
  M.~Agostini {\it et al.}  [GERDA Collaboration],
arXiv:1307.4720 [nucl-ex].  

\bibitem{KlapdorKleingrothaus:2000sn}
  H.~V.~Klapdor-Kleingrothaus, A.~Dietz, L.~Baudis, G.~Heusser, I.~V.~Krivosheina, S.~Kolb, B.~Majorovits and H.~Pas {\it et al.},
Eur.\ Phys.\ J.\ A {\bf 12}, 147 (2001)  [hep-ph/0103062].  

\bibitem{Aalseth:2004wf}
  C.~E.~Aalseth, F.~T.~Avignone, R.~L.~Brodzinski, S.~Cebrian, E.~Garcia, D.~Gonzales, W.~K.~Hensley and I.~G.~Irastorza {\it et al.},
Phys.\ Rev.\ D {\bf 70}, 078302 (2004)  [nucl-ex/0404036].  


\bibitem{Bhang:2012gn}
  H.~Bhang, R.~S.~Boiko, D.~M.~Chernyak, J.~H.~Choi, S.~Choi, F.~A.~Danevich, K.~V.~Efendiev and C.~Enss {\it et al.},
  J.\ Phys.\ Conf.\ Ser.\  {\bf 375}, 042023 (2012).  

\bibitem{Bilenky:2001rz}
  S.~M.~Bilenky, S.~Pascoli and S.~T.~Petcov,
Phys.\ Rev.\ D {\bf 64}, 053010 (2001)  [hep-ph/0102265];  
  S.~Pascoli and S.~T.~Petcov,
Phys.\ Lett.\ B {\bf 544}, 239 (2002)  [hep-ph/0205022];  
S.~Pascoli, S.~T.~Petcov and T.~Schwetz,
Nucl.\ Phys.\ B {\bf 734}, 24 (2006)  [hep-ph/0505226].  


\bibitem{KATRIN}
http://www.katrin.kit.edu/

\bibitem{beta}
C.Giunti and C.W.Kim, Fundamentals of neutrinno Physics and Astrophysics (Oxford University Press, Oxpord, Uk, 2007), ISBN 978-0-19-850871-7.

\bibitem{Wolfenstein:1983yz}
  L.~Wolfenstein,
  Phys.\ Rev.\ Lett.\  {\bf 51}, 1945 (1983).

\bibitem{ckmfitter}
  J.~Charles {\it et al.}  [CKMfitter Group],
  Eur.\ Phys.\ J.\  C {\bf 41}, 1 (2005)
  [arXiv:hep-ph/0406184],
  and updated results from  http://ckmfitter.in2p3.fr.

\bibitem{Jarlskog:1985ht}
  C.~Jarlskog,
  Phys.\ Rev.\ Lett.\  {\bf 55}, 1039 (1985);
  D.~d.~Wu,
  Phys.\ Rev.\  D {\bf 33}, 860 (1986).




\bibitem{KSVZ}
J.~E.~Kim,
Phys.\ Rev.\ Lett.\  {\bf 43}, 103 (1979);  
M.~A.~Shifman, A.~I.~Vainshtein and V.~I.~Zakharov,
  Nucl.\ Phys.\ B {\bf 166}, 493 (1980).  

\bibitem{DFSZ}
M.~Dine, W.~Fischler and M.~Srednicki,
  Phys.\ Lett.\ B {\bf 104}, 199 (1981);  
A.~R.~Zhitnitsky,
  Sov.\ J.\ Nucl.\ Phys.\  {\bf 31}, 260 (1980)  [Yad.\ Fiz.\  {\bf 31}, 497 (1980)].  

\bibitem{Kim:1986ax}
  J.~E.~Kim,
  Phys.\ Rept.\  {\bf 150}, 1 (1987).

\bibitem{Cheng:1987gp}
  H.~-Y.~Cheng,
  Phys.\ Rept.\  {\bf 158}, 1 (1988).

 \bibitem{Peccei}
Peccei, R. D., 1989, in {\it CP Violation}, edited by C. Jarlskog [Adv. Ser. Direct. High Energy Phys. (World Scientific Singapore), pp. 503-551].

 \bibitem{TwoU}
K.~S.~Choi, H.~P.~Nilles, S.~Ramos-Sanchez and P.~K.~S.~Vaudrevange,
Phys.\ Lett.\ B {\bf 675}, 381 (2009)  [arXiv:0902.3070 [hep-th]].

\bibitem{anomaly}
  J.~S.~Bell and R.~Jackiw,
  Nuovo Cim.\ A {\bf 60}, 47 (1969);  
S.~L.~Adler,
  Phys.\ Rev.\  {\bf 177}, 2426 (1969);  
W.~A.~Bardeen,
Phys.\ Rev.\  {\bf 184}, 1848 (1969).  

\bibitem{CASPEr}
D.~Budker, P.~W.~Graham, M.~Ledbetter, S.~Rajendran and A.~Sushkov,
  Phys.\ Rev.\ X {\bf 4}, 021030 (2014)  [arXiv:1306.6089 [hep-ph]];  
  Y.~V.~Stadnik and V.~V.~Flambaum,
  Phys.\ Rev.\ D {\bf 89}, no. 4, 043522 (2014)  [arXiv:1312.6667 [hep-ph]];  
  B.~M.~Roberts, Y.~V.~Stadnik, V.~A.~Dzuba, V.~V.~Flambaum, N.~Leefer and D.~Budker,
  Phys.\ Rev.\ Lett.\  {\bf 113}, 081601 (2014)  [arXiv:1404.2723 [hep-ph]].  

\bibitem{Dine:2000cj}
  M.~Dine,
  hep-ph/0011376.

\bibitem{WB}
 S. Weinberg, Harvard Univ. preprint HUTP-77, A057.

\bibitem{Asztalos:2003px}
  S.~J.~Asztalos, R.~F.~Bradley, L.~Duffy, C.~Hagmann, D.~Kinion, D.~M.~Moltz, L.~JRosenberg and P.~Sikivie {\it et al.},
  Phys.\ Rev.\ D {\bf 69}, 011101 (2004)  [astro-ph/0310042].  

\bibitem{CAPP}
Center for Axion and Precision Physics research, ${\rm http://capp.ibs.re.kr/html/capp\_en/}$.





\end{thebibliography}
\end{document}